\documentclass[11pt, lettersize]{article}
\usepackage{amsmath}
\usepackage{graphicx}
\usepackage{enumerate} 
\usepackage{natbib}
\usepackage{url} 
\usepackage{ulem}
\usepackage{algorithm}
\usepackage{algpseudocode}
\usepackage{soul}
\usepackage{longtable}
\usepackage{booktabs}
\usepackage{fullpage}

\usepackage{comment}

\usepackage[utf8]{inputenc}
\usepackage{longtable}
\usepackage{booktabs}
\usepackage{multirow}
\usepackage{bm}
\usepackage{amsmath, amsfonts, amsthm, amssymb, amsbsy, graphicx, dsfont, mathtools, enumerate, enumitem}
\usepackage{graphicx}
\usepackage[english]{babel}
\usepackage{graphicx}
\usepackage{subfigure, caption}
\usepackage{float}
\usepackage{geometry}
\usepackage{array}
\setcitestyle{authoryear,open={(},close={)}}
\usepackage[CJKbookmarks=true,
            bookmarksnumbered=true,
			bookmarksopen=true,
			colorlinks=true,
			citecolor=blue, 
			linkcolor=blue,
			anchorcolor=red,
			urlcolor=blue]{hyperref}
\usepackage[usenames]{color}

\usepackage{prettyref,soul}
\usepackage[font=small,labelfont=it]{caption}
\RequirePackage[framemethod=TikZ]{mdframed}

\newtheorem{theorem}{Theorem}
 
\newtheorem{lemma}{Lemma}
\newtheorem{proposition}{Proposition}

\theoremstyle{definition}
\newtheorem{definition}{Definition}
\newtheorem{condition}{Condition}
\newtheorem{remark}{Remark}

\newcommand{\argmin}{\mathop{\rm argmin}}

\def\Prob{\mathbb{P}}

\def\wt{\widetilde}

\def\mA{\mathbb{A}}

\def\mE{\mathbb{E}}

\def\mH{\mathbb{H}}

\def\mI{\mathbb{I}}

\def\mP{\mathbb{P}}

\def\mR{\mathbb{R}}

\def\cN{\mathcal{N}}

\def\cS{\mathcal{S}}

\def\cX{\mathcal{X}}

\def\xbi{X_{bi}}
\def\muks{\mu_k^*}
\def\betabks{\beta_{bk}^*}
\def\sigmaks{\Sigma_k^*}
\def\abis{a_{bi}^*}
\def\abisp{a_{b'i}^*}
\def\muk{\mu_k}
\def\betabk{\beta_{bk}}
\def\sigmak{\Sigma_k}
\def\abi{a_{bi}}
\def\abip{a_{b'i}}

\def\hatmubk{\hat{\mu}_{bk}}
\def\hatbetabk{\hat{\beta}_{bk}}
\def\hatsigmak{\hat{\Sigma}_k}
\def\hatabi{\hat{a}_{bi}}

\def\hatmuu{\hat{\mu}_k}
\def\hatmubu{\hat{\mu}_{bk}}
\def\hatbetabu{\hat{\beta}_{bk}}

\def\hatsigmau{\hat{\Sigma}_k}
\def\hatsigmaabis{\hat{\Sigma}_{\abis}}

\def\hatmubabis{\hat{\mu}_{b\abis}}

\def\hatmubv{\hat{\mu}_{bk'}}

\def\hatsigmav{\hat{\Sigma}_{k'}}

\def\betabu{\beta_{bk}}
\def\betabv{\beta_{bk'}}

\def\mubus{\mu_{bk}^*}
\def\mubvs{\mu_{bk'}^*}
\def\mubabis{\mu_{b\abis}^*}

\def\sigmau{\Sigma_k}
\def\sigmav{\Sigma_{k'}}
\def\snr{\mathrm{SNR}}

\newcommand{\method}{MoDaH}

\def\muabi{\mu_{\abi}}
\def\betababi{\beta_{b\abi}}
\def\betababis{\beta_{b\abi^*}^*}

\def\ideal{\zeta_{\mathrm{ideal}}}

\begin{document}

\def\spacingset#1{\renewcommand{\baselinestretch}%
{#1}\small\normalsize} \spacingset{1}


\title{MoDaH~achieves rate optimal batch correction}
\author{Yang Cao\footnote{Email: \texttt{yang.cao.yc2282@yale.edu.}}~~and~~ Zongming Ma\footnote{Email: \texttt{zongming.ma@yale.edu}.}\\
\textit{Yale University}}
\date{December 9, 2025}
\maketitle

\begin{abstract}
Batch effects pose a significant challenge in the analysis of single-cell omics data, introducing technical artifacts that confound biological signals. While various computational methods have achieved empirical success in correcting these effects, they lack the formal theoretical guarantees required to assess their reliability and generalization.
To bridge this gap, we introduce \textbf{M}ixture-M\textbf{o}del-based \textbf{Da}ta \textbf{H}armonization (MoDaH), a principled  {batch correction}
algorithm grounded in a rigorous statistical framework. 

Under a new Gaussian-mixture-model with explicit parametrization of batch effects,
we establish the minimax optimal error rates for batch correction and prove that {MoDaH} achieves this rate {by leveraging the recent theoretical advances in clustering data from anisotropic Gaussian mixtures}.
This constitutes, to the best of our knowledge, the first theoretical guarantee for batch correction.
Extensive experiments on diverse single-cell RNA-seq and spatial proteomics datasets demonstrate that MoDaH not only attains theoretical optimality but also achieves empirical performance comparable to {or even surpassing those of} state-of-the-art heuristics (e.g., Harmony, Seurat-V5, and LIGER), effectively balancing the removal of technical noise with the conservation of biological signal.

\end{abstract}

\section{Introduction}
\label{sec: intro}

{The analysis of single-cell data, whether transcriptomics or proteomics, often involves the integration of multiple datasets that have been processed in different ``batches'', e.g., under distinct biological conditions, in distinct labs, at different times, and/or by different researchers.}
This common practice, combined with inevitable variations in reagents, equipment, and experimental conditions, renders batch effects a {ubiquitous obstacle in single-cell data analysis.}
These systematic, technical variations introduce technical noise that can obscure  
the magnitude of {or even submerge} the underlying biological signals. Consequently, batch effects pose a significant challenge to the analysis and integration of complex single-cell datasets, necessitating robust correction strategies \citep{hicks2018missing, tung2017batch, luecken2019current, zhang2024recovery}.

\begin{figure}[!ht]
\centering
\includegraphics[width=\linewidth]{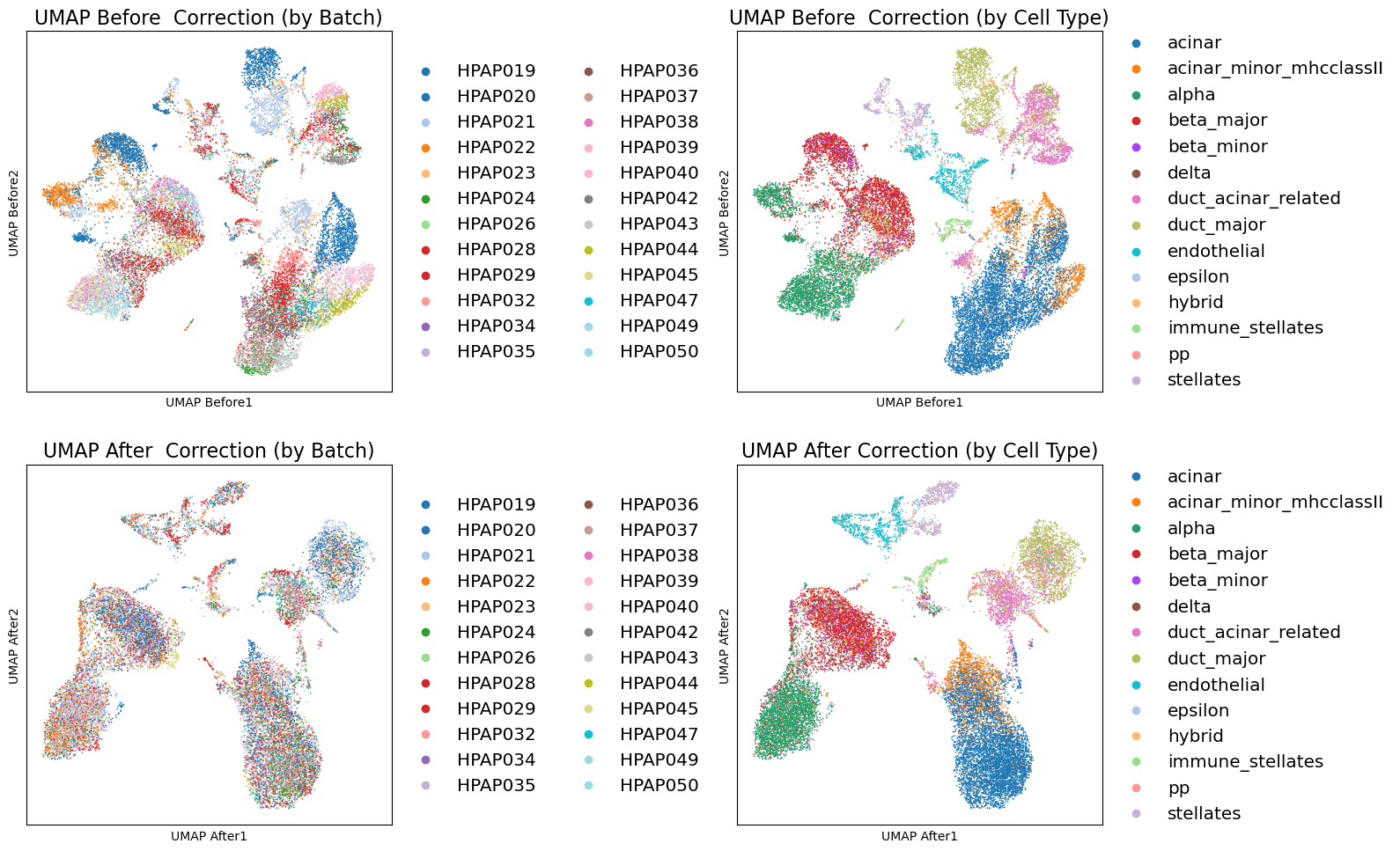}
\caption{UMAP visualizations of a T1D dataset before {(top row)} and after {(bottom row)} batch correction. 
In the left column, individual cells are colored by their source datasets (i.e.~batch indicators), and in the right column, they are colored by annotated cell types. 
Prior to correction, batch effects are clearly visible within each cell type clusters. After correction, the batches are better intermixed while the distinct cell type clusters {are conserved}.}
\label{fig: batch effect intro}
\end{figure}

To illustrate both the challenge posed by batch effects and the objective of their correction, we present here a single-cell RNA sequencing (scRNA-seq) example from a study of type 1 diabetes (T1D) \citep{fasolino2022single}. The dataset comprises samples from 11 healthy individuals, 5 individuals with T1D, and 8 individuals with no clinical
presentation of T1D but positive for beta-cell auto-antibodies (AAB+), where each individual is treated as a distinct batch. Figure \ref{fig: batch effect intro} (top row) visualizes the data {prior to batch correction}, where strong batch effects 
cause cells of the same biological type to cluster incorrectly by their origin {(e.g., the two distinct clusters of alpha cells with little biological difference in the top-right panel)}. The bottom row shows the result after applying our proposed correction method: the technical variation has been substantially reduced {while biologically meaningful differences across cell types have been preserved}, yielding better-mixed batches and biologically more coherent cell type clusters that are {more} suitable for 
downstream analysis. 
{For illustration purpose, we only present the non-rigorous UMAP visualizations of the data before and after our proposed batch correction method here. More rigorous quantitative examinations are deferred to Section \ref{sec: real world}.}

The detrimental impact of batch effects has spurred the development of a diverse array of computational methods to mitigate their influence. Approaches such as Harmony, Seurat, and LIGER have demonstrated considerable empirical success in reducing technical variation and improving data comparability across batches \citep{stuart2019comprehensive, butler2018integrating, korsunsky2019fast, welch2019single, song2020flexible, luecken2022benchmarking}.

Despite their 
empirical successes, theoretical properties of these methods are 
not 
understood, {making their generalizations to unseen datasets and new technologies ungrounded}. 
{Moreover,}
the absence {in the literature} of a formal statistical framework {for the batch correction problem} makes it difficult to 
characterize 
performance guarantees {for the methods}. 
To address {these critical gaps}, we introduce {a Gaussian mixture model with explicit parameterization of batch effects as a formal statistical framework for studying batch correction, and propose} \textbf{M}ixture-M\textbf{o}del-based \textbf{Da}ta \textbf{H}armonization (\method), a  batch correction algorithm {motivated by the statistical model}.
We provide a comprehensive theoretical analysis of the batch correction problem and prove that {\method} achieves the minimax optimal rate of correction under certain regularity assumptions—the first such guarantee in this field.
{Notably, while {\method} is motivated by a statistical model, it achieves comparable empirical performances to state-of-the-art batch correction methods such as Harmony \citep{korsunsky2019fast}, Seurat V5 \citep{seurat}, and LIGER \citep{welch2019single} on a diverse collection of single-cell transcriptomics and proteomics datasets across different species, tissue types, biological conditions, and measurement technologies.}

\subsection{Model and Problem Formulation}
\label{subsec:model}

\paragraph{Model}
To be specific, we model the batch effect correction problem in the following way. 
Consider a dataset consisting of $B$ batches, with the $b$-th batch containing $n_b$ cells, for $b \in  [B] := \{1,\dots,B\} $. 
We assume that all $n:=\sum_{b}n_b$ cells are drawn from $K$ shared
clusters. 
{For notational simplicity, we further define $n_{bk}$ to be the number of cells belong to cluster $k$ in batch $b$. For the $i$-th cell in the $b$-th batch, let its true cluster label be $\abis \in [K]:= \{1, 2, \dots, K\}$. 
If cell $i$ belongs
to cluster $k$ (i.e., $\abis = k$)}, 
we model
its (preprocessed) expression vector\footnote{{It is a common practice to apply count normalization, log1p transformation ($x\mapsto \log(1+x)$), feature screening, scaling, and PCA dimension reduction (or a subset of the foregoing preprocessing steps) as the standard preprocessing of single-cell omics data \citep{luecken2019current}.}} $\xbi\in\mR^d$ 
as a random sample from
{the Gaussian distribution:
\begin{align}
    \label{eq: model}
    \xbi\stackrel{ind}{\sim} \cN(\muks + \betabks, \sigmaks),
\end{align}
where $\muks$ and $\sigmaks$ are the cluster-specific mean} {vectors} and covariance {matrices}, and $\betabks$ is the batch effect specific to cluster $k$ in batch $b$. 
{For model identifiability, we assume $ \sum_{b}n_{bk}\betabks= 0$ for all $k\in [K]$.}

{
Notably, as we {allow cluster-specific covariance matrices},
model \eqref{eq: model} is equivalent to the following mixed effects model. 
For any couple of $(b, i)$ such that $\abis = k$, let
\begin{equation}
\label{eq: model1}
    X_{bi} = \muks + \beta_{bk} + \epsilon_{bi},
\end{equation}
where 
$\beta_{bk} \sim \cN(\betabks, \Sigma_{0,k}^*)$ and $\epsilon_{bi}\sim \cN(0, \Sigma_{1,k}^*)$ are mutually independent.
Then \eqref{eq: model1} is equivalent to \eqref{eq: model} with $\sigmaks = \Sigma_{0,k}^* + \Sigma_{1,k}^*$.
In other words, model \eqref{eq: model} (and equivalently \eqref{eq: model1}) assumes that given an unobserved cluster label that determines an overall cell state, the observed expression vector consists of three components: (i) a cell state average expression vector $\muks$, (ii) a cell state and batch co-determined average batch effect $\betabks$, and (iii) an idiosyncratic noise vector to which both the fluctuation around average expression (i.e.~$\epsilon_{bi}$) and the fluctuation around average batch effect (i.e.~$\beta_{bk} - \betabks$) contribute.
}


{While raw single-cell data are often counts, after preprocessing, especially after transformation and PCA dimension reduction, it is reasonable to model the preprocessed data as approximately following Gaussian distributions \citep{paul2007asymptotics,benaych2012singular,zhong2022empirical}.}

\paragraph{Batch correction as an inference problem}
Under the foregoing modeling framework,
we define batch correction 
as the procedure to find estimators $\hatabi$ and $\hatbetabk$ of $\abis$ and $\betabks$, respectively, for all $i\in [n_b]$,  $k\in [K]$, $b\in [B]$,  and to subsequently remove the estimated batch effects in data through
\begin{align*}
    \wt{X}_{bi} = \xbi - \hat{\beta}_{b\hatabi}.
\end{align*}
The major hurdle
in achieving this goal is the accurate estimation of
batch-specific terms $\{\betabks:k\in[K],\,b\in[B]\}$, which are latent components in the overall means and are unconventional estimands.

\paragraph{Loss function}
Let $a^*=\{a^*_{bi}, i\in [n_b], b\in [B]\}$, 
$\beta^*=\{\beta_{bk}^*: k\in [K], b\in [B]\}$, and $a$ and $\beta$ denote their estimates.
To evaluate batch correction accuracy, we consider the 
mean squared error loss when compared to \textit{oracle} batch correction (i.e.~the best correction an oracle that knows $(a^*,\beta^*)$ can do, namely $\wt{X}^*_{bi} = \xbi-\betababis$):
\begin{equation}
    \label{eq: loss function}	
\begin{aligned}
    h(a, \beta, a^*, \beta^*):=~ & \frac{1}{n}\sum_{b=1}^B\sum_{i=1}^{n_b}
    \|(\xbi - \betababi)-(\xbi-\betababis)\|^2\\
	 =~ & \frac{1}{n}\sum_{b=1}^B\sum_{i=1}^{n_b}\|\betababi-\betababis\|^2.
\end{aligned}
\end{equation}
Here and after, for any vector $x$, $\|x\|$ denotes its Euclidean norm.

\paragraph{Asymptotic regime and minimax rates}
In the ensuing theoretical analysis, we focus on the asymptotic regime where the sample sizes in all batches tend to infinity while $B$, $K$, other parameters, and the ambient dimension remain unchanged.
The assumption of fixed ambient dimension is reasonable in the current context as batch correction of single-cell data is often performed on a fixed number of principal component scores as opposed to original features.
{Leveraging the recent theoretical advances in clustering data from anisotropic Gaussian mixtures in \cite{chen2024achieving},}
we will show later in Section \ref{sec: theory} that under mild regularity conditions, the minimax lower bound on the expectation of the loss function \eqref{eq: loss function} under  model \eqref{eq: model} is
\begin{align}
\label{eq: rate}
    \exp\left(-(1+o(1))\frac{\snr^2}{8}\right)+ \exp\left(-(1+o(1))\log n\right),
\end{align}
and the loss of our method to be proposed, {\method}, is bounded from above by the same rate with high probability. Here, $\snr$ is the signal-to-noise ratio {governing the separation of the data distributions of different clusters}.
Its formal definition is deferred to Definition \ref{def: snr} below.

\medskip

The remainder of this paper is organized as follows. Section \ref{sec: method} details our proposed batch correction method, {\method}. In Section \ref{sec: theory}, we establish the theoretical underpinnings of our approach by deriving the minimax lower bound for the batch correction problem and proving that {\method} achieves this optimal rate. We then validate the empirical performance of {\method} through comprehensive experiments in Section \ref{sec: simu} using simulated data and in Section \ref{sec: real world} using a diverse set of real-world single-cell datasets.
We discuss the connections and differences between {\method} and Harmony and potential future works in Section \ref{sec: discussion}.
Proofs of theorems and additional numerical experiment details are deferred to the appendix.

\section{Methodology}
\label{sec: method}

In this section, we propose {\method}  to conduct batch correction.
To 
remove the batch effect in each cell, say cell $i$ in batch $b$, we need to estimate its cluster assignment $\abis$  with $\hatabi$, and its associated batch effect $\betabks$ 
with $\hatbetabk$. 
For any event $E$, let $\mI\{E\}$ be its indicator.
A statistically intuitive way
to perform such an estimation task is to find the constrained maximum likelihood estimator, which is equivalent to the following optimization problem
\begin{align}
    \label{eq: mle target}
    (\hatabi, \hatbetabk) = & \argmin_{\abi, \betabk, \muk, \sigmak}{\sum_{b=1}^B \sum_{k=1}^K \sum_{\substack{i: i\in [n_b]\\ a_{bi} = k}}}\left[(\xbi-\muabi-\betababi)^\top\sigmak^{-1}(\xbi-\muabi-\betababi)+\log |\sigmak |\right],\nonumber\\
    & \text{subject to }\sum_{b=1}^B\sum_{i=1}^{n_b}\betabk\mI\{\abi = k\} = 0,\quad \text{{for all $k\in [K]$.}}
\end{align}
Here, the constraint $\sum_{b, i}\betabk\mI\{\abi = k\} = 0$ is imposed to 
accord with 
the assumption $\sum_{b}\betabks n_{bk} = 0, {\forall k\in [K]}$ for model identifiability.

However, the optimization problem 
\eqref{eq: mle target} is computationally intractable. To address this issue, we propose the following EM-algorithm, {\method}, for batch effect correction based on any reasonable initialization.
See Algorithm \ref{alg:em}.

\begin{algorithm}[ht]
    \caption{{\method} for Batch Correction}\label{alg:em}
    \begin{algorithmic}
        \State \textbf{Input:} Data $\{\xbi\}_{i=1}^{n_b}$ for all $b\in [B]$, total number of clusters $K$, and a clustering method $\Phi$ for initialization.
        \State \textbf{Initialization:} Initialize cluster assignments $a^{(0)}$ based on $\Phi$.
        \While{$\exists$ $(b, i)$, such that $\abi^{(t)}\neq \abi^{(t-1)}$}
            \State[1] Update the centers and covariance matrices
            \begin{align*}
                \muk^{(t+1)} &= \frac{\sum_{b, i}\xbi\mI\{\abi^{(t)}=k\}}{\sum_{b, i}\mI\{\abi^{(t)}=k\}},\\
                \betabk^{(t+1)} &=  \frac{\sum_{i}\xbi\mI\{\abi^{(t)}=k\}}{\sum_{i}\mI\{\abi^{(t)}=k\}} - \muk^{(t+1)},\\
                \sigmak^{(t+1)} &= \frac{\sum_{b, i}\mI\{\abi^{(t)}=k\} (\xbi-\muk^{(t+1)}-\betabk^{(t+1)} ) (\xbi-\muk^{(t+1)}-\betabk^{(t+1)} )^\top}{\sum_{b, i}\mI\{\abi^{(t)}=k\}}.
            \end{align*}
            \State[2] Update the cluster assignment
            \begin{align*}
                \abi^{(t+1)} &=  \argmin_{k}  (\xbi-\muk^{(t+1)}-\betabk^{(t+1)} )^\top  (\sigmak^{(t+1)} )^{-1} (\xbi-\muk^{(t+1)}-\betabk^{(t+1)} )+\cdots\nonumber\\
                &\cdots+\log |\sigmak^{(t+1)} |.
            \end{align*}
            
        \EndWhile
        \State\textbf{Output:} After convergence in $T$ iterations, output the final estimations $\hatbetabk:=\betabk^{(T)}$, $\hatabi:= \abi^{(T)}$ and the batch effect corrected dataset $\wt{X}_{bi}$ for all $b, k, i$:
        \begin{align*}
            \wt{X}_{bi} = \xbi - \hat{\beta}_{b\hatabi}.
        \end{align*}
    \end{algorithmic}
\end{algorithm}

{In Algorithm \ref{alg:em}, we assume that the true number of clusters, $K$, is given. When it is unknown in practice, we can replace it with an estimate $\hat{K}$. Candidate estimators and robustness of Algorithm \ref{alg:em} performance with respect to 
potential misspecification of $K$ 
are discussed in Sections \ref{subsec:Khat}.}
In addition, 
Algorithm \ref{alg:em} {requires} an initial cluster assignment $a^{(0)}$ for each cell, {which can be} provided by some reasonable {but potentially crude} clustering method. 
The selection of this base method is flexible, encompassing approaches such as k-means clustering and Leiden clustering \citep{traag2019louvain}. 
{For all results on simulated and real-world datasets reported in this manuscript, we use k-means clustering as initialization for {\method}.}
Our theory in Section \ref{sec: theory} shows that, as long as the initialization $a^{(0)}$ is {reasonable}
({within} a certain proximity of $a^*$), 
{{\method}} achieves the minimax optimal rate of batch correction loss \eqref{eq: loss function} under mild regularity conditions.

\section{Theoretical Results}
\label{sec: theory}

In this section, we first introduce the 
signal-to-noise ratio in 
\eqref{eq: model} and key regularity conditions for our theory. 
With these preliminaries,
we first present
the minimax lower bound of the batch correction problem 
with loss function \eqref{eq: loss function}. 
We conclude this section with showing that Algorithm \ref{alg:em} achieves 
the rate in the minimax lower 
bound with high probability. 

{The main technical novelty in our theoretical arguments lies in the decomposition of the minimax lower bounds as the sum of a clustering error term and an estimator error term. After this decomposition, we leverage the arguments in \cite{chen2024achieving} for bounding the clustering error term from below and for proving achievability.}

\subsection{Preliminaries}
We first define the signal-to-noise ratio quantity and explain the intuition behind it.
\begin{definition}[Signal-to-Noise Ratio (SNR)]
\label{def: snr}
    Let $\mubus = \mu_k^*+\betabu^*$ for all $b$ and $k$. 
    Let
    \begin{align*}
        A_{b, k, k'} = \bigg\{x\in\mR^d &: x^\top(\sigmau^*)^{\frac{1}{2}}(\sigmav^*)^{-1}(\mubvs-\mubvs)+\frac{1}{2}x^\top\big((\sigmau^*)^{\frac{1}{2}}(\sigmav^*)^{-1}(\sigmau^*)^{\frac{1}{2}}-I_d\big)x\\
        &\le -\frac{1}{2}(\mubus-\mubvs)^\top(\sigmav^*)^{-1}(\mubus-\mubvs)+\frac{1}{2}\log|\sigmau^*|-\frac{1}{2}\log|\sigmav^*|\bigg\},
    \end{align*}
    and $\cX_{b, k, k'}=2\min_{x\in A_{b, k, k'}}\|x\|$. The signal-to-noise ratio is defined as 
    \begin{align*}
        \snr = \min_{b\in [B]}\min_{k\neq k':k, k'\in[K]}\cX_{b, k, k'}.
    \end{align*}
\end{definition}
The signal-to-noise ratio in Definition \ref{def: snr} is closely related to the quadratic discriminant analysis (QDA) problem of clustering two multivariate Gaussian distributions with different means and covariance matrices. Signal-to-noise ratios in a similar spirit were defined in \cite{chen2024achieving} for the clustering of data generated from Gaussian Mixtures.
In this paper, the SNR represents the difficulty in distinguishing whether a cell in a specific batch belongs to a specific cluster or not in the least favorable scenario. 
To be specific, the set $A_{b, k, k'}$ represents the local difficulty in distinguishing between clusters $k$ and $k'$ within a single batch $b$. 
This task can be framed as testing
$\mH_0: y\sim \cN(\mubus, \sigmau^*)$ versus $\mH_1: y\sim \cN(\mubvs, \sigmav^*)$. 
Applying a change of variables $x=(\sigmau^*)^{-\frac{1}{2}}(y-\mubus)$ transforms this into a testing problem with a standard Gaussian null hypothesis: $\mH_0: x\sim \cN(0_d, I_d)$ versus $\mH_1: x\sim \cN((\sigmau^*)^{-\frac{1}{2}}(\mubvs-\mubus), (\sigmau^*)^{-\frac{1}{2}}\sigmav^*(\sigmau^*)^{-\frac{1}{2}})$. 
The likelihood ratio test for this transformed problem is given by:
\begin{align*}
    \phi:=\mI\left\{-\frac{1}{2}x^\top x\right. &\le -\frac{1}{2}(x-(\sigmau^*)^{-\frac{1}{2}}(\mubvs-\mubus))^\top (\sigmau^*)^{\frac{1}{2}}(\sigmav^*)^{-1}(\sigmau^*)^{\frac{1}{2}}(x-(\sigmau^*)^{-\frac{1}{2}}(\mubvs-\mubus))\\
    &~~~~
    -\left.\frac{1}{2}\log|(\sigmau^*)^{-\frac{1}{2}}\sigmav^*(\sigmau^*)^{-\frac{1}{2}})|\right\}.
\end{align*}
With some algebra, one can check that the test is equivalent to checking membership of $x$ in the set $A_{b, k, k'}$, i.e. $\phi=1$ 
if and only if $x \in A_{b, k, k'}$. Therefore, the quantity $\cX_{b, k, k'} = 2\min_{x\in A_{b, k, k'}}\|x\|$ measures the difficulty of this specific hypothesis testing. 
Since the overall difficulty of batch correction is related to the most challenging within-batch cluster distinction, the SNR is defined as the minimum of these local measures in Definition \ref{def: snr}.


The following proposition connects $\snr$ in Definition \ref{def: snr} to more  interpretable parameters.
For any symmetric matrix $A$, let $\lambda_{\max}(A)$ and $\lambda_{\min}(A)$ be its largest and smallest eigenvalues, respectively.
\begin{proposition}
    \label{prop: snr}
    Let $\omega:=\min_{b, k\neq k'}\|\mubus-\mubvs\|^2$, $\lambda_{\max}:=\max_k\lambda_{\max} (\sigmau^*)$,
    and
    $\lambda_{\min}:=\min_k \lambda_{\min} (\sigmau^*)$.
    Assume that $\omega>0$ and $\lambda_{\min}>0$. 
    If
    $\omega\ge 2d \lambda_{\max} \log\frac{\lambda_{\max}}{ \lambda_{\min}}$, there holds
    \begin{align*}
        \frac{1}{3}\frac{\lambda_{\min}}{ \lambda_{\max}}\frac{\sqrt{\omega}}{\sqrt{\lambda_{\max}}}\le \snr\le 2\frac{\sqrt{\omega}}{\sqrt{\lambda_{\min}}}.
    \end{align*}
\end{proposition}
Proposition \ref{prop: snr} states that when the minimum cluster separation, $\omega$, is sufficiently large, the $\snr$ is on the order of $\sqrt{\omega}$, as long as the eigenvalues of all cluster-specific covariance matrices are bounded away from zero and infinity.
The proof of Proposition \ref{prop: snr} is provided in Section \ref{sec: proof prop snr} of the appendix.

We now switch to 
introducing
key
regularity conditions needed in this section. 
\begin{condition}[Regularity Conditions] ~
\label{cond: regularity}
    \begin{enumerate}
        \item Assume that there exists $\alpha > 0$, such that 
        $\frac{\min_{b,k}n_{bk}}{\max_{b,k}n_{bk}}\ge\alpha$.
        \item Assume that there exists $\gamma>0$, such that $\min_{b, k\neq k'}\|\betabu^*-\betabv^*\|^2\ge \gamma$.
        \item Assume that there exists $\Gamma>0$, such that $\max_{b, k}\|\betabu^*\|^2\le \Gamma$.
    \end{enumerate}
\end{condition}
The first 
part
of Condition \ref{cond: regularity} ensures that the batches and the clusters are of comparable sizes, while the second and the third 
parts are imposed
to avoid singularities. 
In other words, 
we assume that $(a^*, \beta^*)$ belongs to the following parameter space:
\begin{equation}
\label{eq:para-space}
\begin{aligned}
\cS(\alpha, \gamma, \Gamma) = \{(a, \beta):& \min_{b, k}\sum_{i\in[n_b]}\mI\{\abi = k\}\ge \alpha \max_{b, k}\sum_{i\in[n_b]}\mI\{\abi = k\}, \\
& \min_{b, k\neq k'}\|\betabu-\betabv\|^2\ge \gamma, \max_{b, k}\|\betabu^*\|^2\le \Gamma\}.    
\end{aligned}
\end{equation}

\subsection{Minimax Lower Bound of Batch Correction}

The following theorem characterizes the minimax lower bound of the batch correction problem with loss function \eqref{eq: loss function}.


\begin{theorem}
    \label{thm: lower bound}
Suppose that as $n\to\infty$, 
    $B, K, d, {1/\gamma} = O(1)$, 
    $\max_{k}\lambda_{\max}(\sigmau^*), \max_k 1/\lambda_{\min}(\sigmau^*)=O(1)$,
    and
$\snr\to\infty$. 
Then for large values of $n$ and any $\alpha\in (0,\frac{1}{2})$, there holds
    \begin{align}
    \label{eq: lower bound}
        \inf_{(a, \beta)}\sup_{(a^*, \beta^*)\in \cS(\alpha, \gamma, \Gamma)}\mE h(a, \beta, a^*, \beta^*)\ge \exp\left(-(1+o(1))\frac{\snr^2}{8}\right)+ \exp\left(-(1+o(1))\log n\right).
    \end{align}
    If $\snr=O(1)$ as $n\to\infty$, then ${\liminf_{n\to\infty}}\inf_{(a, \beta)}\sup_{(a^*, \beta^*)\in \cS(\alpha, \gamma, \Gamma)}\mE h(a, \beta, a^*, \beta^*)\ge c$ for some constant $c>0$.
\end{theorem}

There are two terms on the right side of \eqref{eq: lower bound}, resulting from the need to estimate
two parameters $(a^*, \beta^*)$ for batch correction.
The first term results from the clustering error in estimating the parameter $a^*$.
The second term is induced from the estimation error of the parameter $\beta^*$.
The detailed proof of Theorem \ref{thm: lower bound} is provided in Section \ref{sec: proof lower bound} of the appendix.
{As we have mentioned, the main novelty in the proof of Theorem \ref{thm: lower bound} lies in the arguments showing that the minimax risk of misclustering bounds that of batch correction from below.
After establishing this key intermediate result, we use the technique in the proof of Theorem 3.1 in \cite{chen2024achieving} to further bound the minimax risk of misclustering from below to obtain the first term on the right side of \eqref{eq: lower bound}. The second term on the right side of \eqref{eq: lower bound} is established from the two-point argument.}

\subsection{High Probability Error Bounds for {\method}}

In this section, we establish a high probability upper bound on the loss function \eqref{eq: loss function} for the {\method} method in Algorithm \ref{alg:em}. 
In parallel to $\omega:=\min_{b, k\neq k'}\|\mubus-\mubvs\|^2$,
we define 
$\Omega=\max_{b, k}\|\mubus\|^2$.
For any cluster assignment $a$, define
\begin{equation}
\label{eq:l}
\ell(a, a^*) = \sum_{b, i}\|\mu_{\abi}^*+\beta_{b\abi}^*-(\mu_{\abis}^*+\beta_{b\abis}^*)\|^2.    
\end{equation}

\begin{theorem}
    \label{thm: upper bound}
    Suppose that as $n\to\infty$, $B, K, d = O(1)$, $\max_{k}\lambda_{\max}(\sigmau^*), \max_k 1/\lambda_{\min}(\sigmau^*)=O(1)$, $\log \Omega = o(\omega)$, $(a^*, \beta^*)\in \cS(\alpha, \gamma, \Gamma)$ and $\snr\to\infty$. 
If $\ell(a^{(0)}, a^*) =  o(n)$ holds with probability at least $1-\eta$, then
    \begin{align}
    \label{eq: upper bound}
        h(a^{(t)}, \beta^{(t)}, a^*,\beta^*)\le 
        \max\{\Gamma, 1\}
        \exp\left(-(1+o(1))\frac{\snr^2}{8}\right)+\exp\left(-(1+o(1))\log n\right)
    \end{align}
    with probability at least $1-\eta-7n^{-1}-\exp(-\snr)$ for all iterate $t\ge\log n+1$ in Algorithm \ref{alg:em}.
\end{theorem}
\begin{remark}
    Theorem \ref{thm: upper bound} requires a decent initialization that is sufficiently close to the ground truth such that $\ell(a^{(0)}, a^*) =  o(n)$. 
    An example is Lloyd's algorithm on clustering Gaussian mixtures, whose performance has been studied in \citep{lu2016statistical, chen2024achieving}. 
    To obtain a decent initialization, Lloyd's algorithm can be applied to each batch to generate $K$ local clusters. 
    We then merge the local clusters to $K$ global cluster by matching local clusters in different batches. 
    By Proposition \ref{prop: snr}, when $\snr \to \infty$, the minimum separation $\omega := \min_{b, k \neq k'} \|\mubus - \mubvs\|^2$ diverges to infinity, whereas the batch effect sizes remain upper bounded by $\Gamma$ {which scales at a slower rate than $\omega$}. 
    Under this regime, the global centers must satisfy $\min_{k \neq k'} \|\mu_k^* - \mu_{k'}^*\|^2 \to \infty$ and so the probability of mismatching local clusters becomes negligible.
\end{remark}

For the proof of Theorem \ref{thm: upper bound}, see Section \ref{sec: proof upper bound} in the appendix.
The rate in the upper bound \eqref{eq: upper bound} matches the rate in the minimax lower bound \eqref{eq: lower bound} {when $\log\Gamma = o(\omega)$}, showing that {\method} achieves rate optimal correction of batch effects under model \eqref{eq: model} and loss function \eqref{eq: loss function}. 
The error bound in Theorem \ref{thm: upper bound} is corroborated by simulation results in Section \ref{sec: simu}.

\section{Simulation Studies}

\label{sec: simu}

This section evaluates the performance of {\method} through simulation studies. 
{Throughout this section, we use k-means clustering for initializing  {\method}.} 
First, we demonstrate that the loss function in \eqref{eq: loss function} converges to zero as the Signal-to-Noise Ratio ($\snr$) and sample size increase, corroborating the upper bounds in Theorem \ref{thm: upper bound}. 
In addition, we show that the performance of {\method} is robust with respect to the mild misspecification of the total number of clusters.
{Moreover, we propose a data-based cluster number estimator $\hat{K}$ based on Leiden clustering.}

\subsection{Simulation Settings}

We simulate synthetic data according to model \eqref{eq: model}\footnote{We refer interested readers to
Section \ref{sec: sup simu} in the appendix for additional simulation studies that examine the robustness of {\method} with respect to deviations of data distributions from model \eqref{eq: model} and the associated assumptions, such as non-Gaussian data distributions,  large numbers of clusters, and missing clusters in some batches.}.
In particular, we set the number of batches at $B=3$, the number of clusters at $K=4$, and the dimension at $d=10$. 
The total sample size for each batch is controlled by a parameter $u>0$, with $n_1 = \lfloor 1000u\rfloor$, $n_2 = \lfloor 1500u\rfloor$, and $n_3 = \lfloor 2000u\rfloor$. The number of cells for a specific cluster $k$ in batch $b$, $n_{bk}$, is determined by the proportions $\Pi_{bk}$ from the following matrix $\Pi\in\mR^{B\times K}$, 
\begin{align*}
    \Pi = \begin{bmatrix}
        0.4 & 0.3 & 0.2 & 0.1 \\
        0.1 & 0.2 & 0.3 & 0.4 \\
        0.25 & 0.25 & 0.25 & 0.25
    \end{bmatrix},
\end{align*}
such that $n_{bk} = n_b \cdot \Pi_{bk}$ before rounding and $\sum_k n_{bk} = n_b$.
The cluster means are defined as $\muks = v \cdot e_k$ for $k \in [K]$, where $v>0$ controls the separation and $\{e_k\}_{k=1}^K$ are the first $K$ standard basis vectors in $\mR^d$. The batch effects $\betabks$ are sampled i.i.d. from a $d$-dimensional standard Gaussian distribution $\cN(0_d, I_d)$. 
The covariance matrices $\sigmaks$ are generated as $A_k^\top A_k + I_d$, where each $A_k \in \mR^{d\times d}$ has entries sampled i.i.d. from $\cN(0, 1)$.

In this simulation framework, the parameter $u$ directly controls the overall sample size, while the parameter $v$ controls the separation between cluster means. As established in Proposition \ref{prop: snr}, increasing $v$ leads to an increase in the $\snr$. 
We focus on examining the performance of {\method} by analyzing the trend of its loss \eqref{eq: loss function} as a function of these two key parameters, $u$ and $v$.

See Section \ref{sec: sup simu} in the appendix for additional simulation studies with increased total number of clusters, non-Gaussian data, and missing clusters in certain batches.

\subsection{Performance of {\method}}
\label{sec: simu1}

In this section, we examine the behavior of loss \eqref{eq: loss function} of {\method} when $u$, $v$ varies. To be specific, we conduct experiments in the following {four} simulation experiments.
\begin{enumerate}
    \item Fix $v = 5$. We examine the  loss \eqref{eq: loss function} of {\method} when $\log(u)$ is taken from a sequence starting from -1 ending at 3 with a stepsize 0.2. 
    \item The same as setting 1 but fixing $v = 10$. 
    \item Fix $u = \exp(1)$. We examine the  loss \eqref{eq: loss function} of {\method} when $v$ is taken from a sequence starting from 2 ending at 20 with a stepsize 1.
    \item The same as setting 3 but fixing $u = \exp(2)$. 
\end{enumerate}

\begin{figure}[!ht]
\centering
\subfigure[$\log$ loss vs. $\log(u)$, $v=5$]{
\begin{minipage}[t]{0.45\textwidth}
\centering
\includegraphics[width=\textwidth]{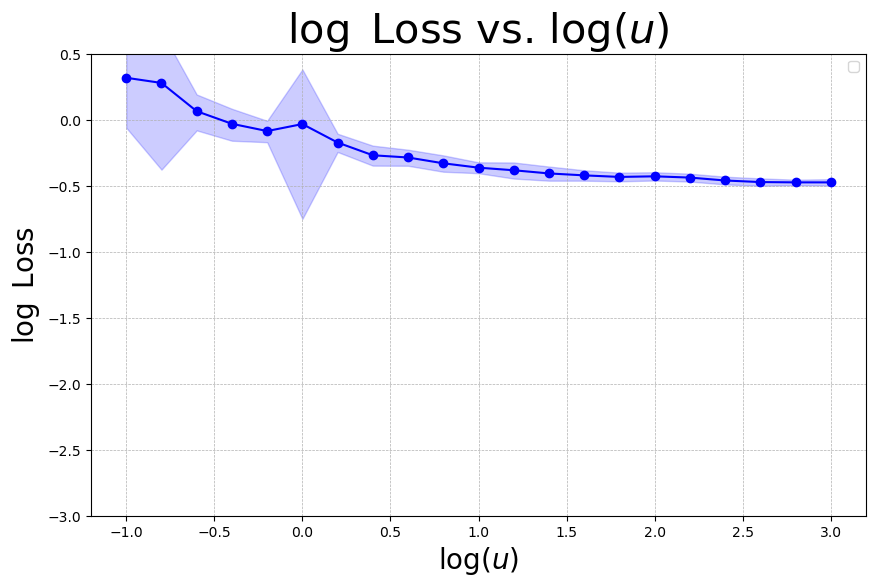}
\end{minipage}%
}%
\subfigure[$\log$ loss vs. $\log(u)$, $v=10$]{
\begin{minipage}[t]{0.45\textwidth}
\centering
\includegraphics[width=\textwidth]{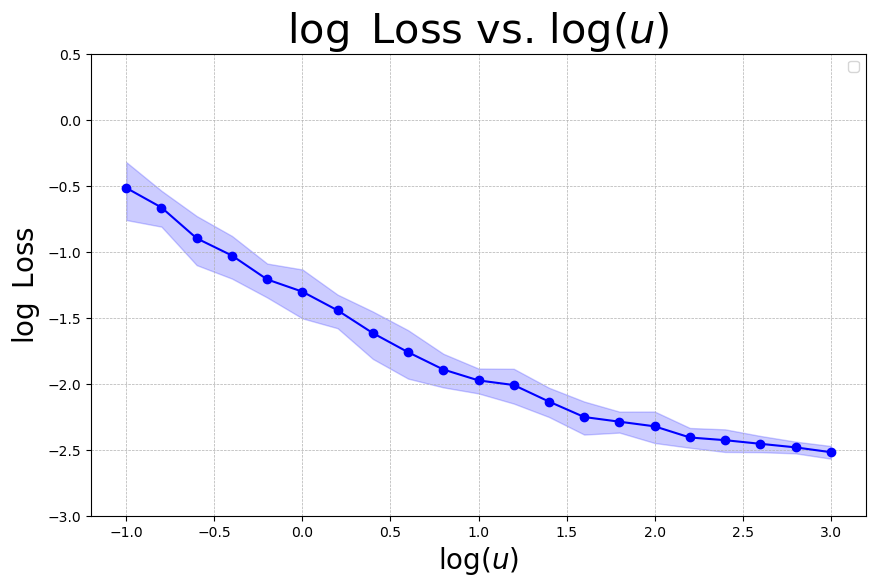}
\end{minipage}%
}%

\subfigure[$\log$ loss vs. $v$, $u=\exp(1)$]{
\begin{minipage}[t]{0.45\textwidth}
\centering
\includegraphics[width=\textwidth]{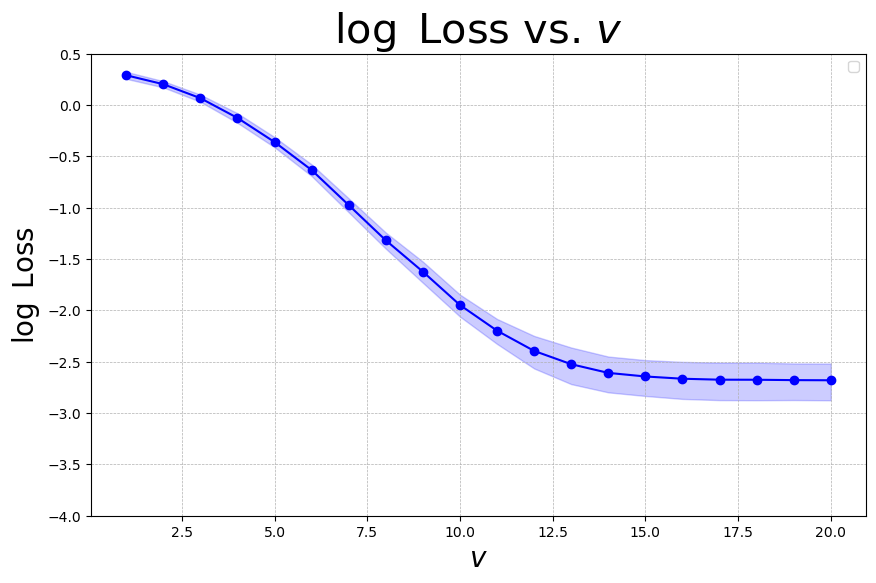}
\end{minipage}%
}%
\subfigure[$\log$ loss vs. $v$, $u=\exp(2)$]{
\begin{minipage}[t]{0.45\textwidth}
\centering
\includegraphics[width=\textwidth]{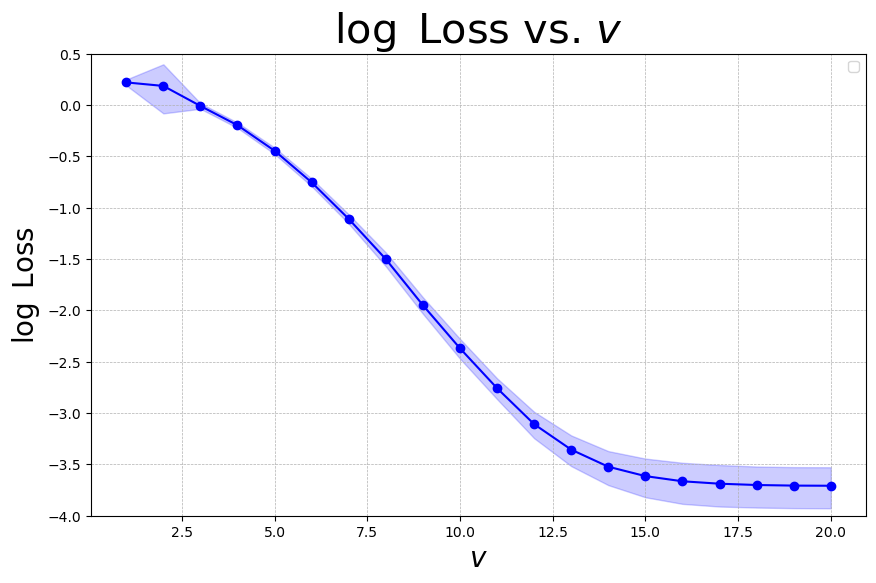}
\end{minipage}%
}%
\caption{{The average batch correction} loss of {\method} when $u$, $v$ varies. Each data point in the figures represent the logarithm of the average loss in 20 simulation instances, and the two ends of the shaded area around each point 
correspond to the logarithms of mean $\pm$ one standard deviation. 
In the top row, we plot the logarithm of the average loss against $\log u$ when $v$ is fixed at $5$ or $10$. 
In the bottom row, we plot the logarithm of the average loss against $u$ when $v$ is fixed at $\exp(1)$ or $\exp(2)$.}
\label{fig: simulation with u v changes}
\end{figure}

We summarize the results in the four simulation experiments in Figure \ref{fig: simulation with u v changes}. 
For better visualization, we plot the logarithm of the average loss \eqref{eq: loss function} over 20 simulation instances in each setting
against $\log(u)$ and $v$, respectively, 
together with the logarithms of the average loss $\pm$ one standard deviation across the $20$ instances at each parameter configuration.

When $v$ is fixed at 5 or 10, the  loss \eqref{eq: loss function} of {\method} 
decreases as $u$ increases until it reaches a plateau the height of which is lower when $v = 10$ compared to when $v=5$.
This is in alignment with the error bound
in Theorem \eqref{thm: upper bound}, which predicts that the error 
decreases with the increase of the sample size (monotone in $u$ in settings 1 and 2), until the sample size is so large that the first term in the error bound, i.e.~$\exp\left(-(1+o(1))\frac{\snr^2}{8}\right)$, dominates. 
Between settings 1 and 2, a larger $v$ corresponds to a larger SNR, and so we expect the plateau height to be lower when $v=10$.
On the other hand, when $u$ is fixed at $\exp(1)$ or $\exp(2)$, the  loss \eqref{eq: loss function} of {\method} decreases as $v$ increases until it reaches a plateau the height of which is lower when $u = \exp(2)$ compared to when $u = \exp(1)$. 
As larger $v$ 
leads to larger $\snr$, the simulation results in settings 3 and 4 also match the prediction by the error bound in Theorem \eqref{thm: upper bound}: The error decreases with $\snr$ increases, until the $\snr$ is sufficiently large such that the second term in the error bound, i.e.~$\exp\left(-(1+o(1))\log n\right)$, dominates.
Between settings 3 and 4, $u = \exp(2)$ corresponds to a larger sample size, and hence a lower plateau in the plot. 

\subsection{Robustness of {\method} under Cluster Number Misspecification}
\label{subsec:Khat}
In practice, the true number of clusters $K$ is often unknown and needs to be estimated. In this section, we show that our method can achieve {robust}
performance when the true number of clusters $K$ in Algorithm \ref{alg:em} is replaced by 
{a potentially misspecified value}. 
We first plot the performance of our method when {the input total number of clusters}
varies, and then we provide a practical estimator $\hat{K}$ for this input.

To evaluate the performance of {\method} beyond the batch  correction loss in \eqref{eq: loss function}, we employ {nine performance metrics in} the \texttt{scib-metrics} package \citep{luecken2022benchmarking}, a standardized benchmarking pipeline for single-cell data {batch correction} that has been widely adopted in the single-cell genomics community. 
{The nine metrics can be grouped into two categories:
(1) \texttt{Isolated labels}, \texttt{Leiden NMI}, \texttt{Leiden ARI}, \texttt{Sihouette label}, and \texttt{cLISI} in the
bio-conservation category, which exam the preservation of true biological heterogeneity and penalize over-correction of batch effects, and 
(2) \texttt{Sihouette batch}, \texttt{iLISI}, \texttt{KBET}, and \texttt{Graph connectivity}, 
in the batch-correction category,
which check the removal of technical variation and guard against under-correction of batch effects.}
For ease of comparison, all metrics have been rescaled to a common $[0, 1]$ range, where a score of 1 represents the best possible performance and 0 the worst. 
{The precise definitions of these metrics 
can be found} in Section \ref{sec: metrics} of the appendix.




\begin{figure}[!ht]
\centering
\subfigure[Batch correction loss]{
\begin{minipage}[t]{0.32\textwidth}
\centering
\includegraphics[width=\textwidth]{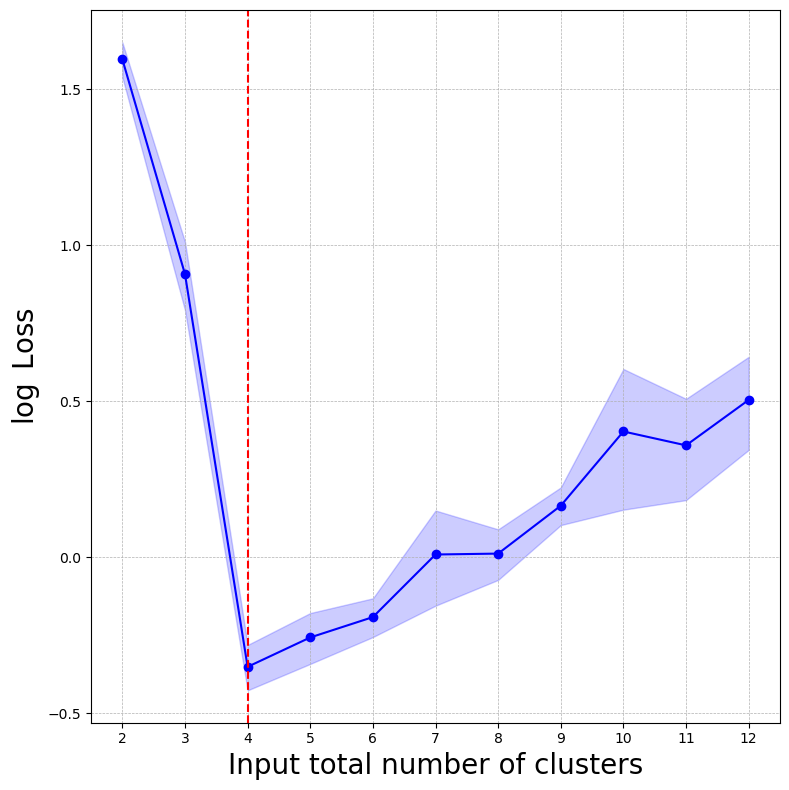}
\end{minipage}%
}%
\subfigure[Bio-conservation metrics]{
\begin{minipage}[t]{0.32\textwidth}
\centering
\includegraphics[width=\textwidth]{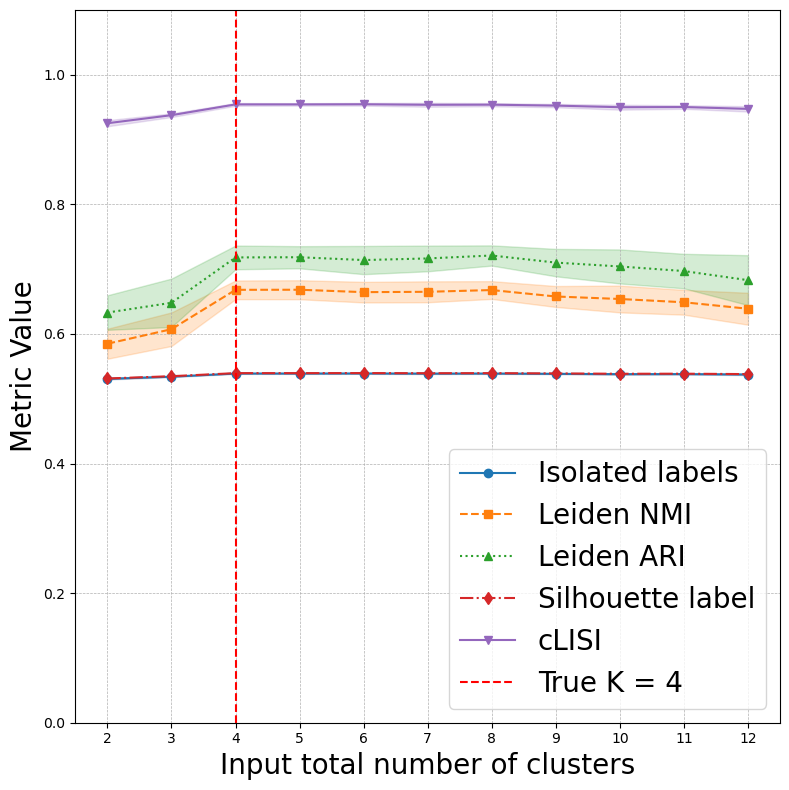}
\end{minipage}%
}%
\subfigure[Batch-correction metrics]{
\begin{minipage}[t]{0.32\textwidth}
\centering
\includegraphics[width=\textwidth]{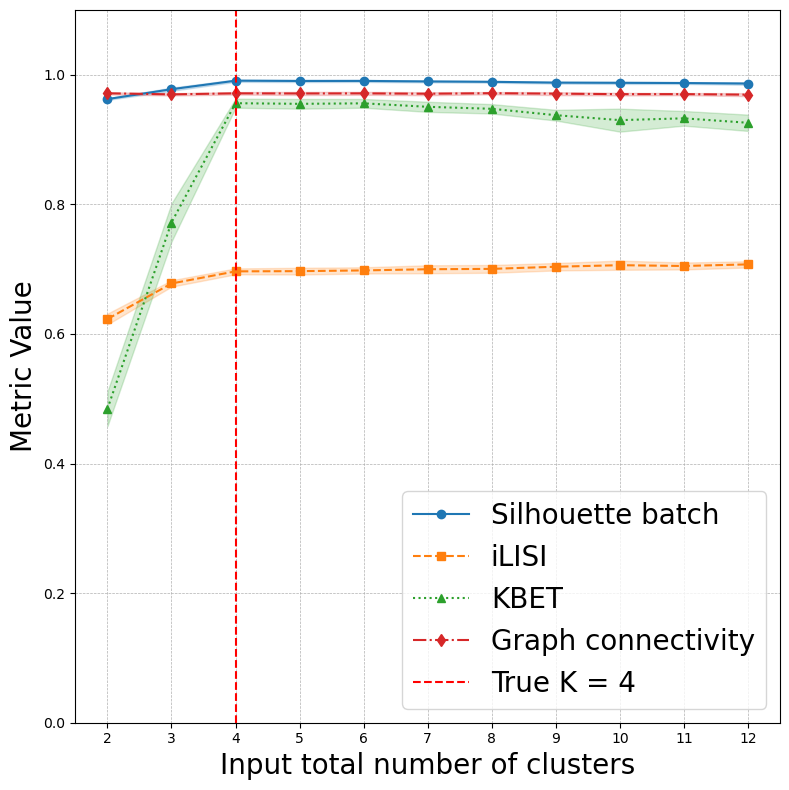}
\end{minipage}%
}%
\caption{Performance of {\method} as the input total number of clusters varies, evaluated by the batch correction loss (the left panel) and bio-conservation (the middle panel) and batch-correction metrics (the right panel). 
In the left panel, we plot the logarithm of the average loss versus the input number (with the ends of the shaded area corresponding to the logarithms of the average loss $\pm$ one standard deviation). 
In the middle and the right panels, we plot the average scores of \texttt{scib-metrics} metrics versus the input number of clusters, and the shaded area around each data point represents $\pm$ one standard deviation. 
All results are obtained from 20 simulation instances.}
\label{fig: simulation with k changes}
\end{figure}

In this simulation experiment, we fix $u = \exp(1)$ and $v = 5$.
We conduct 20 simulation instances and report the logarithm of the average  loss in \eqref{eq: loss function} as well as the average scores of the nine metrics in 20 simulation instances in Figure \ref{fig: simulation with k changes}.
The shaded area around each data point in Figure \ref{fig: simulation with k changes} indicates the average $\pm$ one standard deviation of the loss/metric.
As shown in Figure \ref{fig: simulation with k changes}, the average batch correction loss in \eqref{eq: loss function} is minimized when the 
{input}
number of clusters equals the true value 
and subsequently increases as 
{the input total cluster number}
grows. 
In addition,
all metrics
achieve their maximum value {when the total number of clusters is correctly specified}.
Notably,
these metrics remain stable (or only worsen slowly) over a sizable range of 
misspecified input values of the total number of clusters, especially when the specified value is larger than true $K$.
These results demonstrate  that {\method} is robust with respect to the batch correction loss and various bio-conservation and batch-correction metrics {under misspecification of $K$ as its input, especially when the input is larger than the truth}.

In practice, we propose to estimate  
$K$ using Leiden clustering \citep{traag2019louvain}. 
Across these simulation examples, setting the number of neighbors to $20$ when constructing the neighborhood graph and the resolution level to $0.25$ in the subsequent Leiden clustering on the graph consistently identified the correct number of clusters (i.e., $\hat{K} = K$) in all the simulated instances. 

\section{Batch Correction in Single-cell Datasets}
\label{sec: real world}

In this section, we apply {\method} to correct  batch effects in a diverse collection of real-world single-cell datasets. 
{Throughout this section, we use k-means clustering to initialize {\method}.} 
We evaluate its performance across five distinct scenarios, 
including scRNA-seq datasets {from human pancreas samples in a type-1 diabetes study (T1D) \citep{fasolino2022single}, from mouse blood samples (PBMC) \citep{han2018mapping}},
and from cell lines (Cell Line) \citep{korsunsky2019fast},
{single-cell expression profiles from spatial} proteomics datasets {from healthy human intestine samples measured by the CODEX technology (CODEX) \citep{hickey2023organization} and from human colorectal cancer samples measured by the CyCIF technology (CyCIF) \citep{lin2023multiplexed}}.
{These single-cell datasets collectively allow us to perform a comprehensive performance assessment of {\method} across different species, tissues, biological conditions, and measurement technologies.}

We benchmarked {\method} against state-of-the-art methods recommended by a recent comparative study \citep{tran2020benchmark}: Harmony \citep{korsunsky2019fast}, LIGER \citep{welch2019single}, and Seurat \citep{stuart2019comprehensive}. 
In our comparisons, we {further replaced Seurat V3 studied in \cite{tran2020benchmark} with}
Seurat V5 (version 5.3.1) \citep{seurat} as 
{the latter}
demonstrates improved performance over V3. 
{For each dataset, we apply a common preprocessing and supply the preprocessed data to all batch correction methods in comparison\footnote{{We retained top $20$ PCs as the last step in preprocessing for all results reported in this section. See Section \ref{sec: sup real world} for results with top $25$ PCs.}},}
with the sole exception of LIGER which requires the raw data matrix. 

\begin{figure}[!ht]
\centering
\includegraphics[width=\textwidth]{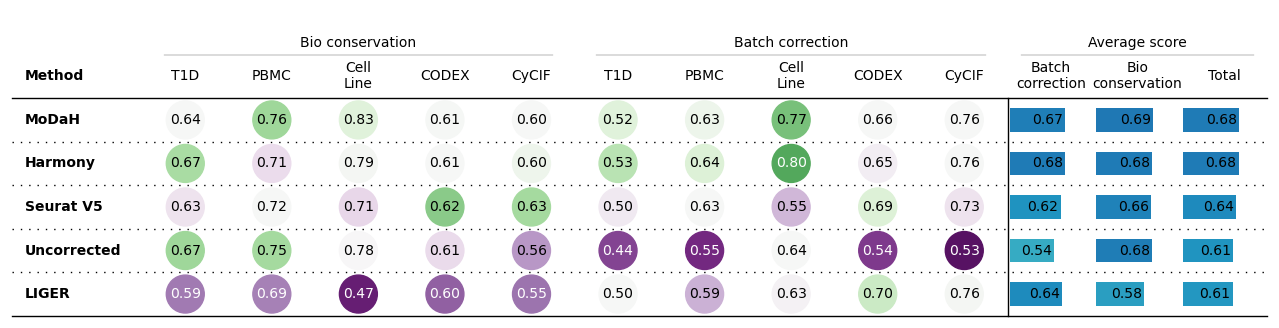}
\caption{Summary of performance scores of all methods in comparison on five datasets. 
Scores in circles represent the average scores of five ``bio conservation'' metrics and of four ``batch correction'' metrics in individual datasets. 
Scores in bars show the final averages of metric scores within and across the two categories over datasets. All scores are rescaled to a [0, 1] range, with higher scores corresponding to better performances.}
\label{fig: average scores}
\end{figure}

To {quantitatively benchmark the performances of all methods in comparison},
we {use the same nine metrics from}
the \texttt{scib-metrics} package \citep{luecken2022benchmarking} as in the simulation studies {reported in Section \ref{subsec:Khat}}. 
{Recall that all metrics are rescaled to have a $[0, 1]$ range, where 1 represents the best possible performance and 0 represents the worst.}
Figure \ref{fig: average scores} provides a cross-dataset high-level summary by averaging the performance metrics within the ``bio conservation'' category and within the ``batch correction'' category of the methods in comparison over all five real-world datasets.
The ``bio conservation'' and ``batch correction'' scores for each dataset (shown in the circles) are averages of {\texttt{Isolated labels}, \texttt{Leiden NMI}, \texttt{Leiden ARI}, \texttt{Sihouette label}, and \texttt{cLISI}, and of \texttt{Sihouette batch}, \texttt{iLISI}, \texttt{KBET}, and \texttt{Graph connectivity}, respectively}.
{The average ``bio conservation'' score and the average ``batch correction'' score of each method across all five datasets are reported as the first two entries in the ``Average score'' column, and their weighted average ($5/9$ ``bio conservation'' + $4/9$ ``batch correction'') is reported as the ``Total'' score of each method. Here, the weights for ``bio conservation'' and ``batch correction'' scores are chosen such that the nine original metrics have equal weights in the ``Total'' score.}
{For each dataset, we also calculate the metrics on the preprocessed data without batch correction as the baseline, which corresponds to the ``Uncorrected'' line in Figure \ref{fig: average scores} and subsequent figures.}

Results in Figure \ref{fig: average scores} 
show that {{\method} is a not only model-based method with rigorous decision-theoretic justification, but it empirical performance on real-world datasets also reaches state of the art}: it ties with Harmony as the best methods in the overall performance across all metrics on the  five datasets and shows strong results in both  batch correction and bio conservation categories. 
Both outperform Seurat V5 and LIGER.
{In the rest of this section, we present detailed benchmarking results on individual datasets.}

\subsection{Single-Cell RNA-seq Data}

We first evaluate the performance of {\method} alongside Harmony, LIGER, and Seurat V5 on three single-cell RNA sequencing (scRNA-seq) datasets. 
The first, introduced in Section \ref{sec: intro} as an illustrative example (Figure \ref{fig: batch effect intro}), is a human type 1 diabetes (T1D) dataset from \citet{fasolino2022single}, consisting of samples from 11 healthy individuals, 5 individuals with T1D, and 8 individuals with no clinical
presentation of T1D but positive for beta-cell auto-antibodies (AAB+),  where each individual constitutes a batch. 
The second is a mouse peripheral blood mononuclear cell (PBMC) dataset from \citet{han2018mapping}, comprising six batches.
{In this dataset, sample sizes across different batches vary drastically and certain cell types are missing in some batches.
See Table \ref{tab: pbmc composition} in the appendix for detailed cell counts of different cell types in different batches for this dataset.
The third scRNA-seq dataset is the cell line (Cell Line) dataset from \citet{korsunsky2019fast}, which contains a ultra challenging batch composition: pure 293T cells in batch $1$, pure Jurkat cells in batch $2$, and a 50:50 mixture of both in batch $3$. 
The deviation from the assumptions of comparable batch sample sizes and comparable cluster sizes in Section \ref{sec: theory} in the latter two datasets
allows us to test the performance of {\method} on real data beyond the scope of our theory. }

For all datasets, we preprocess the data by count normalization, log1p transformation, selection of $1000$ highly-variable genes, rescaling of the selected highly-variable genes, and dimension reduction to top $20$ PCs.
We estimate the number of clusters from the data using Leiden clustering \citep{traag2019louvain}. Consistent with the simulation experiments, the Leiden algorithm was parameterized with $20$ neighbors and a resolution of $0.25$.



\begin{figure}[!ht]
\centering
\subfigure[Human T1D dataset]{
\begin{minipage}[t]{\textwidth}
\centering
\includegraphics[width=\textwidth]{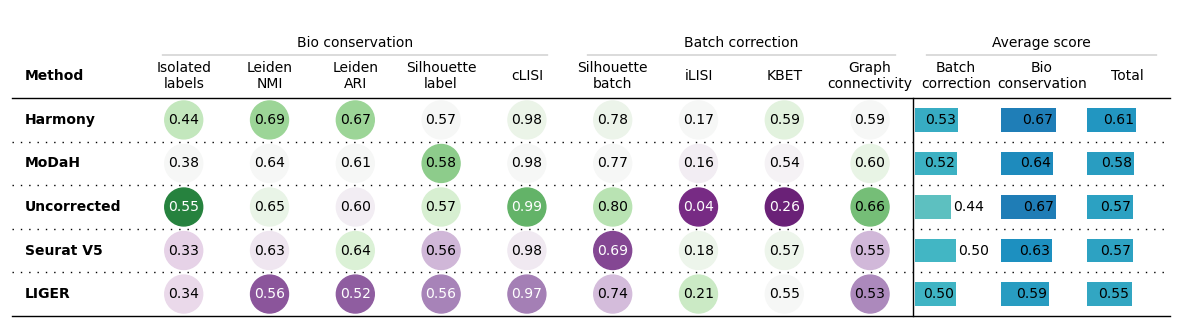}
\end{minipage}%
}%

\subfigure[Mouse PBMC dataset]{
\begin{minipage}[t]{\textwidth}
\centering
\includegraphics[width=\textwidth]{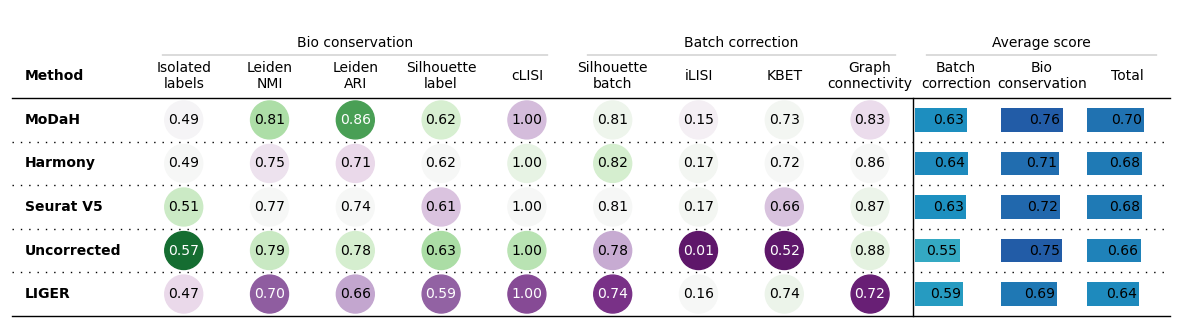}
\end{minipage}%
}%

\subfigure[Cell line dataset]{
\begin{minipage}[t]{\textwidth}
\centering
\includegraphics[width=\textwidth]{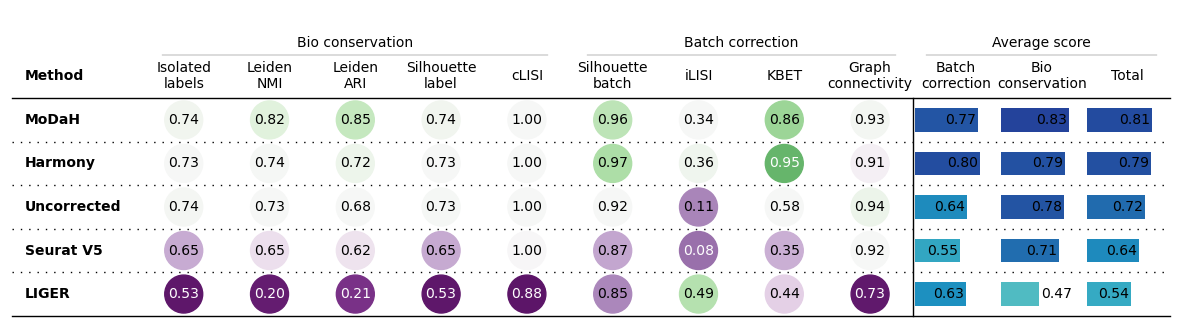}
\end{minipage}%
}%

\caption{Performance scores of all methods in comparison on three scRNA-seq datasets. 
Scores in bars show the averages of metric scores within and across the ``bio conservation'' and the ``batch-correction'' categories.
All scores are rescaled to a [0, 1] range, with higher scores indicating better performances. }
\label{fig: scrnaseq}
\end{figure}


{The performance scores of different methods in comparison on the three scRNA-seq datasets, reported in Figure \ref{fig: scrnaseq} show that {\method} and Harmony consistently outperform. 
The performances of Seurat V5 and LIGER are comparable on the human T1D and the mouse PBMC datasets, but lag behind on the most challenging cell line dataset.
UMAP visualizations of these datasets before and after batch corrections by different methods are included as Figures \ref{fig: t1d umap}--\ref{fig: cell line umap} in the appendix.
These results confirm that {\method} provides a robust and balanced performance comparable to the top-performing heuristic batch correction methods for scRNA-seq data. }


\subsection{Spatial Proteomics Data}

In this section, we further {benchmark the batch correction performance of {\method} against Harmony, LIGER, and Seurat V5}
on two {spatial} proteomics datasets. 
{For benchmarking purpose, we focus on correcting batch effects in the single-cell expression profiles of these datasets and ignore the spatial aspect.}
The first dataset is a co-detection by indexing (CODEX) dataset of the healthy human intestine \citep{hickey2023organization}, which comprises samples from three donors, with each donor as a batch. 
The second is a cyclic immunofluorescence (CyCIF) dataset of colorectal cancer (CRC)  \citep{lin2023multiplexed}. 
We selected five sections (Section 20, 39, 59, 78, and 102) from the CRC1 sample, with each section as a distinct batch. 
For benchmarking convenience, both datasets are randomly subsampled to $100000$ cells {across all batches} while keeping the relative sizes of different batches.

\begin{figure}[!ht]
\centering
\includegraphics[width=\linewidth]{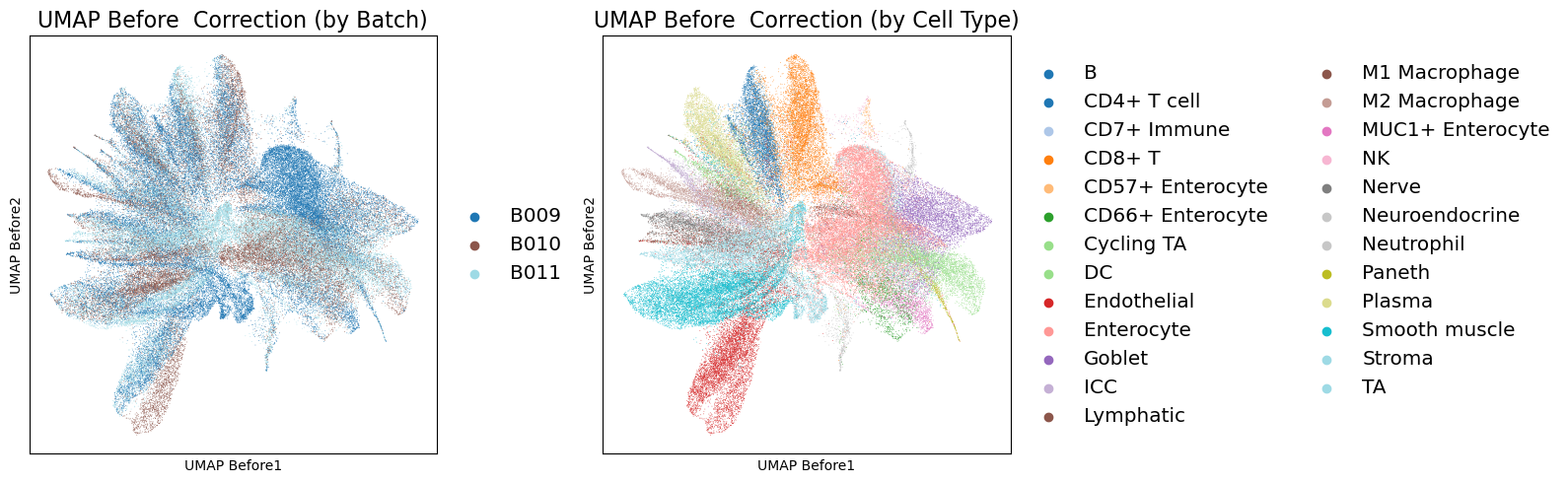}
\caption{UMAP visualizations of the uncorrected CODEX proteomics dataset. Cells are colored by batch (left) to show the initial technical variations across batches and by annotated cell type (right) to show the underlying biological structure.}
\label{fig: protein data}
\end{figure}

{We preprocess the proteomics datasets by log1p transformation\footnote{We obtain the CODEX dataset in a preprocessed format, and so we omit the log1p transformation step on this dataset.}, rescaling of features, and dimension reduction to top 20 PCs.}
Compared with scRNA-seq data, the cell populations in these proteomics datasets exhibit a less distinct separation among cell type clusters.
See Figure \ref{fig: protein data} for an illustration with the human intestine CODEX dataset (cf.~Figure \ref{fig: batch effect intro} top row). 
To {adapt to the different data characteristic},
the default parameters {used in Leiden clustering} for estimating total number of clusters were modified {for spatial proteomic datasets}:  
the number of neighbors was set to $5$ {in neighborhood graph construction} and the resolution to $1$ {in clustering}.


\begin{figure}[!ht]
\centering
\subfigure[Healthy human intestine CODEX dataset]{
\begin{minipage}[t]{\textwidth}
\centering
\includegraphics[width=\textwidth]{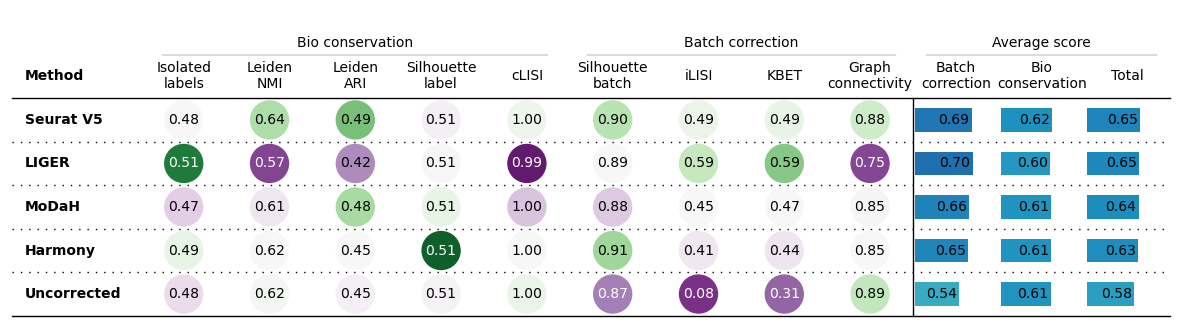}
\end{minipage}%
}%

\subfigure[Human colorectal cancer CyCIF dataset]{
\begin{minipage}[t]{\textwidth}
\centering
\includegraphics[width=\textwidth]{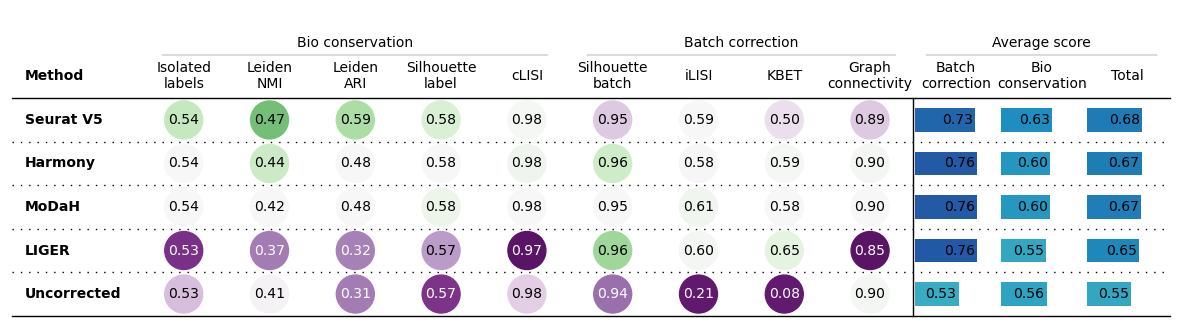}
\end{minipage}%
}%
\caption{Performance scores of all methods in comparison on two spatial proteomics datasets. 
Scores in bars show the averages of metric scores within and across the ``bio conservation'' and the ``batch-correction'' categories.
All scores are rescaled to a [0, 1] range, with higher scores indicating better performances. }
\label{fig: protein}
\end{figure}


The scorecards for batch correction on the proteomics datasets, presented in Figure \ref{fig: protein}, provide further insights into the performances of the methods in comparison. 
On the CODEX dataset, {all four methods achieve comparable performances, with Seurat V5 and LIGER (tie at $0.65$) having slightly better total score than MoDaH and Harmony ($0.64$ and $0.63$, respectively).}
On the CyCIF dataset, {Seurat V5, MoDaH, and Harmony are essentially tied in their total scores ($0.68$, $0.67$, and $0.67$, respectively), while LIGER's performance ($0.65$) 
{is slightly behind when compared with the other three methods.}}
UMAP visualizations of these datasets before and after batch corrections by different methods are included as Figures \ref{fig: codex umap}--\ref{fig: cycif umap} in the appendix.

These benchmarking results underscore the suitability of {\method} for effectively analyzing complex spatial proteomics datasets by continuing to maintain strong balance between removing technical artifacts and preserving biological signals.

\section{Discussion}
\label{sec: discussion}


\paragraph{The connection between {\method} and Harmony}
In Section \ref{sec: real world}, we have observed that the performance of {\method} is comparable to that of Harmony. We now briefly discuss the  connection between the two methods to account for this similarity in performance.

The Harmony algorithm integrates single-cell datasets by minimizing an objective function that jointly encourages cell clustering based on biological similarity while maximizing the diversity of batches within each cluster. 
The {main Harmony} objective function, {with the notation of this manuscript}, can be written as:
\begin{align}
\label{eq: harmony loss}
    & \min_{R,\mu} \sum_{b=1}^B\sum_{i=1}^{n_b}\sum_{k=1}^K \left[R_{kbi} \|X_{bi} - \mu_k\|^2 + \sigma R_{kbi}\log R_{kbi}+ \sigma\theta R_{kbi} \log\left(\frac{O_{kb}}{E_{kb}}\right) \right],\\
    &\mbox{subject~to~~}  R_{kbi} \ge 0,~~ \forall b,i,k, ~~\mbox{ and }~~  \sum_{k=1}^{K} R_{kbi} = 1,~~  \forall b,i. \nonumber
\end{align}
where $\mu_k\in\mR^{d}$ represents the cluster centroids, $R_{kbi}\in[0,1]$ is the soft assignment probability of cell $i$ in batch $b$ to cluster $k$,  $X_{bi}\in \mR^{d}$ is the input embedding (i.e.~preprocessed expression profile) for cell $i$ in batch $b$, $O\in [0,1]^{K\times B}$ is the observed co-occurrence matrix of cells in clusters and
batches, $E\in [0,1]^{K\times B}$ is the expected co-occurrence matrix of cells in clusters and batches,
under the assumption of independence between cluster and batch assignment, 
and 
$\sigma>0, \theta>0$ are two tuning parameters.

The similarity in performance between {\method} and Harmony can be understood by drawing a conceptual parallel between the objective function \eqref{eq: harmony loss} and our probabilistic model \eqref{eq: model}. 
The primary term in the Harmony objective function \eqref{eq: harmony loss} is a soft-assignment k-means objective, i.e.~$\sum_{b,i,k}R_{kbi}\|X_{bi}-\mu_k\|^2$,
the hard-assignment counterpart of which
can be interpreted as the maximum likelihood estimation objective for the cluster assignment parameter in a Gaussian mixture model with a shared spherical covariance structure: $X_{bi}\sim \mathcal{N}(\mu_k, \tau^2 I_d)$ when $R_{kbi}=1$.
In comparison, model \eqref{eq: model} is a more flexible data generative model
that allows for arbitrary, cluster-specific covariance matrices. 
This additional flexibility is crucial for the success of {\method}: 
By modeling the heterogeneous, non-spherical geometry of each cluster, {\method} could correct for batch effects that could only be detected by within cluster local geometry.
In contrast, Harmony compensates for the rigidity in covariance matrix modeling with a soft assignment strategy with an entropic penalty.

Therefore, the two methods converge to similar empirical performances by improving upon hard-assignment k-means clustering in different ways.
{\method} employs a more expressive statistical modeling approach, while Harmony takes a pure algorithmic route.

\paragraph{Potential future works} 
There are two theoretical settings beyond the scope of the current manuscript that are of great interest.
First, we have demonstrated that {\method} exhibits empirical robustness even when a number of clusters are missing in some batches. 
Establishing conditions for the consistency (and, if possible, the optimality) of {\method}  in this setting could further informs its practical applications.
In addition, since consortia efforts have led to ever increasing numbers of single-cell datasets collected for the same tissues \citep{czi2025cz}, the asymptotic regime of a divergent number of batches, i.e.~$B\to \infty$, becomes relevant.
This regime offers the opportunity to identify more complex batch effect structures, and developing rigorous inference procedures for such high-dimensional settings presents a fertile ground for future research. 



\section*{Code and Data Availability}
The code and the data for generating results reported in this manuscript are available at \url{https://doi.org/10.5281/zenodo.17873392}.

\section*{Acknowledgment}
We would like to acknowledge support from the National Science Foundation (DMS-2245575 to Z.M.) and the National Institutes of Health (U01CA294514 to Y.C. and Z.M.).

\bibliography{ref}

\newpage

\appendix

\begin{center} 
    \Large \textbf{Appendix of ``MoDaH~achieves rate optimal batch correction''}
\end{center}

\section{Proof of Theorem \ref{thm: lower bound}}

\label{sec: proof lower bound}

Since when $n\to \infty$, $\max\left\{\exp\left(-(1+o(1))\frac{\snr^2}{8}\right), \exp\left(-(1+o(1))\log n\right)\right\}$ is of the same order as $\exp\left(-(1+o(1))\frac{\snr^2}{8}\right)+  \exp\left(-(1+o(1))\log n\right)$, it is sufficient to show
that: the following two statements hold if $\snr\to\infty$:
\begin{align}
    &\inf_{(a, \beta)}\sup_{(a^*, \beta^*)\in \cS(\alpha, \gamma, \Gamma)}\mE h(a, \beta, a^*, \beta^*)\ge \exp\left(-(1+o(1))\frac{\snr^2}{8}\right),\label{eq: a error}\\
    &\inf_{(a, \beta)}\sup_{(a^*, \beta^*)\in \cS(\alpha, \gamma, \Gamma)}\mE h(a, \beta, a^*, \beta^*)\ge \exp\left(-(1+o(1))\log n\right);\label{eq: beta error}
\end{align}
and the following statement holds if $\snr=O(1)$:
\begin{align}
    \inf_{(a, \beta)}\sup_{(a^*, \beta^*)\in \cS(\alpha, \gamma, \Gamma)}\mE h(a, \beta, a^*, \beta^*)\ge c,\label{eq: aerror low snr}
\end{align}
for some constant $c>0$.

\paragraph*{Step 1} To establish \eqref{eq: a error} and \eqref{eq: aerror low snr}, we first define the mapping $\Phi_{(a, b)}: [K] \to [K]$ as
\begin{align*}
    \Phi_{(a, b)}(k) = \min\left\{ \operatorname*{argmax}_{k'} \sum_{i \in [n_b]} \mI\{ \abi = k, \abis = k' \} \right\},
\end{align*}
where $\Phi_{(a, b)}(k)$ denotes the value $k'$ that maximizes $\sum_{i \in [n_b]} \mI\{ \abi = k, \abis = k' \}$. In cases of ties, the smallest index $k'$ is chosen.
 {By definition, $\Phi_{(a,b)}(k)$ provides the most represented true cluster label $k'$ among the cells estimated as belonging to the $k$-th cluster in batch $b$.}
Furthermore, we define the secondary mapping $\Phi'_{(a, b)}: [K] \to [K]$ by
\begin{align*}
    \Phi'_{(a, b)}(k) = \min\left\{ \operatorname*{argmax}_{k'' \neq \Phi_{(a, b)}(k)} \sum_{i \in [n_b]} \mI\{ \abi = k, \abis = k'' \} \right\},
\end{align*}
which represents the index $k'$ (distinct from $\Phi_{(a, b)}(k)$) that maximizes $\sum_{i \in [n_b]} \mI\{ \abi = k, \abis = k' \}$, again resolving ties by selecting the minimum value. 
 {By definition, it is the most represented true cluster label after $\Phi_{(a,b)}(k)$ among cells clustered to cluster $k$ in batch $b$. 
When all cells estimated to be in cluster $k$ in batch $b$ have the same true cluster label, we set $\Phi'_{(a, b)}(k) = 0$.}

Then, we have
\begin{align*}
    &\frac{1}{n}\sum_{i\in [n_b]}\|\betababi-\betababis\|^2\mI\{\abi = k\}\\
    &\ge \frac{1}{n}\sum_{i\in [n_b]}\|\betababi-\betababis\|^2\mI\{\abi = k, \abis = \Phi_{(a, b)}(k)\} \\
    &~~~ + \frac{1}{n}\sum_{i\in [n_b]}\|\betababi-\betababis\|^2\mI\{\abi = k, \abis = \Phi'_{(a, b)}(k)\}\\
    &\ge \frac{1}{2}\|\beta^*_{b\Phi_{(a, b)}(k)}- \beta^*_{b\Phi'_{(a, b)}(k)}\|^2\cdot \frac{1}{n}\sum_{i\in [n_b]}\mI\{\abi = k, \abis = \Phi'_{(a, b)}(k)\},\\
    &\ge \frac{\gamma}{2}\frac{1}{K-1}\frac{1}{n}\sum_{i\in [n_b]}\mI\{\abi = k, \abis \neq \Phi_{(a, b)}(k)\}.
\end{align*}
 {Here, the first inequality holds as we are decomposing the term $\mI\{\abi=k\}$.
The second inequality holds as 
\begin{equation*}
m \|a\|^2 + n \|b\|^2 \geq \frac{n}{2}\|a-b\|^2
\end{equation*}
for all $a,b$ whenever $m\geq n$.
Finally, the last equality holds by the definitions of $\gamma$ and $\Phi_{(a,b)}(k)$.}
Therefore, there further holds
\begin{align}
    h(a, \beta, a^*, \beta^*)&=\frac{1}{n}\sum_b\sum_{i\in [n_b]}\|\betababi-\betababis\|^2,\nonumber\\
    &\ge \frac{\gamma}{2(K-1)}\frac{1}{n}\sum_b\sum_k\sum_{i\in [n_b]}\mI\{\abi = k, \abis \neq \Phi_{(a, b)}(k)\},\nonumber\\
    &\ge \frac{\gamma}{2(K-1)}\frac{1}{n}\sum_b\inf_{\Phi:[K]\to [K]}\sum_{i\in [n_b]}\mI\{\Phi(\abi) \neq \abis\}.\label{eq: transfer}
\end{align}
For simplicity of notation, define $f(a, a^*) = \frac{1}{n}\sum_b\inf_{\Phi:[K]\to [K]}\sum_{i\in [n_b]}\mI\{\Phi(\abi) \neq \abis\}$.

Now we construct a reduced parameter space for $a^*$. 
Without loss of generality, assume that the minimum in the definition of $\snr$ (see Definition \ref{def: snr}) is attained at $(b, k, k') = (1, 1, 2)$. 
Construct sets $T_{bk}$ for $b\in [B]$ and $k\in [K]$ such that for any $b\neq 1$, $\{T_{bk}\}_{k=1}^K$ is a decomposition of $[n_b]$ such that $|T_{bk}|\ge \lfloor \frac{n_b}{K}\rfloor$, and for $b=1$, $\{T_{1k}\}_{k=1}^K$ are disjoint subsets of $[n_1]$ such that $|T_{1k}| = \lfloor\frac{n_1}{K}-\frac{n_1}{8K^2}\rfloor$. 
 {Here, $T_{11}$ and $T_{12}$ may not contain all cells belonging to cluster $1$ or $2$ in batch $1$.}
Define a parameter space
\begin{align*}
    \mA = \{a: \forall b\in[B],  \abi=k \text{ for all } i\in T_{bk}, \text{ and } a_{1i}\in\{1, 2\} \text{ for }i\in T^c\},
\end{align*}
where $T = \cup_k T_{1k}$. 
It is straightforward to verify that for any $\alpha \in (0, \frac{1}{2})$, $a^* \in \mA$ satisfies the restriction on $a^*$ imposed by $(a^*, \beta^*) \in \cS(\alpha, \gamma, \Gamma)$. Therefore, by \eqref{eq: transfer}, there holds
\begin{align*}
    \inf_{(a, \beta)}\sup_{(a^*, \beta^*)\in \cS(\alpha, \gamma, \Gamma)}\mE h(a, \beta, a^*, \beta^*)\ge\frac{\gamma}{2(K-1)}\inf_a\sup_{a^*\in \mA}\mE f(a, a^*).
\end{align*}
 {In other words, the foregoing arguments show that the minimax lower bound on batch correction can be bounded from below by a minimax lower bound on mis-clustering in a reduced parameter space.}

Meanwhile, for any $a\neq \tilde{a}\in \mA$, there holds for sufficiently large $n_1$ that
\begin{align*}
    \sum_{i\in n_1}\mI\{a_{1i} \neq \tilde{a}_{1i}\}\le |T^c| \le (\frac{n_1}{8K^2}+2)\cdot K \le \frac{n_1}{7K},
\end{align*}
and for any $\Phi$ that is not an identity transformation, there holds
\begin{align*}
    \sum_{i\in n_1}\mI\{\Phi(a_{1i}) \neq \tilde{a}_{1i}\}\ge\min_{k\in[K]}|T_{1k}| -  |T^c|\ge \frac{n_1}{K}-\frac{n_1}{8K^2} - 1-\frac{n_1}{7K}\ge \frac{n_1}{2K}.
\end{align*}
Therefore, $f(a, a^*) = \frac{1}{n}\sum_b\sum_{i\in [n_b]}\mI\{\abi \neq \tilde{a}_{bi}\}$ for $a, \tilde{a}\in \mA$. Then 
\begin{align*}
    \inf_a\sup_{a^*\in \mA}\mE f(a, a^*)&\ge \inf_a\frac{1}{|\mA|}\sum_{a^*\in \mA}\mE f(a, a^*),\\
    &\ge \frac{1}{n}\sum_{i\in T^c}\inf_a \frac{1}{|\mA|}\sum_{a^*\in \mA} \mP_{a^*}(a_{1i}\neq a^*_{1i}).
\end{align*}
For any $i\in T^c$, define $\mA_k = \{a\in \mA: a_{1i}=k\}$ for $k\in\{1, 2\}$. Then there exists a  bijection $g:\mA_1\to \mA_2$ such that $g$ only changes $a_{1i}$ from 1 to 2. Then 
\begin{align*}
    \inf_a \frac{1}{|\mA|}\sum_{a^*\in \mA} \mP_{a^*}(a_{1i}\neq a^*_{1i}) &= \inf_a \frac{1}{|\mA|}\sum_{a^*\in \mA_1}( \mP_{a^*}(a_{1i}\neq 1)+\mP_{g(a^*)} (a_{1i}\neq 2)),\\
    &\ge \frac{1}{2}\inf_{\hat\phi} ( \mP_{\mH_0}(\hat\phi=1)+\mP_{\mH_1} (\hat\phi=0)),
\end{align*}
where $\mH_0:X\sim \cN(\mu_{11}^*, \Sigma_1^*)$, $\mH_1:X\sim \cN(\mu_{12}^*, \Sigma_2^*)$, and $\hat\phi$ is a testing procedure between $\mH_0$ and $\mH_1$. By Lemma 3.1 in \cite{chen2024achieving}, a lower bound for $\frac{1}{2}\inf_{\hat\phi} ( \mP_{\mH_0}(\hat\phi=1)+\mP_{\mH_1} (\hat\phi=0))$, there further holds 
\begin{align*}
    \inf_a \frac{1}{|\mA|}\sum_{a^*\in \mA} \mP_{a^*}(a_{1i}\neq a^*_{1i})\ge \exp\left(-(1+o(1))\frac{\snr^2}{8}\right),
\end{align*}
if $\snr\to \infty$, and $\inf_a \frac{1}{|\mA|}\sum_{a^*\in \mA} \mP_{a^*}(a_{1i}\neq a^*_{1i})\ge c$ for some constant $c>0$ if $\snr = O(1)$.

Therefore, if $\snr\to \infty$, there holds for a positive constant $\geq C(\gamma,K)$ that depends only on $\gamma$ and $K$ that
\begin{align*}
    \inf_{(a, \beta)}\sup_{(a^*, \beta^*)\in \cS(\alpha, \gamma, \Gamma)}\mE h(a, \beta, a^*, \beta^*)\geq C(\gamma,K)\inf_a\sup_{a^*\in\mA}\mE f(a, a^*) \ge\exp\left(-(1+o(1))\frac{\snr^2}{8}\right),
\end{align*}
and if $\snr=O(1)$, there exists some constant $c>0$, such that
\begin{align*}
    \inf_{(a, \beta)}\sup_{(a^*, \beta^*)\in \cS(\alpha, \gamma, \Gamma)}\mE h(a, \beta, a^*, \beta^*)\geq C(\gamma,K)\inf_a\sup_{a^*\in\mA}\mE f(a, a^*) \ge c.
\end{align*}

\paragraph{Step 2} To prove \eqref{eq: beta error}, consider the following parameter space which is a subset of $\cS(\alpha, \gamma, \Gamma)$: take an arbitrary element $(a^0, \beta^0)\in \cS(\alpha, \gamma+1, \Gamma+1)\subset \cS(\alpha, \gamma, \Gamma)$, let $\beta^1$ be defined as: $\beta^1$ takes the same value as $\beta^0$ with the only exception that $(\beta^1_{11})_1 = (\beta^0_{11})_1 + \frac{1}{\sqrt{n}}$, where $(v)_1$ represents the first element of a vector $v$. Then it is straightforward to validate $(a^0, \beta^1)\in \cS(\alpha, \gamma, \Gamma)$, and therefore
\begin{align*}
    \inf_{(a, \beta)}\sup_{(a^*, \beta^*)\in \cS(\alpha, \gamma, \Gamma)}\mE h(a, \beta, a^*, \beta^*)&\ge \inf_{(a, \beta)}\sup_{(a^*, \beta^*)\in \{(a^0,\beta^0), (a^0, \beta^1)\}}\mE h(a, \beta, a^*, \beta^*),\\
    &\ge \frac{n_{11}}{n}\inf_{\hat\theta}\sup_{\theta\in\{0, \frac{1}{\sqrt{n}}\}}\mE|\hat\theta-\theta|^2,
\end{align*}
where $\hat\theta$ is an estimator of $\theta$ based on $n_{11}$ observations $(x_1, x_2, \cdots, x_{n_{11}})$ of $\cN(\theta, \sigma^2)$, and $\sigma$ is a constant related with $\Sigma_1^*$. Consider the setting where $\theta$ follows the prior distribution that $\Prob(\theta = 0) = 0.5$ and $\Prob(\theta = \frac{1}{\sqrt{n}}) = 0.5$, then the minimax risk is greater than the Bayesian risk:
\begin{align*}
    \inf_{\hat\theta}\sup_{\theta\in\{0, \frac{1}{\sqrt{n}}\}}\mE|\hat\theta-\theta|^2\ge \inf_{\hat\theta}\mE_\theta|\hat\theta-\theta|^2.
\end{align*}
The Bayesian estimator is given by $\theta_B = \frac{1}{\sqrt{n}}\frac{p_1(\overline x)}{p_0(\overline x)+ p_1(\overline x)}$, where $\overline x = \sum_{i=1}^{n_{11}}x_i/n_{11}$, $p_0$ is the probability density function of $P_0 = \cN(0, \frac{1}{n_{11}\sigma^2})$, $p_1$ is the probability density function of $P_1 = \cN(\frac{1}{n}, \frac{1}{n_{11}\sigma^2})$. Therefore, the Bayesian risk can be lower bounded by
\begin{align*}
    \mE_\theta|\theta_B-\theta|^2 &= \frac{1}{2n}\int\frac{p_1(x)p_0(x)}{p_1(x)+p_0(x)}dx \ge \frac{1}{4n} \int\min\{p_0(x), p_1(x)\}dx ,\\
    &= \frac{1}{4n}(1-d_{TV}(P_0, P_1)) \succeq\frac{1}{n} = \exp\left(-(1+o(1))\log n\right),
\end{align*}
where $d_{TV}(P_0, P_1)$ represents the TV distance between $P_0$ and $P_1$, which is a constant related with $\frac{n_{11}}{n}$ and $\sigma$.

Therefore
\begin{align*}
    \inf_{(a, \beta)}\sup_{(a^*, \beta^*)\in \cS(\alpha, \gamma, \Gamma)}\mE h(a, \beta, a^*, \beta^*)&\ge \inf_{(a, \beta)}\sup_{(a^*, \beta^*)\in \{(a^0,\beta^0), (a^0, \beta^1)\}}\mE h(a, \beta, a^*, \beta^*),\\
    &\ge \exp\left(-(1+o(1))\log n\right).
\end{align*}

Combining the above points (a) and (b) together, then if $\snr\to\infty$, there holds
\begin{align*}
    \inf_{(a, \beta)}\sup_{(a^*, \beta^*)\in \cS(\alpha, \gamma, \Gamma)}\mE h(a, \beta, a^*, \beta^*)\ge \exp\left(-(1+o(1))\frac{\snr^2}{8}\right)+ \exp\left(-(1+o(1))\log n\right),
\end{align*}
and $\liminf_{n\to\infty}\inf_{(a, \beta)}\sup_{(a^*, \beta^*)\in \cS(\alpha, \gamma, \Gamma)}\mE h(a, \beta, a^*, \beta^*)\ge c$ for some constant $c$ if $\snr=O(1)$.

\section{Proof of Theorem \ref{thm: upper bound}}
\label{sec: proof upper bound}

Let $t$ be any integer satisfying $t\ge \log(n)$, then the following holds for the output $(a^{(t+1)}, \beta^{(t+1)})$ of Algorithm \ref{alg:em} after $t+1$ iterations:
\begin{align*}
    h(a^{(t+1)}, \beta^{(t+1)}, a^*, \beta^*)&\le \frac{2}{n}\sum_{b, i}\|\beta_{b\abi^{(t+1)}}^*-\betababis\|^2+\frac{2}{n}\sum_{b, i}\|\beta_{b\abi^{(t+1)}}^{(t+1)} - \beta_{b\abi^{(t+1)}}^*\|^2\\
    &\le\max_{b, k\neq k'}2\|\betabu^*-\betabv^*\|^2\frac{1}{n}\sum_{b, i}\mI\{\abi^{(t+1)}\neq\abis\} + 2\max_{b, k}\|\betabu^{(t+1)}-\betabu^*\|^2 \\
    & = \frac{8\Gamma }{n}\sum_{b, i}\mI\{\abi^{(t+1)}\neq\abis\} + 2\max_{b, k}\|\betabu^{(t+1)}-\betabu^*\|^2.
\end{align*}
 {Here, the first inequality holds as $\|a+b\|^2\leq 2\|a\|^2 + 2\|b\|^2$. 
The second inequality holds as $\sum_{i}a_ib_i \leq \max_{i}a_i\sum_{i}b_i$ when $a_i,b_i\geq 0$ for all $i$.}
For notational simplicity, define 
\begin{align*}
    \hatbetabk(a):=\frac{\sum_{i}\xbi\mI\{a_{bi}=k\}}{\sum_{i}\mI\{a_{bi}=k\}}-\frac{\sum_{b,i}\xbi\mI\{a_{bi}=k\}}{\sum_{b,i}\mI\{a_{bi}=k\}}
\end{align*}
for any cluster assignment vector $a$. 
Then, the output of Algorithm \ref{alg:em} satisfies $\betabk^{(t+1)} = \hatbetabk(a^{(t)})$ for all $b, k$, and there further holds
\begin{align*}
    h(a^{(t+1)}, \beta^{(t+1)}, a^*, \beta^*)&\le \frac{8\Gamma}{n}\sum_{b, i}\mI\{\abi^{(t+1)}\neq\abis\} + 2\max_{b, k}\|\hatbetabu(a^{(t)})-\betabu^*\|^2,\\
    &\le \frac{8\Gamma}{n}\sum_{b, i}\mI\{\abi^{(t+1)}\neq\abis\} + 4\max_{b, k}\|\hatbetabu(a^{(t)})-\hatbetabu(a^*)\|^2 \\
    &~~~~ +4\max_{b, k}\|\hatbetabu(a^*)-\betabu^*\|^2.
\end{align*}
We analyze the three terms on the right hand side sequentially.

\paragraph*{Term 1}For the first term $\frac{8\Gamma}{n}\sum_{b, i}\mI\{\abi^{(t+1)}\neq\abis\}$, there holds
\begin{align*}
    \frac{8\Gamma}{n}\sum_{b, i}\mI\{\abi^{(t+1)}\neq\abis\}\le \frac{8\Gamma}{\omega}\frac{\ell(a^{(t+1)}, a^*)}{n},
\end{align*}
where $\omega$ is defined in Condition \ref{cond: regularity} and $\ell$ in \eqref{eq:l}.

We now seek to establish an upper bound for $\ell(a^{(t+1)}, a^*)$, which represents the clustering error. Intuitively, our clustering task should be no more difficult than clustering data generated by a standard Gaussian Mixture Model with $K \cdot B$ mixing components. 
Our setting, described in model \eqref{eq: model}, consists of $K \cdot B$ batch-specific clusters in total. However, a key feature of our model is that clusters of the same biological type $k$ across different batches share common covariance matrices. This shared structure allows for more robust parameter estimation than what is possible in a general GMM with $K \cdot B$ fully distinct components. 
Leveraging the techniques in \cite{chen2024achieving}, we formalize this intuition in the following proposition, which provides an upper bound for $\ell(a, a^*)$, whose proof is outlined in Section \ref{sec: proof clustering error}.

\begin{proposition}\label{prop: clustering error}
    Suppose that as $n\to \infty$, $B, K, d = O(1)$, $\max_{k\neq k'}\frac{\lambda_{\max}\sigmau^*}{\lambda_{\min}\sigmav^*}=O(1)$ and $\snr\to\infty$. 
    When $n$ is sufficiently large,
    if $\ell(a^{(0)}, a^*) =  o(n)$ holds with probability at least $1-\eta$, we have
    \begin{align}
        \ell(a^{(t')}, a^*)\le \Omega n\exp\left(-(1+o(1))\frac{\snr^2}{8}\right),\label{eq: event cluster error}
    \end{align}
    with probability at least $1-\eta-5n^{-1}-\exp(-\snr)$ for all integer $t'\ge\log n$.
\end{proposition}
With Proposition \ref{prop: clustering error}, Proposition \ref{prop: snr} and $\log\Omega =o(\omega)$, we further have
\begin{align}
    \frac{8\Gamma}{n}\sum_{b, i}\mI\{\abi^{(t+1)}\neq\abis\}\le \frac{\Gamma\Omega}{\omega}\exp\left(-(1+o(1))\frac{\snr^2}{8}\right) = \Gamma \exp\left(-(1+o(1))\frac{\snr^2}{8}\right),
\end{align}
with probability at least $1-\eta-5n^{-1}-\exp(-\snr)$.

\paragraph*{Term 2} For $\max_{b, k}\|\hatbetabu(a^{(t)})-\hatbetabu(a^*)\|^2$, we first have from the Cauchy--Schwarz inequality that
\begin{align*}
    \max_{b, k}\|\hatbetabu(a^{(t)})-\hatbetabu(a^*)\|^2&\le 2\max_{b, k}\|\hatmubu(a^{(t)})-\hatmubu(a^*)\|^2+2\max_{ k}\|\hatmuu(a^{(t)})-\hatmuu(a^*)\|^2,
\end{align*}
 where $\hatmubu(a)=\frac{\sum_{i}\xbi\mI\{a_{bi}=k\}}{\sum_{i}\mI\{a_{bi}=k\}}$ and $\hatmuu(a) = \frac{\sum_{b, i}\xbi\mI\{a_{bi}=k\}}{\sum_{b, i}\mI\{a_{bi}=k\}}$ for all $a$. 
 We bound the two terms on the right side of the last display separately. 
 
The first term, which concerns the estimation error of a single centroid of the $K \cdot B$ clusters, can be bounded by applying the same arguments used in \cite{chen2024achieving} for Gaussian mixture models. Specifically, it can be bounded by \eqref{eq: mubu accuracy} in Lemma \ref{lemma: estimation bounds}, where Lemma \ref{lemma: estimation bounds} is a technical lemma presented in Section \ref{sec: technical lemmas} which can be derived in exactly the same way as in \cite{chen2024achieving}. 
 
The second term is more complex as it involves multiple centroids, i.e.~all centroids belonging to the same biological cluster type across different batches. The bound for this second term is provided in the following lemma, whose proof is presented in Section \ref{sec: proof lemma muu a astar}.
\begin{lemma}
\label{lemma: bound muu a astar}
    Under all the conditions in Proposition \ref{prop: clustering error}, there exists $C>0$ that depends on $\alpha, \lambda_{\max}=\max_k(\lambda_{\max} \sigmau^*), \lambda_{\min}=\min_k(\lambda_{\min} \sigmau^*)$, such that
    \begin{align*}
        \max_{ k}\|\hatmuu(a^{(t)})-\hatmuu(a^*)\|^2\le \frac{CK^2}{\omega n}\ell(a^{(t)}, a^*),
    \end{align*}
    with probability at least $1-n^{-2}$ on the event where \eqref{eq: event cluster error} holds.
\end{lemma}
With Lemma \ref{lemma: bound muu a astar} and \eqref{eq: mubu accuracy} in Lemma \ref{lemma: estimation bounds}, we have
\eqref{eq: event cluster error} and 
\begin{align}
    \max_{b, k}\|\hatbetabu(a^{(t)})-\hatbetabu(a^*)\|^2&\le 2\max_{b, k}\|\hatmubu(a^{(t)})-\hatmubu(a^*)\|^2+2\max_{k}\|\hatmuu(a^{(t)})-\hatmuu(a^*)\|^2,\nonumber\\
    &\le \frac{CK^2}{\omega n}\ell(a^{(t)}, a^*)\le \exp\left(-(1+o(1))\frac{\snr^2}{8}\right),\label{eq: step2}
\end{align}
simultaneously holds with probability at least $1-\eta-6n^{-1}-\exp(-\snr)$ for all $t\ge\log n$.

\paragraph*{Term 3}
For $\max_{b, k}\|\betabu(a^*)-\betabu^*\|^2$, we first obtain from the Cauchy--Schwarz inequality that
\begin{align*}
    \max_{b, k}\|\betabu(a^*)-\betabu^*\|^2&\le 2\max_{b, k}\|\hatmubu(a^*)-\mubus\|^2+2\max_{ k}\|\hatmuu(a^*)-\mu_k^*\|^2.
\end{align*}

Similar to the previous bound on  $\max_{b, k}\|\hatbetabu(a^{(t)})-\hatbetabu(a^*)\|^2$, the first term can be bounded by \eqref{eq: mubu accuracy given true assignment} in Lemma \ref{lemma: estimation bounds} which can be achieved using the same arguments as in \cite{chen2024achieving}, while the second term can be bounded by the following lemma.
\begin{lemma}
\label{lemma: bound muu astar true}
    Under all the conditions in Proposition \ref{prop: clustering error}, there exists $C>0$ that depends on $\alpha, \lambda_{\max}=\max_k(\lambda_{\max} \sigmau^*), \lambda_{\min}=\min_k(\lambda_{\min} \sigmau^*)$, such that
    \begin{align*}
        \max_{ k}\|\hatmuu(a^*)-\mu_k^*\|^2\le \frac{CK\log n}{n},
    \end{align*}
    with probability at least $1-n^{-2}$ on the event where \eqref{eq: event cluster error} holds.
\end{lemma}
With Lemma \ref{lemma: bound muu astar true} and \eqref{eq: mubu accuracy given true assignment} in Lemma \ref{lemma: estimation bounds}, 
we have
\eqref{eq: event cluster error}, \eqref{eq: step2} and
\begin{align*}
    \max_{b, k}\|\betabu(a^*)-\betabu^*\|^2&\le 2\max_{b, k}\|\hatmubu(a^*)-\mubus\|^2+2\max_{ k}\|\hatmuu(a^*)-\mu_k^*\|^2,\\
    &\le  \frac{CK\log n}{n}=\exp\left(-(1+o(1))\log n\right),
\end{align*}
simultaneously  holds with probability at least $1-\eta-7n^{-1}-\exp(-\snr)$ for all $t\ge\log n$.

\medskip

Assembling the bounds on individual terms, we obtain that 
\begin{align*}
    h(a^{(t+1)}, \beta^{(t+1)}, a^*,\beta^*)\le \max\{\Gamma, 1\}\exp\left(-(1+o(1))\frac{\snr^2}{8}\right)+\exp\left(-(1+o(1))\log n\right),
\end{align*}
with probability at least $1-\eta-7n^{-1}-\exp(-\snr)$ for all $t\ge\log n+1$.

\section{Proof of Proposition \ref{prop: snr}}
\label{sec: proof prop snr}
We will first prove the lower bound. 
Assume that the minimum in the definition of $\snr$ is achieved at $(b, k, k')$. 
By definition, there exists some $x\in \mR^d$, such that $\|x\| = \snr/2$, and 
\begin{align*}
    &x^\top(\sigmau^*)^{\frac{1}{2}}(\sigmav^*)^{-1}(\mubus-\mubvs)+\frac{1}{2}x^\top((\sigmau^*)^{\frac{1}{2}}(\sigmav^*)^{-1}(\sigmau^*)^{\frac{1}{2}}-I_d)x\\
        &~~~~\le -\frac{1}{2}(\mubus-\mubvs)^\top(\sigmav^*)^{-1}(\mubus-\mubvs)+\frac{1}{2}\log|\sigmau^*|-\frac{1}{2}\log|\sigmav^*|.
\end{align*}
Therefore, with some rearrangement, there holds
\begin{align*}
    &-x^\top(\sigmau^*)^{\frac{1}{2}}(\sigmav^*)^{-1}(\mubus-\mubvs) + \frac{1}{2}x^\top x + \frac{1}{2}\log|\sigmau^*|-\frac{1}{2}\log|\sigmav^*|\\
    &~~~~\ge \frac{1}{2}(\mubus-\mubvs)^\top(\sigmav^*)^{-1}(\mubus-\mubvs) + \frac{1}{2}x^\top(\sigmau^*)^{\frac{1}{2}}(\sigmav^*)^{-1}(\sigmau^*)^{\frac{1}{2}}x \\
    &~~~~\ge \frac{1}{2}\frac{\|\mubus-\mubvs\|^2}{\lambda_{\max}}+\frac{1}{8}\frac{\lambda_{\min}}{\lambda_{\max}}\snr^2>0. 
\end{align*}
Thuse, there further holds 
\begin{align*}
    &\frac{1}{2}\frac{\|\mubus-\mubvs\|^2}{\lambda_{\max}}+\frac{1}{8}\frac{\lambda_{\min}}{\lambda_{\max}}\snr^2\\
    &~~~~ \le \|-x^\top(\sigmau^*)^{\frac{1}{2}}(\sigmav^*)^{-1}(\mubus-\mubvs) + \frac{1}{2}x^\top x + \frac{1}{2}\log|\sigmau^*|-\frac{1}{2}\log|\sigmav^*|\| \\
    &~~~~ \le \|x^\top(\sigmau^*)^{\frac{1}{2}}(\sigmav^*)^{-1}(\mubus-\mubvs)\| + \frac{1}{2}x^\top x+\frac{1}{2}\|\log|\sigmau^*|-\log|\sigmav^*|\| \\
    &~~~~ \le \frac{1}{2}\snr\frac{\sqrt{\lambda_{\max}}}{\lambda_{\min}}\|\mubus-\mubvs\| + \frac{1}{8}\snr^2+\frac{1}{2}d\log\frac{\lambda_{\max}}{ \lambda_{\min}}.
\end{align*}
Therefore, when $\|\mubus-\mubvs\|^2\ge \omega\ge 2d \lambda_{\max} \log\frac{\lambda_{\max}}{ \lambda_{\min}}$, it is straightforward to validate that 
\[
\snr\ge \frac{1}{3}\frac{\lambda_{\min}}{ \lambda_{\max}}\frac{\|\mubus-\mubvs\|}{\sqrt{\lambda_{\max}}}\ge \frac{1}{3}\frac{\lambda_{\min}}{ \lambda_{\max}}\frac{\sqrt{\omega}}{\sqrt{\lambda_{\max}}}
\] 
is a necessary condition for the above inequality to hold.

Proceed to the upper bound. 
Assume $\omega=\min_{b, k\neq k'}\|\mubus-\mubvs\|^2$ is achieved at $(b, k, k')$. Take $x = -(\sigmau^*)^{-\frac{1}{2}}(\mubus-\mubvs)$. Then 
\begin{align*}
    &-x^\top(\sigmau^*)^{\frac{1}{2}}(\sigmav^*)^{-1}(\mubus-\mubvs) \\
    &~~~~ = (\mubus-\mubvs)^\top(\sigmav^*)^{-1}(\mubus-\mubvs)\\
    &~~~~ =\frac{1}{2}(\mubus-\mubvs)^\top(\sigmav^*)^{-1}(\mubus-\mubvs) + \frac{1}{2}x^\top(\sigmau^*)^{\frac{1}{2}}(\sigmav^*)^{-1}(\sigmau^*)^{\frac{1}{2}}x.
\end{align*}
Therefore, when $\omega\ge 2d \lambda_{\max} \log\frac{\lambda_{\max}}{ \lambda_{\min}}$, there further holds
\begin{align*}
    \frac{1}{2}x^\top x\ge\frac{1}{2}\frac{\omega}{\lambda_{\max}}\ge \frac{1}{2}d \log\frac{\lambda_{\max}}{ \lambda_{\min}}\ge-\frac{1}{2}(\log|\sigmau^*|-\log|\sigmav^*|),
\end{align*}
which leads to
\begin{align*}
    &x^\top(\sigmau^*)^{\frac{1}{2}}(\sigmav^*)^{-1}(\mubus-\mubvs)+\frac{1}{2}x^\top((\sigmau^*)^{\frac{1}{2}}(\sigmav^*)^{-1}(\sigmau^*)^{\frac{1}{2}}-I_d)x\\
    &~~~~ \le -\frac{1}{2}(\mubus-\mubvs)^\top(\sigmav^*)^{-1}(\mubus-\mubvs)+\frac{1}{2}\log|\sigmau^*|-\frac{1}{2}\log|\sigmav^*|.
\end{align*}
Therefore, $\snr \le 2\|x\|\le 2\frac{\sqrt{\omega}}{\sqrt{\lambda_{\min}}}$.

\section{Proof Outline of Proposition \ref{prop: clustering error}}

\label{sec: proof clustering error}

As we have mentioned in the proof of Theorem \ref{thm: upper bound}, 
our clustering problem should not be harder than that on data from a standard Gaussian mixture with $K \cdot B$ components, 
since the shared covariance matrices in model \eqref{eq: model} provide additional information.

Here, we outline the proof of Proposition \ref{prop: clustering error}, adapting a similar approach to that used in the proof of  \cite[Theorem 3.2]{chen2024achieving}. 
Due to technical similarity, we only focus on the main steps of the arguments.
 
To be specific, for an arbitrary cluster assignment $a$, consider a one-step iteration in Algorithm \ref{alg:em}, and define
\begin{align}
    &\hatmubk(a) = \frac{\sum_{i}\xbi\mI\{\abi=k\}}{\sum_{i}\mI\{\abi=k\}},\\
    &\hatsigmak(a) = \frac{\sum_{b, i}\mI\{\abi = k\} (\xbi-\hatmubk(a) ) (\xbi-\hatmubk(a) )^\top}{\sum_{b, i}\mI\{\abi = k\}},\\
    &\hatabi(a) = \argmin_k  (\xbi-\hatmubk(a) )^\top (\hatsigmak(a) )^{-1} (\xbi-\hatmubk(a) )+\log |\hatsigmak(a) |.
\end{align}
When $\abis = k$, mis-clustering of  cell $i$ in the $b$-th batch occurs if
\begin{align}
\label{eq: likelihood compare}
     (\xbi-\hatmubu(a) )^\top (\hatsigmau(a) )^{-1}& (\xbi-\hatmubu(a) )+\log |\hatsigmau(a) |\nonumber\\
    &\ge  (\xbi-\hatmubv(a) )^\top (\hatsigmav(a) )^{-1} (\xbi-\hatmubv(a) )+\log |\hatsigmav(a) |,
\end{align}
for some $k'\neq k$. Define $\mubus = \mu_k^*+\betabu^*$ and $\epsilon_{bi} = \xbi-\mubus$, then $\epsilon_{bi}\sim\cN(0, \sigmau^*)$. 
With some rearrangement, \eqref{eq: likelihood compare} is equivalent to
\begin{align}
    \label{eq: rearranged likelihood compare}
    &\langle\epsilon_{bi}, \hatsigmav(a^*)^{-1}(\mubus-\hatmubv(a^*))\rangle-\langle\epsilon_{bi}, \hatsigmau(a^*)^{-1}(\mubus-\hatmubu(a^*))\rangle\nonumber\\
    &~~~~~~~~ +\frac{1}{2} \langle\epsilon_{bi},  (\hatsigmav(a^*)^{-1}-\hatsigmau(a^*)^{-1} )\epsilon_{bi} \rangle-\frac{1}{2}\log |\hatsigmau(a^*) |+\frac{1}{2}\log |\hatsigmav(a^*) |\nonumber\\
    &~~~~ \le -\frac{1}{2} \langle\mubus-\mubvs,  (\sigmav^* )^{-1}(\mubus-\mubvs) \rangle+F_{bi}+Q_{bi}+G_{bi}+H_{bi}+K_{bi}+L_{bi},
\end{align}
where
\begin{align*}
    F_{bi}(a, k, k') & =  \langle\epsilon_{bi}, \hatsigmav(a)^{-1}(\hatmubv(a)-\hatmubv(a^*)) \rangle- \langle\epsilon_{bi}, \hatsigmau(a)^{-1}(\hatmubu(a)-\hatmubu(a^*)) \rangle \nonumber\\
    &~~~~ - \langle\epsilon_{bi},  (\hatsigmav(a)^{-1}-\hatsigmav(a^*)^{-1} )(\mubus-\hatmubv(a^*)) \rangle \nonumber\\
    &~~~~ + \langle\epsilon_{bi},  (\hatsigmau(a)^{-1}-\hatsigmau(a^*)^{-1} )(\mubus-\hatmubu(a^*)) \rangle,\\
    Q_{bi}(a, k, k') & =-\frac{1}{2} \langle\epsilon_{bi},  (\hatsigmav(a)^{-1}-\hatsigmav(a^*)^{-1} )\epsilon_{bi} \rangle+\frac{1}{2} \langle\epsilon_{bi},  (\hatsigmau(a)^{-1}-\hatsigmau(a^*)^{-1} )\epsilon_{bi} \rangle,\\
    G_{bi}(a, k, k') & = \frac{1}{2}\langle\mubus-\hatmubu(a),\hatsigmau(a)^{-1}(\mubus-\hatmubu(a))\rangle -\frac{1}{2}\langle\mubus-\hatmubu(a^*),\hatsigmau(a)^{-1}(\mubus-\hatmubu(a^*))\rangle\nonumber\\
    &~~~~ +\frac{1}{2}\langle\mubus-\hatmubu(a^*), (\hatsigmau(a)^{-1}-\hatsigmau(a^*)^{-1})(\mubus-\hatmubu(a^*))\rangle \nonumber\\
    &~~~~ -\frac{1}{2}\langle\mubus-\hatmubv(a),\hatsigmav(a)^{-1}(\mubus-\hatmubv(a))\rangle +\frac{1}{2}\langle\mubus-\hatmubv(a^*),\hatsigmav(a)^{-1}(\mubus-\hatmubv(a^*))\rangle\nonumber\\
    &~~~~ -\frac{1}{2}\langle\mubus-\hatmubv(a^*), (\hatsigmav(a)^{-1}-\hatsigmav(a^*)^{-1})(\mubus-\hatmubv(a^*))\rangle,\\
    H_{bi}(a, k, k') & = \frac{1}{2}\langle\mubus-\hatmubu(a^*), \hatsigmau(a^*)^{-1}(\mubus-\hatmubu(a^*))\rangle \nonumber\\
    &~~~~ -\frac{1}{2}\langle\mubus-\hatmubv(a^*), \hatsigmav(a^*)^{-1}(\mubus-\hatmubv(a^*))\rangle+\frac{1}{2}\langle\mubus-\mubvs, \hatsigmav(a^*)^{-1}(\mubus-\mubvs)\rangle \nonumber\\
    &~~~~ -\frac{1}{2}\langle\mubus-\mubvs, (\hatsigmav(a^*)^{-1}-(\sigmav^*)^{-1})(\mubus-\mubvs)\rangle,\\
    K_{bi}(a, k, k') & = \frac{1}{2}(\log |\hatsigmau(a) |-\log |\hatsigmau(a^*) |)-\frac{1}{2}(\log |\hatsigmav(a) |-\log |\hatsigmav(a^*) |),\\
    L_{bi}(a, k, k') & = \frac{1}{2}(\log |\hatsigmau(a^*) |-\log |\sigmau^* |)-\frac{1}{2}(\log |\hatsigmav(a^*) |-\log |\sigmav^* |).
\end{align*}
Using similar arguments as Lemma B.3 in \cite{chen2024achieving} and applying Lemma \ref{lemma: estimation bounds}, we have the following lemma stating that $F_{bi}+Q_{bi}+G_{bi}+H_{bi}+K_{bi}+L_{bi}$ can be absorbed into the term $\frac{\delta_n}{2}\langle\mubus-\mubvs,  (\sigmav^* )^{-1}(\mubus-\mubvs) \rangle$ for some $0<\delta_n = o(1)$.
\begin{lemma}
\label{lemma: delta estimation}
    Under the conditions of Proposition \ref{prop: clustering error}, for any $C'>0$, there exist some $\delta_n=o(1)$, such that
    \begin{align*}
        &\max_{b\in [B]}\max_{a:\ell(a, a^*)\le \tau}\sum_{i}\max_{k'\in [k]\setminus\{\abis\}}\frac{F_{bi}(a, \abis, k')^2\|\mubabis-\mubvs\|^2}{\langle\mubabis-\mubvs,(\sigmav^*)^{-1}(\mubabis-\mubvs)\rangle^2\ell(a, a^*)}\le 
        \frac{\delta_n^2}{288},
        \\
        &\max_{b\in [B]}\max_{a:\ell(a, a^*)\le \tau}\max_{i\in[n_b]}\max_{k'\in [k]\setminus\{\abis\}}\frac{|H_{bi}(a, \abis, k')|}{\langle\mubabis-\mubvs,(\sigmav^*)^{-1}(\mubabis-\mubvs)\rangle}\le \frac{\delta_n}{12},
        \\
        &\max_{b\in [B]}\max_{a:\ell(a, a^*)\le \tau}\max_{i\in[n_b]}\max_{k'\in [k]\setminus\{\abis\}}\frac{|G_{bi}(a, \abis, k')|}{\langle\mubabis-\mubvs,(\sigmav^*)^{-1}(\mubabis-\mubvs)\rangle}\le \frac{\delta_n}{12},
        \\
        &\max_{b\in [B]}\max_{a:\ell(a, a^*)\le \tau}\sum_{i}\max_{k'\in [k]\setminus\{\abis\}}\frac{Q_{bi}(a, \abis, k')^2\|\mubabis-\mubvs\|^2}{\langle\mubabis-\mubvs,(\sigmav^*)^{-1}(\mubabis-\mubvs)\rangle^2\ell(a, a^*)}\le \frac{\delta_n^2}{288},
        \\
        &\max_{b\in [B]}\max_{a:\ell(a, a^*)\le \tau}\sum_{i}\max_{k'\in [k]\setminus\{\abis\}}\frac{K_{bi}(a, \abis, k')^2\|\mubabis-\mubvs\|^2}{\langle\mubabis-\mubvs,(\sigmav^*)^{-1}(\mubabis-\mubvs)\rangle^2\ell(a, a^*)}\le \frac{\delta_n^2}{288},
        \\
        &\max_{b\in [B]}\max_{a:\ell(a, a^*)\le \tau}\max_{i\in[n_b]}\max_{k'\in [k]\setminus\{\abis\}}\frac{|L_{bi}(a, \abis, k')|}{\langle\mubabis-\mubvs,(\sigmav^*)^{-1}(\mubabis-\mubvs)\rangle}\le \frac{\delta_n}{12},
    \end{align*}
    holds with probability at least $1-n^{-C'}-\frac{4}{nd}$.
\end{lemma}

Define the ideal error $\ideal(\delta)$ for any $\delta>0$
\begin{align*}
    \ideal(\delta) =\sum_{b, i}\sum_{k'\in[K]\setminus\{\abis\}} & \|\mubabis-\mubvs\|^2\mI\{\langle\epsilon_{bi}, \hatsigmav(a^*)^{-1}(\mubabis-\hatmubv(a^*))\rangle \nonumber\\
    & -\langle\epsilon_{bi}, \hatsigmaabis(a^*)^{-1}(\mubabis-\hatmubabis(a^*))\rangle+\frac{1}{2} \langle\epsilon_{bi},  (\hatsigmav(a^*)^{-1}-\hatsigmaabis(a^*)^{-1} )\epsilon_{bi} \rangle\nonumber\\
    & -\frac{1}{2}\log |\hatsigmaabis(a^*) |+\frac{1}{2}\log |\hatsigmav(a^*) |\\
    & \hskip -7em
    \le -\frac{1-\delta}{2} \langle\mubabis-\mubvs,  (\sigmav^* )^{-1}(\mubabis-\mubvs) \rangle\}.
\end{align*}
Applying Lemma \ref{lemma: delta estimation} on \eqref{eq: rearranged likelihood compare}, we have that when $\ell(a^{(0)}, a^*)\le \tau =  o(n)$ for some $\tau>0$, for any $C'>0$, there exists some $0<\delta_n=o(1)$, such that
\begin{align}
    \ell(a^{(t+1)}, a^*)\le \ideal(\delta_n) + \frac{1}{2}\ell(a^{(t)}, a^*),\label{eq: induction}
\end{align}
holds  with probability of at least $1-n^{-C'}-\frac{4}{nd}$ for all $t\ge 1$,
Furthermore, similar to Lemma B.4 in \cite{chen2024achieving}, for any $0<\delta_n=o(1)$, we have
\begin{align}
    \ideal(\delta_n)\le \Omega n\exp\left(-(1+o(1))\frac{\snr^2}{8}\right),\label{eq: ideal}
\end{align}
with probability at least $1-n^{-C'}-\exp(-\snr)$.

Finally, by mathematical induction on \eqref{eq: induction} and \eqref{eq: ideal}, there further holds 
\begin{align*}
    \ell(a^{(t)}, a^*)\le \Omega n\exp\left(-(1+o(1))\frac{\snr^2}{8}\right),
\end{align*}
with probability at least $1-\eta-5n^{-1}-\exp(-\snr)$ for all $t\ge\log n$.

\section{Proof of Lemma \ref{lemma: bound muu a astar}}
\label{sec: proof lemma muu a astar}

Let $p_{bk} = \frac{\sum_{i}\mI\{\abi^{(t)} = k\}}{\sum_{b', i}\mI\{\abip^{(t)} = k\}}$ and $q_{bk} = \frac{n_{bk}}{\sum_{b'}n_{b'k}}$, then there holds
\begin{align}
    \|\hatmuu(a^{(t)})-\hatmuu(a^*)\|^2 & = \left\|\sum_b p_{bk}\hatmubu(a^{(t)}) - \sum_b q_{bk}\hatmubu(a^*)\right\|^2\nonumber\\
    &  \le 2B\left(\sum_bq_{bk}^2\left\|\hatmubu(a^{(t)})-\hatmubu(a^*)\right\|^2 \right.\nonumber\\
    &~~~~~~~~~~~~~~\left. +\sum_b(p_{bk}-q_{bk})^2\|\hatmubu(a^{(t)})\|^2\right),\label{eq: decomposition muu a astar}
\end{align}
where the inequality holds due to the Cauchy–-Schwarz inequality and the fact that $p_{bk}\hatmubu(a^{(t)}) - q_{bk}\hatmubu(a^*) = q_{bk}(\hatmubu(a^{(t)})-\hatmubu(a^*)) + (p_{bk}-q_{bk})\hatmubu(a^{(t)})$.

For the first term in the parentheses on the rightmost side of \eqref{eq: decomposition muu a astar}, 
by the first part in Condition \ref{cond: regularity}, there holds
\begin{align*}
    q_{bk} = \frac{n_{bk}}{\sum_{b'}n_{b'k}}\le \frac{1}{\alpha B}.
\end{align*}
Therefore, 
\begin{align*}
    \sum_bq_{bk}^2\left\|(\hatmubu(a^{(t)})-\hatmubu(a^*))\right\|^2\le \frac{1}{\alpha^2 B}\max_{b, u}\|\hatmubu(a^{(t)})-\hatmubu(a^*)\|^2.
\end{align*}

For the second term in the parentheses on the rightmost side of \eqref{eq: decomposition muu a astar}, note that on the event that \eqref{eq: event cluster error} holds, there holds $\frac{1}{n}\sum_{b, i}\mI\{\abi^{(t)}\neq\abis\}\le \exp\left(-(1+o(1))\frac{\snr^2}{8}\right) = o(1)\le \frac{1}{2n}\sum_{b', i}\mI\{\abisp = k\} = O(1)$. Therefore, there exists $C>0$, such that
\begin{align*}
    |p_{bk} - q_{bk}| & = \left|\frac{\sum_{i}\mI\{\abi^{(t)} = k\}}{\sum_{b', i}\mI\{\abip^{(t)} = k\}} - \frac{\sum_{i}\mI\{\abis = k\}}{\sum_{b', i}\mI\{\abisp = k\}}\right| \\
    & \le \left|\frac{\sum_{i}\mI\{\abis = k\}+\sum_{b', i}\mI\{\abip^{(t)}\neq \abisp\}}{\sum_{b', i}\mI\{\abisp = k\}-\sum_{b',i}\mI\{\abip^{(t)}\neq \abisp\}} - \frac{\sum_{i}\mI\{\abis = k\}}{\sum_{b', i}\mI\{\abisp = k\}}\right| \\
    & \le 2\sum_{b',i}\mI\{\abip^{(t)}\neq \abisp\}\frac{1}{\sum_{b', i}\mI\{\abisp = k\}-\sum_{b',i}\mI\{\abip\neq \abisp\}} \\
    & \le 4\sum_{b',i}\mI\{\abip^{(t)}\neq \abisp\} \frac{1}{\sum_{b', i}\mI\{\abisp = k\}} \\
    &\le \exp\left(-(1+o(1))\frac{\snr^2}{8}\right),
\end{align*}
where the first inequality is due to $|\sum_{i}\mI\{\abi^{(t)} = k\}-\sum_{i}\mI\{\abis = k\}|\le \sum_{b',i}\mI\{\abip^{(t)}\neq \abisp\}$ and $|\sum_{b',i}\mI\{\abip^{(t)}= k\}-\sum_{b',i}\mI\{\abisp = k\}|\le \sum_{b',i}\mI\{\abip^{(t)}\neq \abisp\}$. 
Therefore $(p_{bk} - q_{bk})^2\|\hatmubu(a^{(t)})\|^2\le \exp\left(-(1+o(1))\frac{\snr^2}{4}\right)$ can be absorbed into the first term in the parentheses
due to \eqref{eq: mubu accuracy} in Lemma \ref{lemma: estimation bounds}.

Summarizing the above two points and applying \eqref{eq: mubu accuracy} in Lemma \ref{lemma: estimation bounds}, we have for \eqref{eq: decomposition muu a astar}, there exists $C>0$ that depends on $\alpha, \lambda_{\max}=\max_k(\lambda_{\max} \sigmau^*), \lambda_{\min}=\min_k(\lambda_{\min} \sigmau^*)$, such that
\begin{align*}
    \|\hatmuu(a^{(t)})-\hatmuu(a^*)\|^2& \le \frac{2}{\alpha^2 B} \max_{b, u}\|\hatmubu(a^{(t)})-\hatmubu(a^*)\|^2\le \frac{CK^2}{\omega n}\ell(a^{(t)}, a^*),
\end{align*}
with probability at least $1-n^{-2}$ on the event where \eqref{eq: event cluster error} holds.

\section{Proof of Lemma \ref{lemma: bound muu astar true}} 
\label{sec: proof lemma muu astar true}

Let $q_{bk} = \frac{n_{bk}}{\sum_{b'}n_{b'k}}$, then there holds
\begin{align}
    \|\hatmuu(a^*)-\mu_k^*\|^2 & = \left\|\sum_b q_{bk}\hatmubu(a^*) - \sum_b q_{bk}\mubus\right\|^2 \nonumber\\
    & \le 2B\sum_bq_{bk}^2\left\|(\hatmubu(a^*)-\mubus)\right\|^2,\label{eq: decomposition muu astar true}
\end{align}
where the inequality holds due to the Cauchy–Schwarz inequality.
Note that 
\begin{align*}
    q_{bk} = \frac{n_{bk}}{\sum_{b'}n_{b'k}}\le \frac{1}{\alpha B}.
\end{align*}
Therefore,
\begin{align*}
    \sum_bq_{bk}^2\left\|(\hatmubu(a^*)-\mubus)\right\|^2\le \frac{1}{\alpha^2 B}\max_{b, k}\|\hatmubu(a^*)-\mubus\|^2.
\end{align*}


By \eqref{eq: mubu accuracy given true assignment} in Lemma \ref{lemma: estimation bounds}, there exists $C>0$ that depends on $\alpha, \lambda_{\max}=\max_k(\lambda_{\max} \sigmau^*), \lambda_{\min}=\min_k(\lambda_{\min} \sigmau^*)$, such that
\begin{align*}
    \|\hatmuu(a^*)-\mu_u^*\|^2 \le  \frac{CK\log n}{n},
\end{align*}
with probability at least $1-n^{-2}$  on the event where \eqref{eq: event cluster error} holds.

\section{Technical Lemmas}
\label{sec: technical lemmas}
In this section, for the manuscript to be self-contained, we introduce lemmas whose proofs follow  the same arguments as in Section C of \cite{chen2024achieving}. 
Hence, their proofs are omitted. 

We first introduce a technical lemma bounding terms involving random noises $\epsilon_{bi}$ in \eqref{eq: model1} with high probability.
\begin{lemma}
    \label{lemma: technical}
    For any $C'>0$, there exists some constant $C>0$ that depends on $\alpha, \lambda_{\max}:=\max_k(\lambda_{\max} \sigmau^*), \lambda_{\min}:=\min_k(\lambda_{\min} \sigmau^*), C'$, such that the following statements hold with probability of at least $1-n^{-C'}$.
    \begin{enumerate}
        \item 
        There hold
    \begin{align*}
        &\max_{k\in[K]}\left\|\frac{\sum_{i}\mI\{\abis = k\}\epsilon_{bi}}{\sqrt{\sum_{i}\mI\{\abis = k\}}}\right\|\le C\sqrt{d+\log(n/B)},\\
        &\max_{k\in[K]}\frac{1}{d+\sum_{i}\mI\{\abis = k\}}\left\|\sum_{i}\mI\{\abis = k\}\epsilon_{bi}\epsilon_{bi}^\top\right\|\le C,\\
        &\max_{T\subset[n_b]}\left\|\frac{1}{\sqrt{|T|}}\sum_{i\in T}\epsilon_{bi}\right\|\le C\sqrt{d+n/B},\\
        &\max_{k\in[K]}\max_{T\subset\{i:\abis = k\}}\left\|\frac{1}{\sqrt{|T|(d+\sum_{i}\mI\{\abis = k\})}}\sum_{i\in T}\epsilon_{bi}\right\|\le C.
    \end{align*}
    \item 
    There hold
    \begin{align*}
        \max_{k\in[K]}\left\|\frac{1}{\sum_{b, i}\mI\{\abis = k\}}\sum_{b, i}\mI\{\abis = k\}\epsilon_{bi}\epsilon_{bi}^\top-\sigmau^*\right\|\le C\sqrt{\frac{K(d+\log n)}{n}}.
    \end{align*}
    \item 
    There hold for any $s=o(n)$
    \begin{align*}
        \max_{T\subset\cup_{b=1}^B\{(b, i): i\in[n_b]\}: |T|\le s}\frac{1}{|T|\log(n/|T|)+\min\{1, \sqrt{|T|}d\}}\left\|\sum_{(b, i)\in T}\epsilon_{bi}\epsilon_{bi}^\top\right\|\le C.
    \end{align*}
    \end{enumerate}
\end{lemma}
The bounds in Lemma \ref{lemma: technical} follow from the same arguments as those for Lemmas C.3--C.5 of \cite{chen2024achieving}.

With Lemma \ref{lemma: technical}, following the same arguments as in Lemma C.7 in \cite{chen2024achieving}, we have the following lemma.

\begin{lemma}
    \label{lemma: estimation bounds}
    For any $\tau = o(n)$ and any $C'>0$, there exists some constant $C>0$ that depends on $\alpha, \lambda_{\max}, \lambda_{\min}, C'$, such that the following statements hold given $\ell(a, a^*)\le \tau$ with probability  at least $1-n^{-C'}$.
    \begin{align}
        &\max_{k\in[K]}\|\hatmubu(a^*)-\mubus\|\le C\sqrt{\frac{BK(d+\log(n/B))}{n}},\label{eq: mubu accuracy given true assignment}\\
        &\max_{k\in[K]}\|\hatmubu(a)-\hatmubu(a^*)\|\le C\left(\frac{BK}{n\sqrt{\omega}}\ell(a, a^*)+\frac{BK\sqrt{d+n/B}}{n\sqrt{\omega}}\sqrt{\ell(a, a^*)}\right),\label{eq: mubu accuracy}\\
        &\max_{k\in[K]}\|\hatsigmau(a^*)-\sigmau^*\|\le C\sqrt{\frac{K(d+\log n)}{n}},\\
        &\max_{k\in[K]}\|\hatsigmau(a^*)-\sigmau^*\|\le C\left(\frac{BK}{n\sqrt{\omega}}\ell(a, a^*)+\frac{BK(\sqrt{d+n/B})+Kd}{n\sqrt{\omega}}\sqrt{\ell(a, a^*)}\right),\\
        &\max_{k\in[K]} \|\hatsigmau(a^*)^{-1}-(\sigmau^*)^{-1}\|\le C\left(\frac{BK}{n\sqrt{\omega}}\ell(a, a^*)+\frac{BK(\sqrt{d+n/B})+Kd}{n\sqrt{\omega}}\sqrt{\ell(a, a^*)}\right).
    \end{align}
\end{lemma}

\section{Definitions of Benchmarking Metrics}

\label{sec: metrics}

 {In this section, we define the metrics in \cite{luecken2022benchmarking} that have been used in Sections \ref{sec: simu} and \ref{sec: real world} to assess the performance of batch correction methods. }
As suggested in \cite{luecken2022benchmarking}), all metrics are rescaled to the unit interval $[0, 1]$, where higher scores indicate better performances.

\subsection{Biological Conservation Metrics}

\paragraph{Isolated Labels Score.}
This metric assesses the method's ability to capture rare or batch-specific cell identities. For each isolated cell label $c \in C_{\text{iso}}$ (where $C_{\text{iso}}$ is the set of labels present in a minimal number of batches), we perform Leiden clustering on the integrated graph. We then identify the cluster $k$ that maximizes the $F_1$ score for label $c$. The final score is the mean $F_1$ score across all isolated labels:
\begin{equation*}
    \mathrm{average}(F_1)=\frac{1}{|C_{\text{iso}}|} \sum_{c \in C_{\text{iso}}} \max_{k} \left( \frac{2 \cdot P(c, k) \cdot R(c, k)}{P(c, k) + R(c, k)} \right).
\end{equation*}
Here, $P(c, k)$ denotes the precision of cluster $k$ with respect to label $c$ (the fraction of cells in cluster $k$ that belong to label $c$), and $R(c, k)$ denotes the recall (the fraction of all cells of label $c$ that are assigned to cluster $k$).

\paragraph{Global Clustering Fidelity (Leiden NMI and Leiden ARI).}
We evaluate the global clustering structure by applying Leiden clustering to the integrated data and comparing the resulting partition $C_{\text{leiden}}$ to the ground-truth cell type labels $C_{\text{bio}}$.
\begin{itemize}
    \item \textbf{Normalized Mutual Information (NMI):}
    \begin{equation*}
        \text{NMI}(C_{\text{bio}}, C_{\text{leiden}}) = \frac{2 \cdot I(C_{\text{bio}}; C_{\text{leiden}})}{H(C_{\text{bio}}) + H(C_{\text{leiden}})},
    \end{equation*}
    where $I(\cdot;\cdot)$ is the mutual information and $H(\cdot)$ is the entropy.
    \item \textbf{Adjusted Rand Index (ARI):}
    \begin{equation*}
        \text{ARI} = \frac{\sum_{ij} \binom{n_{ij}}{2} - [\sum_i \binom{a_i}{2} \sum_j \binom{b_j}{2}] / \binom{n}{2}}{\frac{1}{2} [\sum_i \binom{a_i}{2} + \sum_j \binom{b_j}{2}] - [\sum_i \binom{a_i}{2} \sum_j \binom{b_j}{2}] / \binom{n}{2}}.
    \end{equation*}
    This metric relies on a contingency table describing the overlap between the two partitions, where $n_{ij}$ represents the number of cells common to biological label $i$ and cluster $j$. The terms $a_i = \sum_j n_{ij}$ and $b_j = \sum_i n_{ij}$ are the marginal sums, representing the total number of cells in biological label $i$ and cluster $j$, respectively.
\end{itemize}

\paragraph{Local Neighborhood Purity (Silhouette Label).}
To measure local cluster compactness, we compute the Average Silhouette Width (ASW) based on cell identity labels. For a cell $i$, the silhouette width is $s(i) = \frac{b(i) - a(i)}{\max\{a(i), b(i)\}}$, where $a(i)$ is the mean intra-cluster distance and $b(i)$ is the mean nearest-cluster distance. The raw scores are rescaled to $[0, 1]$:
\begin{equation*}
    \text{ASW}_{\text{label}} = \frac{1}{N} \sum_{i=1}^{N} \frac{s(i) + 1}{2}.
\end{equation*}
A score of 1 indicates that cells are optimally co-located with cells of the same type and well-separated from distinct types.

\paragraph{Cell-type Local Inverse Simpson's Index (cLISI).}
The cLISI metric quantifies the local label diversity in the $k$-nearest neighbor ($k$NN) graph. For each cell $i$, we calculate the raw LISI score with respect to cell type labels, denoted as $x_i$, which represents the effective number of cell types in the local neighborhood. The final metric is derived by computing the median of these raw scores, $\tilde{x} = \text{median}(\{x_i\}_{i=1}^N)$, and linearly rescaling it such that 1 indicates a pure neighborhood:
\begin{equation*}
    \text{cLISI} = \frac{|C| - \tilde{x}}{|C| - 1},
\end{equation*}
where $|C|$ is the total number of cell types.

\subsection{Batch Effect Removal Metrics}
These metrics quantify the removal of batch effect after integration.

\paragraph{Silhouette Batch (ASW Batch).}
We utilize the absolute silhouette width to assess mixing. Here, the silhouette score $s_{\text{batch}}(i)$ is computed using the standard silhouette formula, but with batch assignments as the cluster labels rather than cell types. Since ideal mixing corresponds to a raw silhouette score of 0 (indicating cells are equidistant to cells of other batches as they are to their own), we rescale the absolute ASW:
\begin{equation*}
    \text{ASW}_{\text{batch}} = 1 - \frac{1}{N} \sum_{i=1}^{N} |s_{\text{batch}}(i)|.
\end{equation*}
A final score of 1 indicates that cells are indistinguishable by batch (perfect mixing).

\paragraph{Integration Local Inverse Simpson's Index (iLISI).}
To assess local mixing, we calculate the raw LISI score for each cell $i$ using batch labels, denoted as $y_i$. This value represents the effective number of batches in the local neighborhood. The final iLISI score is obtained by linearly rescaling the median of the raw scores, $\tilde{y} = \text{median}(\{y_i\}_{i=1}^N)$, such that 1 signifies perfect mixing (equal representation of all batches):
\begin{equation*}
    \text{iLISI} = \frac{\tilde{y} - 1}{|B| - 1},
\end{equation*}
where $|B|$ is the total number of batches.

\paragraph{k-Nearest-Neighbor Batch Effect Test (kBET).}
kBET compares the local batch distribution in $k$NN neighborhoods to the global batch distribution using a $\chi^2$ test. The raw metric is the average rejection rate (RR) of the null hypothesis (that local and global distributions are identical). The scaled score is:
\begin{equation*}
    \text{kBET}_{\text{score}} = 1 - \overline{\text{RR}},
\end{equation*}
where a score of 1 indicates ideal mixing (0\% rejection rate).

\paragraph{Graph Connectivity.}
This metric evaluates whether cells of the same type remain connected across the integration graph. For each cell type $c$, we compute the size of the largest connected component (LCC) in the subgraph $G_c$ restricted to cells of type $c$:
\begin{equation*}
    \text{GC} = \frac{1}{|C|} \sum_{c \in C} \frac{|\text{LCC}(G_c)|}{|N_c|},
\end{equation*}
where $|N_c|$ is the total number of cells of type $c$. A score of 1 implies all cell types are fully connected across batches.


\section{Supplementary Simulation Experiments}

\label{sec: sup simu}

In this section, we provide supplementary simulation experiments to assess the robustness of {\method} under deviations from the initial settings presented in Section \ref{sec: simu}. Specifically, we evaluate its performance under three conditions: (1) with an increased number of clusters ($K$); (2) when the data is generated from a heavy-tailed distribution instead of a Gaussian; and (3) in scenarios with unbalanced batch compositions where some clusters are missing from certain batches.

\subsection{Performance with an Increased Number of Clusters}

In this section, we evaluate the performance of {\method} in a simulation where the number of clusters ($K$) is increased to 10, compared to $K$=4 in the original experiment (Section \ref{sec: simu}). For this analysis, the matrix $\Pi\in \mR^{B\times K}$, which determines $n_{bk} \approx n_b \cdot \Pi_{bk}$, is taken as
\begin{align*}
    \Pi = \frac{1}{3}\left[\begin{array}{cccccccccc}
        0.5 & 0.4 & 0.3 & 0.2 & 0.1 & 0.5 & 0.4 & 0.3 & 0.2 & 0.1 \\
        0.1 & 0.2 & 0.3 & 0.4 & 0.5 & 0.1 & 0.2 & 0.3 & 0.4 & 0.5 \\
        0.3 & 0.3 & 0.3 & 0.3 & 0.3 & 0.3 & 0.3 & 0.3 & 0.3 & 0.3
    \end{array}\right].
\end{align*}
In this section, we consider settings 1 and 3 in Section \ref{sec: simu1} and the setting in Section \ref{subsec:Khat} with $u=\exp(1)$ and $v = 5$.

\begin{figure}[!ht]
\centering
\subfigure[$\log$ loss vs. $\log(u)$]{
\begin{minipage}[t]{0.45\textwidth}
\centering
\includegraphics[width=\textwidth]{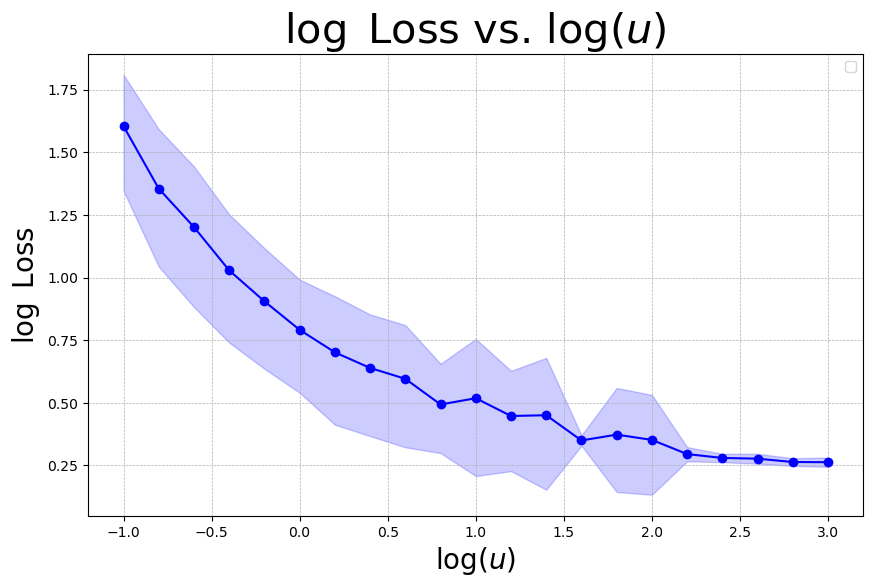}
\end{minipage}%
}%
\subfigure[$\log$ loss vs. $v$]{
\begin{minipage}[t]{0.45\textwidth}
\centering
\includegraphics[width=\textwidth]{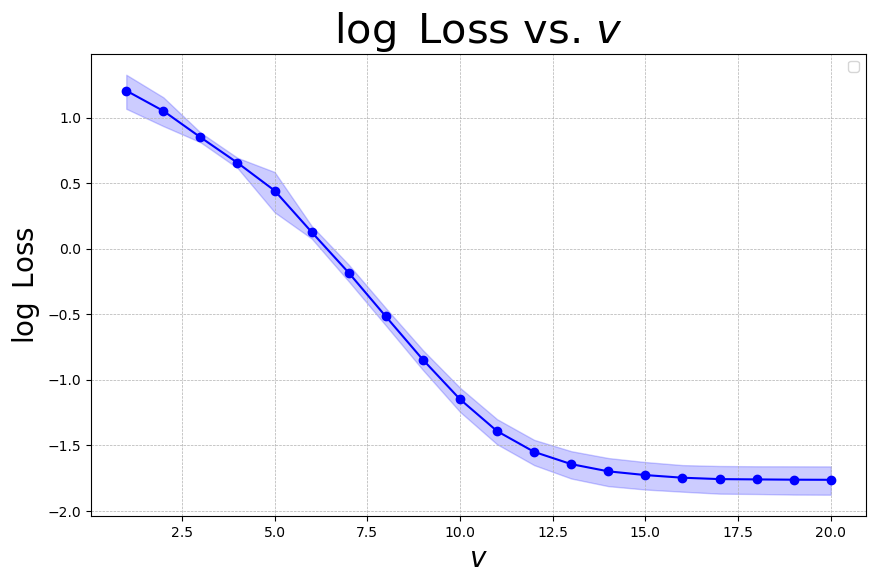}
\end{minipage}%
}%

\subfigure[Batch Correction Loss]{
\begin{minipage}[t]{0.32\textwidth}
\centering
\includegraphics[width=\textwidth]{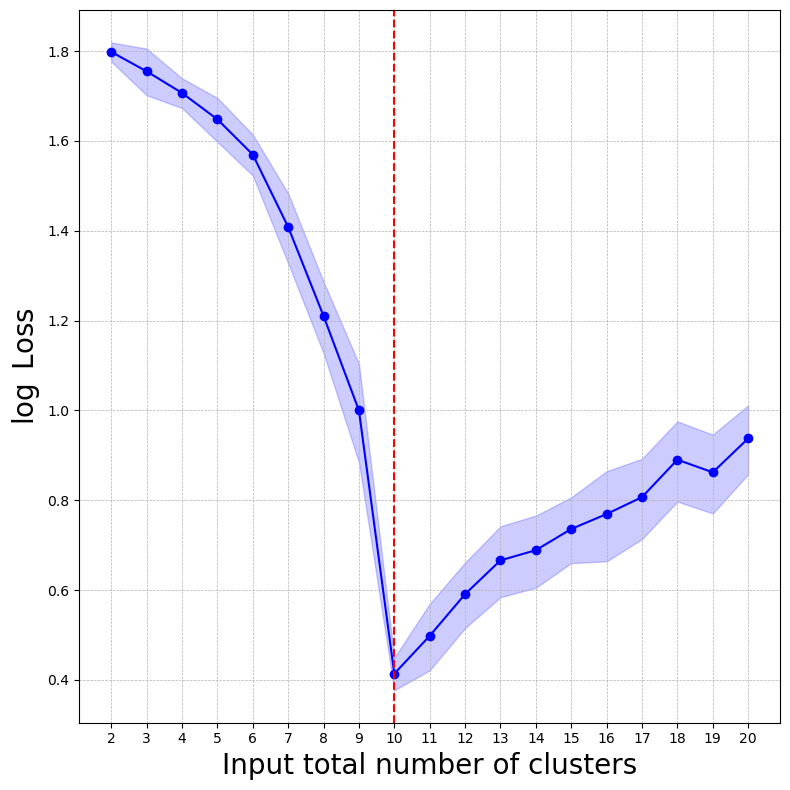}
\end{minipage}%
}%
\subfigure[Bio Conservation Metrics]{
\begin{minipage}[t]{0.32\textwidth}
\centering
\includegraphics[width=\textwidth]{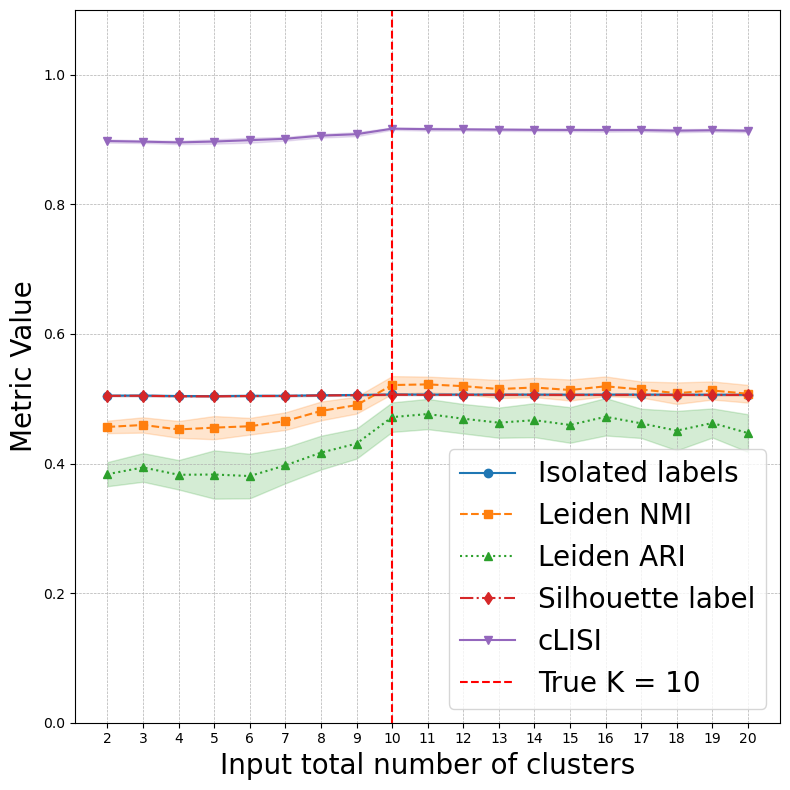}
\end{minipage}%
}%
\subfigure[Batch Correction Metrics]{
\begin{minipage}[t]{0.32\textwidth}
\centering
\includegraphics[width=\textwidth]{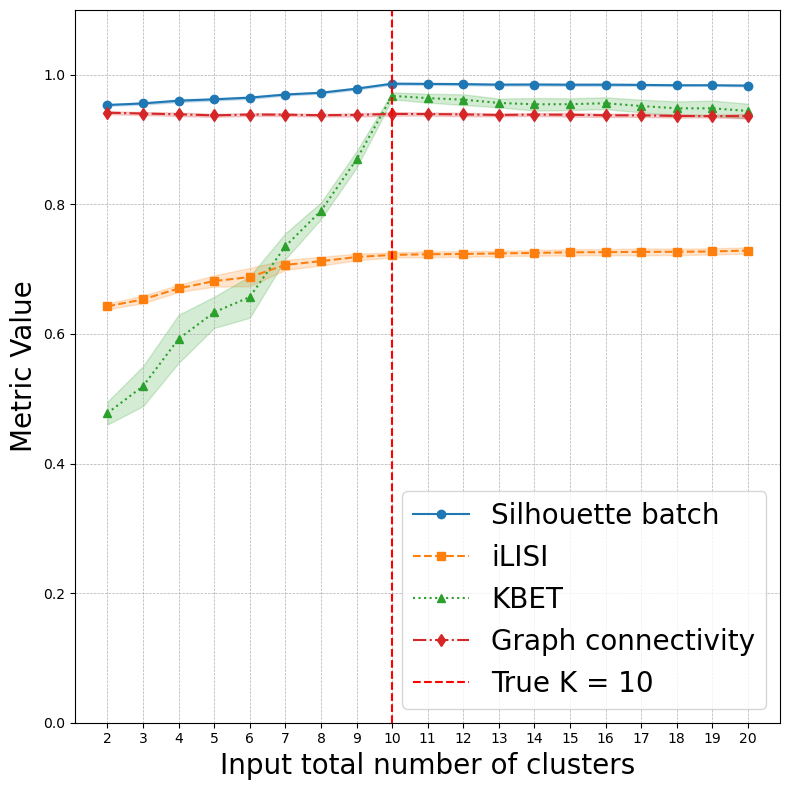}
\end{minipage}%
}%
\caption{The performance of {\method} when $u$, $v$ or input total number of clusters varies with $K=10$. 
In the top row and the left panel in the bottom row, we plot the logarithm of the average loss versus $\log(u)$, $v$, and the input total number of clusters, respectively (with the ends of the shaded area corresponding to the logarithms of the average loss $\pm$ one standard deviation). 
In the middle and the right panels of the bottom row, we plot the average scores of \texttt{scib-metrics} metrics versus the input number of clusters, and the shaded area around each data point represents $\pm$ one standard deviation. 
All results are obtained from 20 simulation instances.
}
\label{fig: simulation with u v changes large k}
\end{figure}

With an increased number of clusters, the performance of {\method} as a function of the parameters $u$, $v$, and as a function of the input total number of clusters is plotted in Figure \ref{fig: simulation with u v changes large k}. The observed trends are consistent with those reported in Section \ref{sec: simu}. 

\subsection{Performance on Non-Gaussian Data}

In this section, we test the robustness of {\method} with respect to deviation from Gaussianity by generating data from a multivariate $t$ distribution.
Specifically, $\xbi \in \mR^d$ for cell $i$ in batch $b$ with $\abis = k$ is sampled from the following multivariate $t$-distribution mixture model:
\begin{equation} \label{eq: model t distribution}
    \xbi \sim t_d(\muks + \betabks, \sigmaks, 5),
\end{equation}
where $t_d(\mu, \Sigma, \nu)$ denotes the multivariate t-distribution with $\nu$ degrees of freedom. 
For the plot of $\log$(average loss) vs. $\log(u)$,  we fix $v = 10$, and for the plot of $\log$(average loss) vs. $v$, we fix $u=\exp(1)$.
For the plots with varying input total number of clusters, we fix $v = 10$ and $u=\exp(1)$.
All other simulation parameters remain the same as in Section \ref{sec: simu}. 


\begin{figure}[!ht]
\centering
\subfigure[$\log$ loss vs. $\log(u)$]{
\begin{minipage}[t]{0.45\textwidth}
\centering
\includegraphics[width=\textwidth]{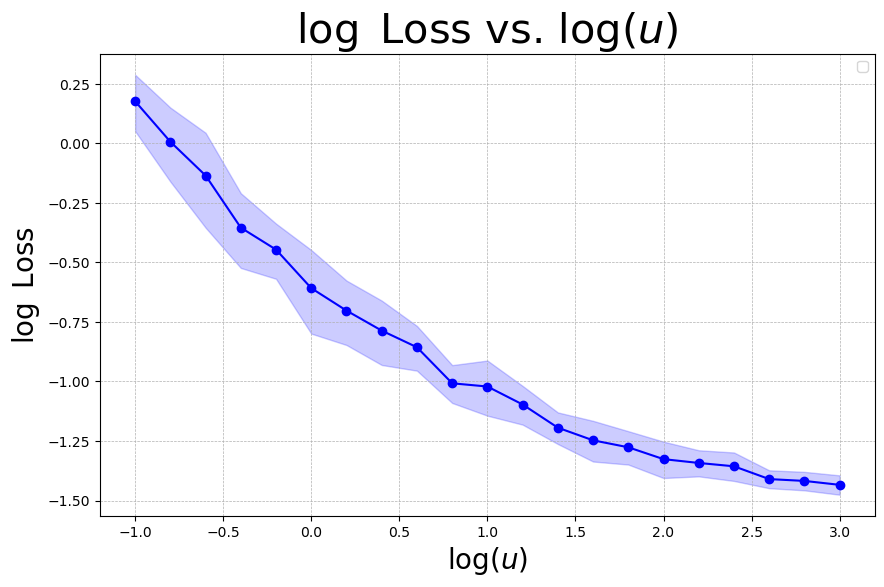}
\end{minipage}%
}%
\subfigure[$\log$ loss vs. $v$]{
\begin{minipage}[t]{0.45\textwidth}
\centering
\includegraphics[width=\textwidth]{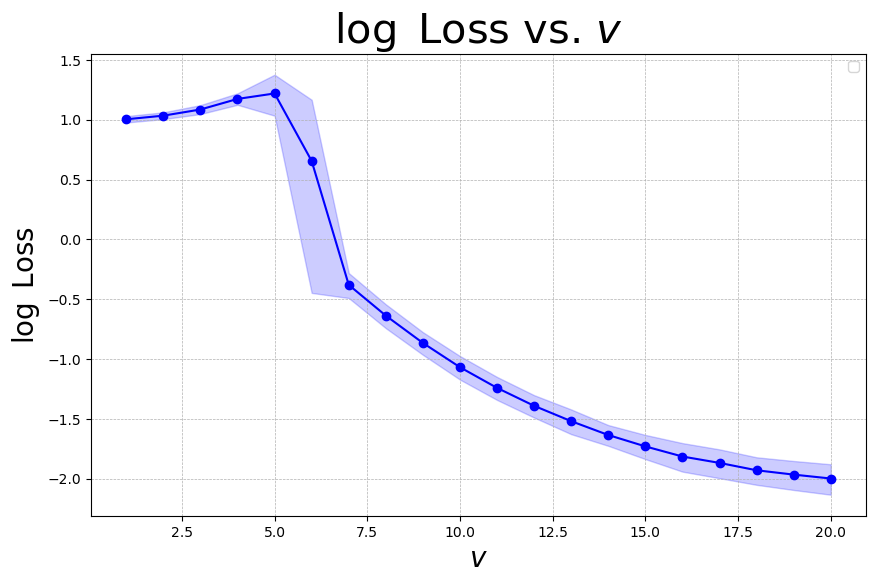}
\end{minipage}%
}%

\subfigure[Batch Correction Loss]{
\begin{minipage}[t]{0.32\textwidth}
\centering
\includegraphics[width=\textwidth]{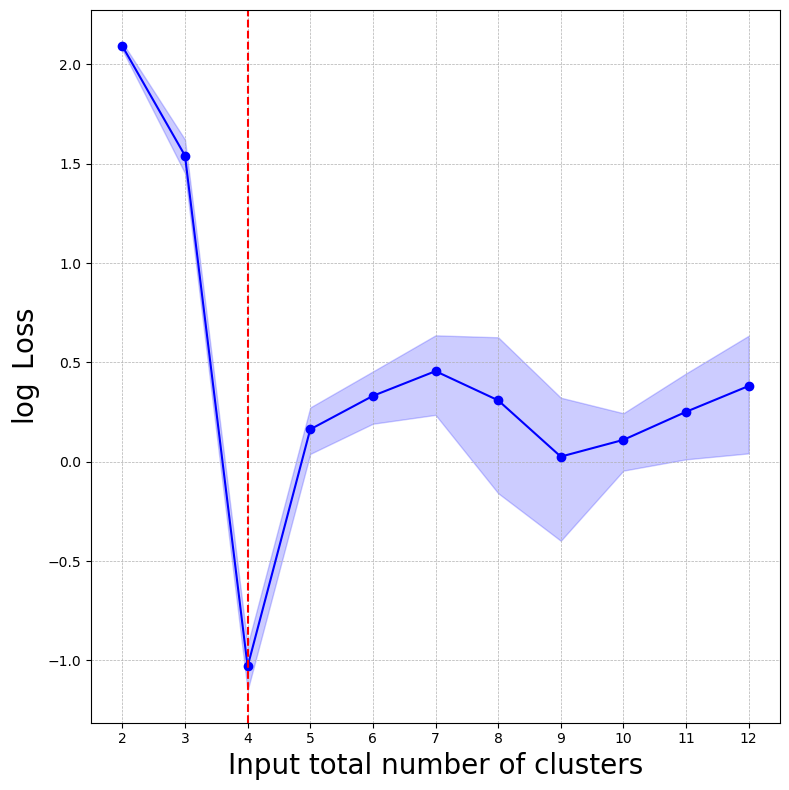}
\end{minipage}%
}%
\subfigure[Bio Conservation Metrics]{
\begin{minipage}[t]{0.32\textwidth}
\centering
\includegraphics[width=\textwidth]{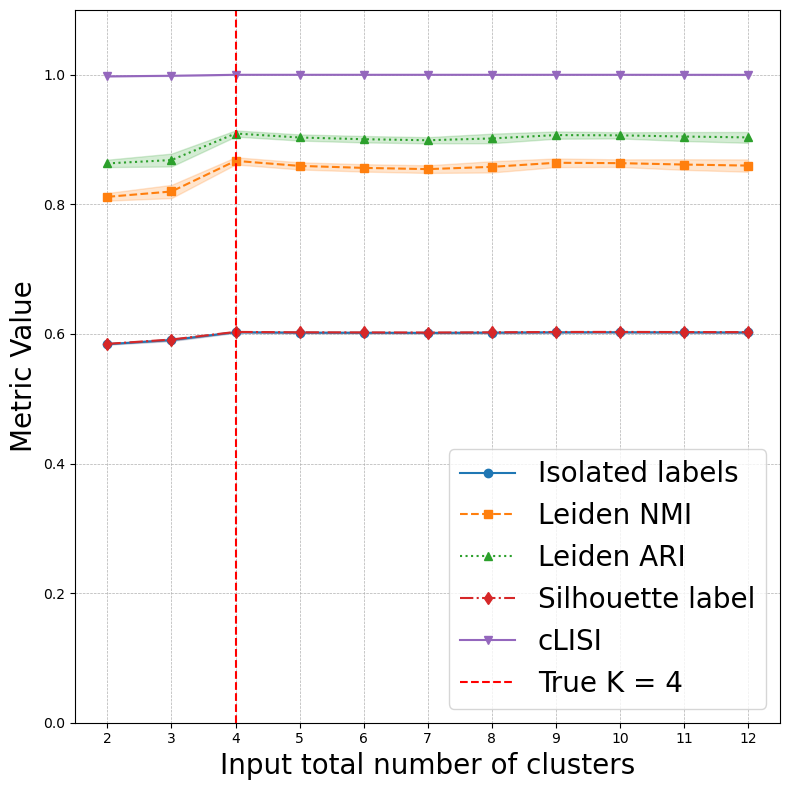}
\end{minipage}%
}%
\subfigure[Batch Correction Metrics]{
\begin{minipage}[t]{0.32\textwidth}
\centering
\includegraphics[width=\textwidth]{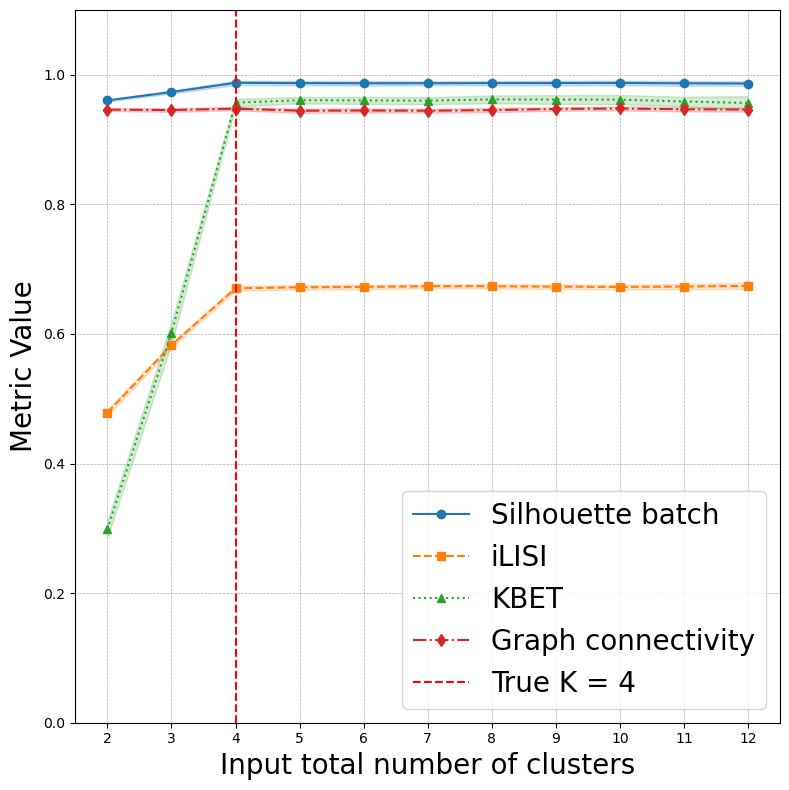}
\end{minipage}%
}%
\caption{The performance of {\method} when $u$, $v$ or input total number of clusters varies with data generated from a multivariate $t$ distribution mixture. 
In the top row and the left panel in the bottom row, we plot the logarithm of the average loss versus $\log(u)$, $v$, and the input total number of clusters, respectively (with the ends of the shaded area corresponding to the logarithms of the average loss $\pm$ one standard deviation). 
In the middle and the right panels of the bottom row, we plot the average scores of \texttt{scib-metrics} metrics versus the input number of clusters, and the shaded area around each data point represents $\pm$ one standard deviation. 
}
\label{fig: simulation with u v changes t distribution}
\end{figure}

The simulation results  are presented in Figure \ref{fig: simulation with u v changes t distribution}. Compared to the Gaussian setting (Section \ref{sec: simu}), the loss still converges to zero, but at a noticeably slower
rate. 
We also observe a sharp phase transition with respect to the parameter $v$; the batch correction loss initially increases for $v < 6$ before decreasing sharply thereafter. 
Despite this, the method demonstrates robustness, as a wide range of input total number of clusters continue to yield good performances.

\subsection{Performance with Missing Clusters}

This section evaluates the performance of {\method} in a simulation setting with an unbalanced batch composition, where one cluster is completely missing in one of the batches. This scenario is defined using the following mixture proportion matrix, $\Pi \in \mR^{B\times K}$:
\begin{equation*}
    \Pi = \begin{bmatrix}
        0.4 & 0.3 & 0.2 & 0.1 \\
        0.1 & 0.2 & 0.3 & 0.4 \\
        0.0 & 0.3 & 0.3 & 0.4
    \end{bmatrix}.
\end{equation*}
In this matrix, the entry $\Pi_{31}=0$ indicates that the first cluster ($k=1$) is entirely absent from the third batch ($b=3$).
For the study of $\log$(average loss) vs. $\log(u)$,  we fix $v = 5$, and for the study of $\log$(average loss) vs. $v$, we fix $u=\exp(1)$.
For the study with varying input total number of clusters, we fix $v = 5$ and $u=\exp(1)$.
All other simulation parameters remain the same as in Section \ref{sec: simu}.

\begin{figure}[!ht]
\centering
\subfigure[$\log$ loss vs. $\log(u)$]{
\begin{minipage}[t]{0.45\textwidth}
\centering
\includegraphics[width=\textwidth]{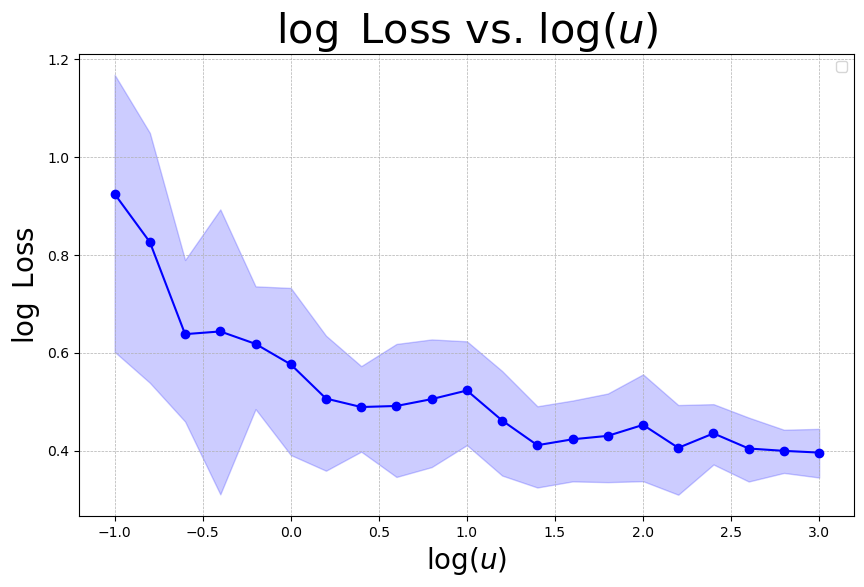}
\end{minipage}%
}%
\subfigure[$\log$ loss vs. $v$]{
\begin{minipage}[t]{0.45\textwidth}
\centering
\includegraphics[width=\textwidth]{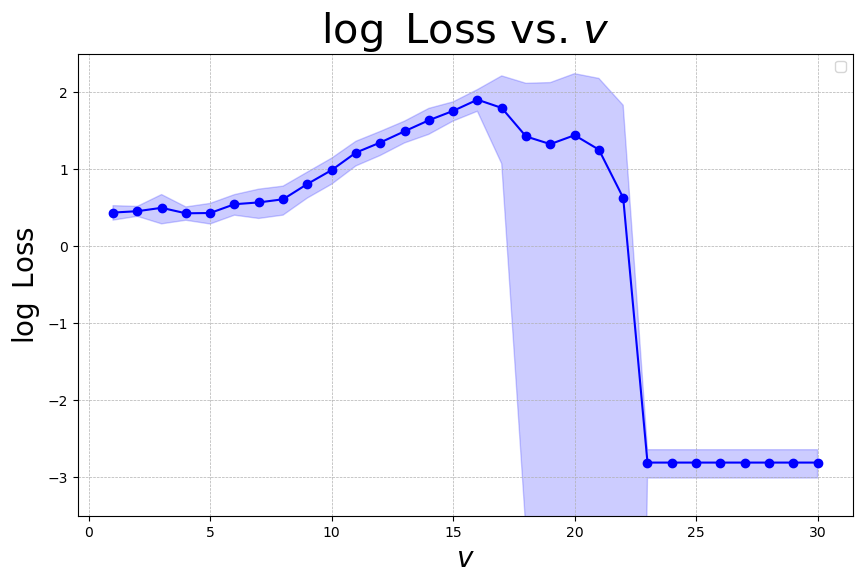}
\end{minipage}%
}%

\subfigure[Batch Correction Loss]{
\begin{minipage}[t]{0.32\textwidth}
\centering
\includegraphics[width=\textwidth]{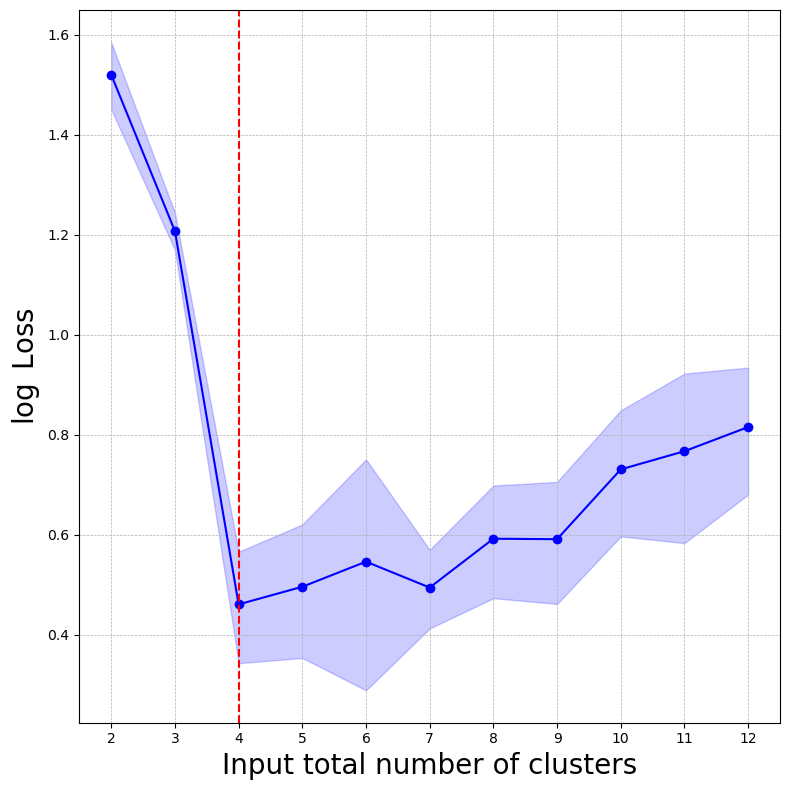}
\end{minipage}%
}%
\subfigure[Bio Conservation Metrics]{
\begin{minipage}[t]{0.32\textwidth}
\centering
\includegraphics[width=\textwidth]{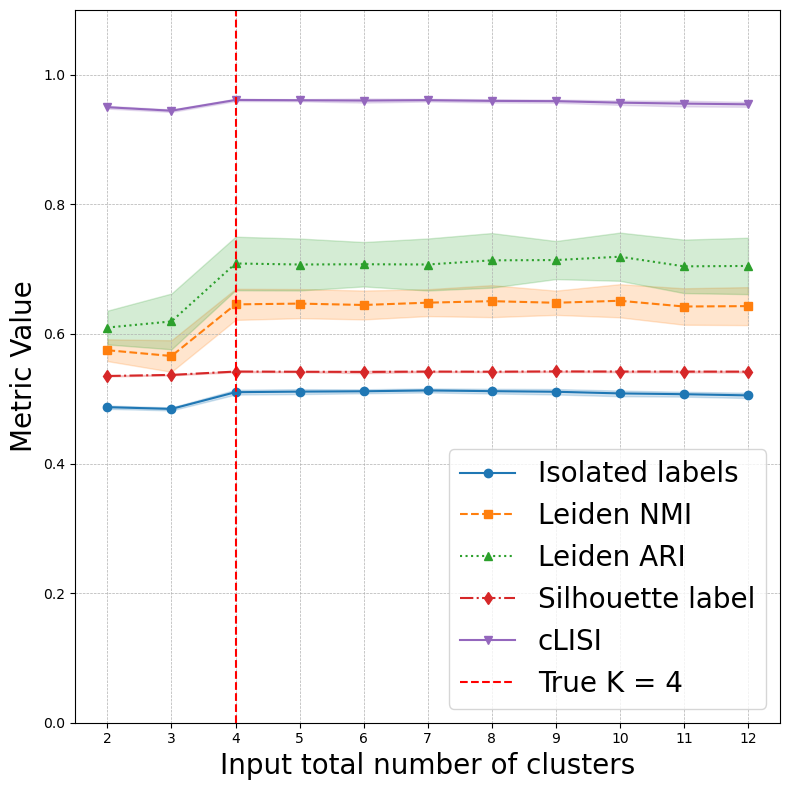}
\end{minipage}%
}%
\subfigure[Batch Correction Metrics]{
\begin{minipage}[t]{0.32\textwidth}
\centering
\includegraphics[width=\textwidth]{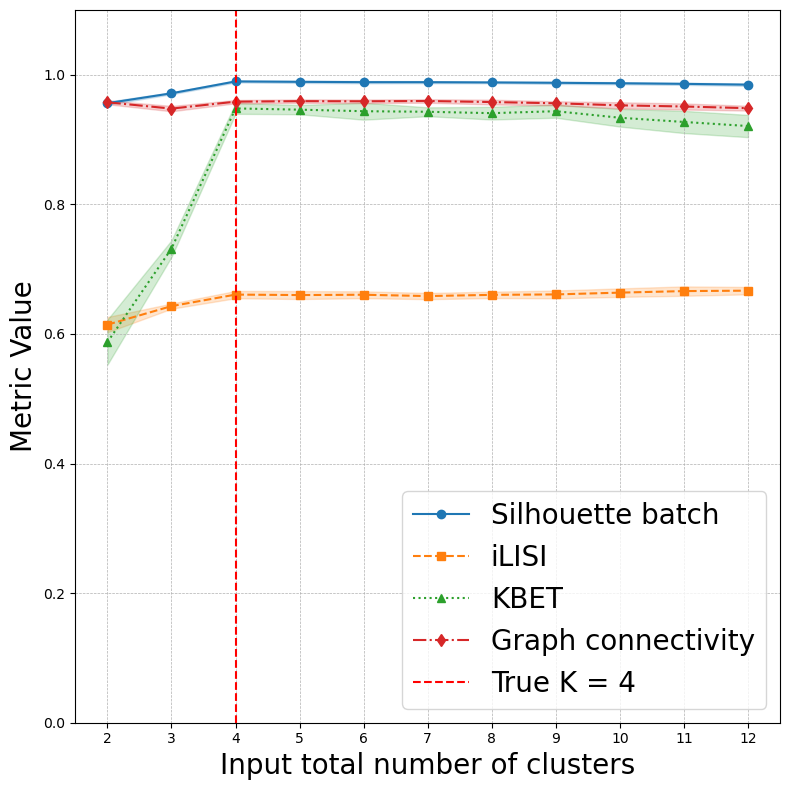}
\end{minipage}%
}%
\caption{The performance of {\method} when $u$, $v$ or input total number of clusters varies with data generated from a scenario with a missing cluster in a batch. 
In the top row and the left panel in the bottom row, we plot the logarithm of the average loss versus $\log(u)$, $v$, and the input total number of clusters, respectively (with the ends of the shaded area corresponding to the logarithms of the average loss $\pm$ one standard deviation). 
In the middle and the right panels of the bottom row, we plot the average scores of \texttt{scib-metrics} metrics versus the input number of clusters, and the shaded area around each data point represents $\pm$ one standard deviation. 
}
\label{fig: simulation with u v changes miss cluster}
\end{figure}

The simulation results for the unbalanced batch composition setting are presented in Figure \ref{fig: simulation with u v changes miss cluster}. Although the loss still converges to zero, its trend is markedly different from the regular settings studied in Section \ref{sec: simu}. 
We observe a complex phase transition with respect to the parameter $v$ governing the SNR: the loss is non-monotonic, increasing for values of $v$ up to approximately 16 before exhibiting a sharp decrease around $v=22$. 
This phase transition suggests that a strong SNR may be necessary to detect the absence of a cluster.
Despite the phase transition behavior, the method remains robust to the input total number of clusters, with a wide range of input values yielding comparable performance.

\section{Supplementary Results for Single-Cell Data Analysis}

\label{sec: sup real world}

This section provides supplementary materials for batch correction in single-cell datasets described in Section \ref{sec: real world}. 
First, we provide the contingency table between cell types and batche labels for the mouse PBMC dataset. 
Second, we show that the performance of {\method} is robust with respect to the choice of the number of principal components in preprocessing. 
Finally, we provide UMAP visualizations of all datasets before and after batch correction by different methods in comparison, augmenting the quantitative comparisons in Section \ref{sec: real world}.

\subsection{Composition Details of the Mouse PBMC Dataset}

We provide the contingency table between cell types and batches in the mouse PBMC dataset in Table \ref{tab: pbmc composition}. 
We can observe from the table that the batch sizes and cell type sizes are highly unbalanced, with some batches much larger in sample sizes and with some cell types completely missing in some batches.

\begin{table}
    \centering
    \begin{tabular}{c|cccccc|c}
    \toprule
          &  Batch 1 &  Batch 2 &  Batch 3 &  Batch 4 &  Batch 5 &  Batch 6 & Total\\
         \hline
        B cell &	82&	225&	1005&	14&	21&	64&	1411\\
        Basophil&	0&	55&	3&	0&	0&	0&	58\\
        Dendritic cell&	0&	71&	1&	0&	2&	0&	74\\
        Erythroblast&	0&	93&	17&	0&	0&	4&	114\\
        Macrophage&	39&	136&	686&	82&	187&	50&	1180\\
        Monocyte&	0&	358&	0&	0&	1&	0&	359\\
        NK cell&	13&	12&	163&	20&	74&	37&	319\\
        Neutrophil&	14&	1469&	346&	0&	17&	46&	1892\\
        T cell&	135&	47&	980&	19&	50&	457&	1688\\
        \hline
        Total&	283&	2466&	3201&	135&	352&	658&	7095\\
        \bottomrule
    \end{tabular}
    \caption{Cell counts by cell type annotations and batches in the mouse PBMC dataset.}
    \label{tab: pbmc composition}
\end{table}

\subsection{Robustness to The Number of Principal Components Retained}

\begin{figure}[!ht]
\centering
\includegraphics[width=\textwidth]{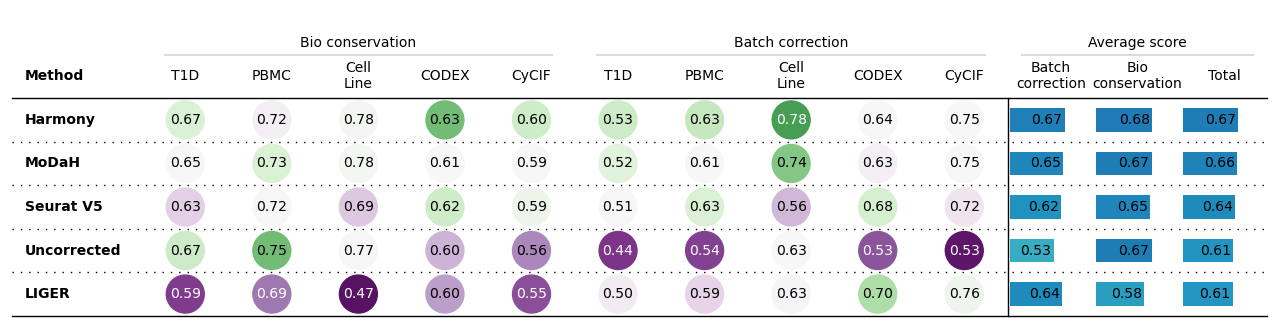}
\caption{Summary of performance scores of all methods in comparison on five datasets with top $25$ PCs retained after preprocessing. 
Scores in circles represent the average scores of five ``bio conservation'' metrics and of four ``batch correction'' metrics in individual datasets. 
Scores in bars show the final averages of metric scores within and across the two categories over datasets. All scores are rescaled to a [0, 1] range, with higher scores corresponding to better performances.
}
\label{fig: average scores 25}
\end{figure}

\begin{figure}[ht]
\centering
\subfigure[T1D dataset]{
\begin{minipage}[t]{.8\textwidth}
\centering
\includegraphics[width=\textwidth]{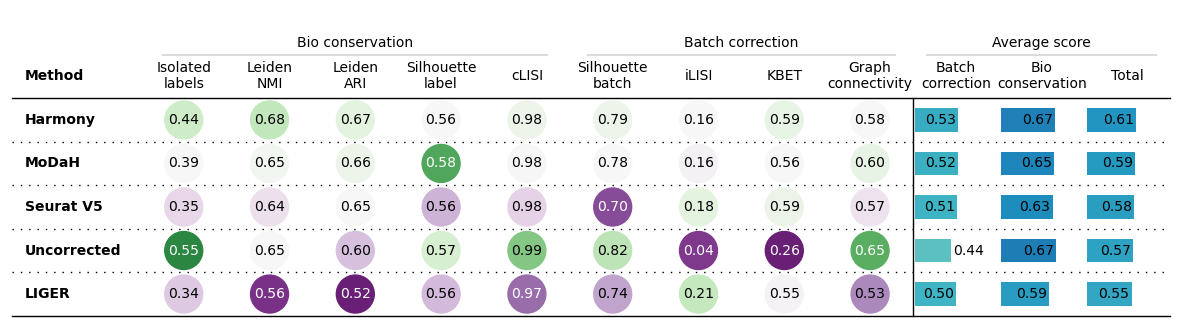}
\end{minipage}%
}%

\subfigure[PBMC dataset]{
\begin{minipage}[t]{.8\textwidth}
\centering
\includegraphics[width=\textwidth]{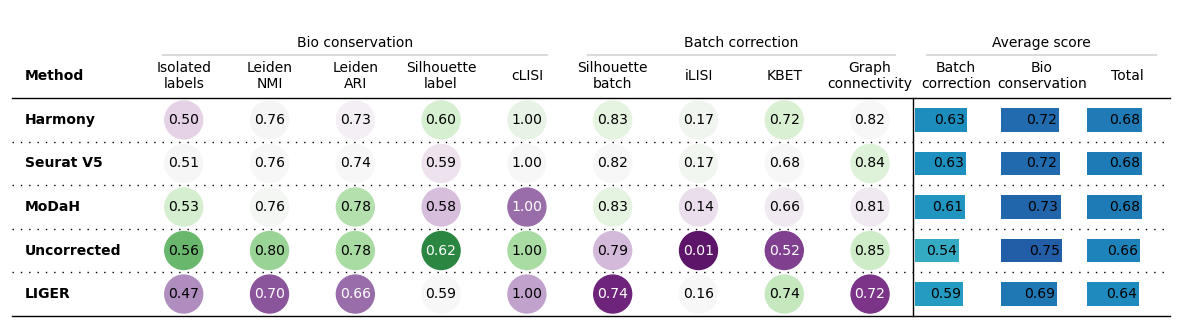}
\end{minipage}%
}%

\subfigure[Cell Line dataset]{
\begin{minipage}[t]{.8\textwidth}
\centering
\includegraphics[width=\textwidth]{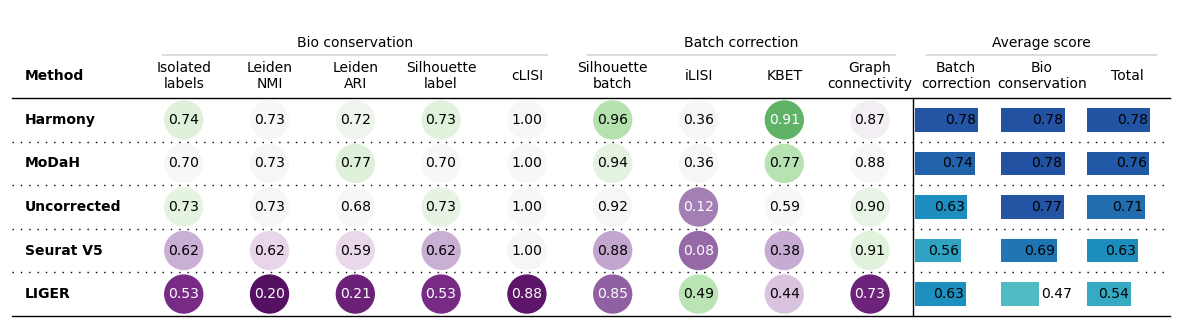}
\end{minipage}%
}%

\subfigure[CODEX dataset]{
\begin{minipage}[t]{.8\textwidth}
\centering
\includegraphics[width=\textwidth]{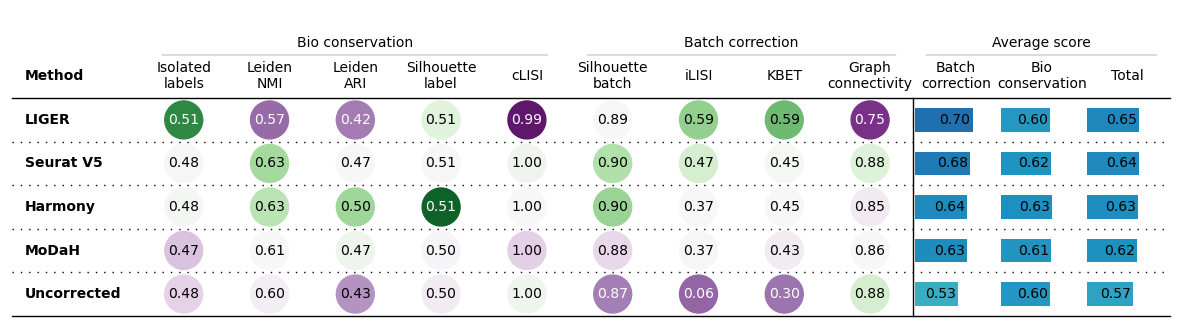}
\end{minipage}%
}%

\subfigure[CyCIF dataset]{
\begin{minipage}[t]{.8\textwidth}
\centering
\includegraphics[width=\textwidth]{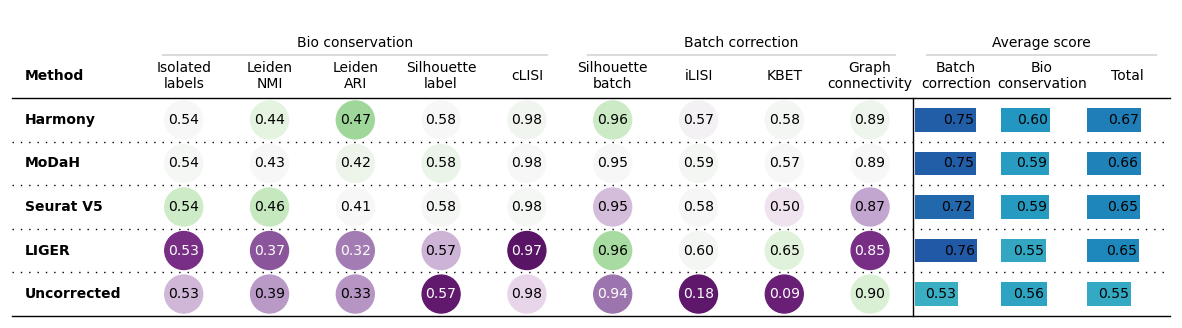}
\end{minipage}%
}%

\caption{
Performance scores of all methods in comparison on five datasets with top $25$ PCs retained after preprocessing. 
Scores in bars show the averages of metric scores within and across the ``bio conservation'' and the ``batch-correction'' categories.
All scores are rescaled to a [0, 1] range, with higher scores indicating better performances. 
}
\label{fig: 25 pcs}
\end{figure}

Here, we check if the methods' performance is robust with respect to the number of principal components (PCs) used for the analysis. 
We re-ran the experiments in Section \ref{sec: real world} on all five datasets, this time using the top $25$ PCs instead of the top $20$.
The results are summarized in Figures \ref{fig: average scores 25} and \ref{fig: 25 pcs}. 
The results are 
similar to those in Section \ref{sec: real world}, with Harmony and {\method} leading in overall performance, further demonstrating the robustness of {\method}.

\subsection{UMAP Visualizations}

\begin{figure}[ht]
\centering
\includegraphics[width=\linewidth]{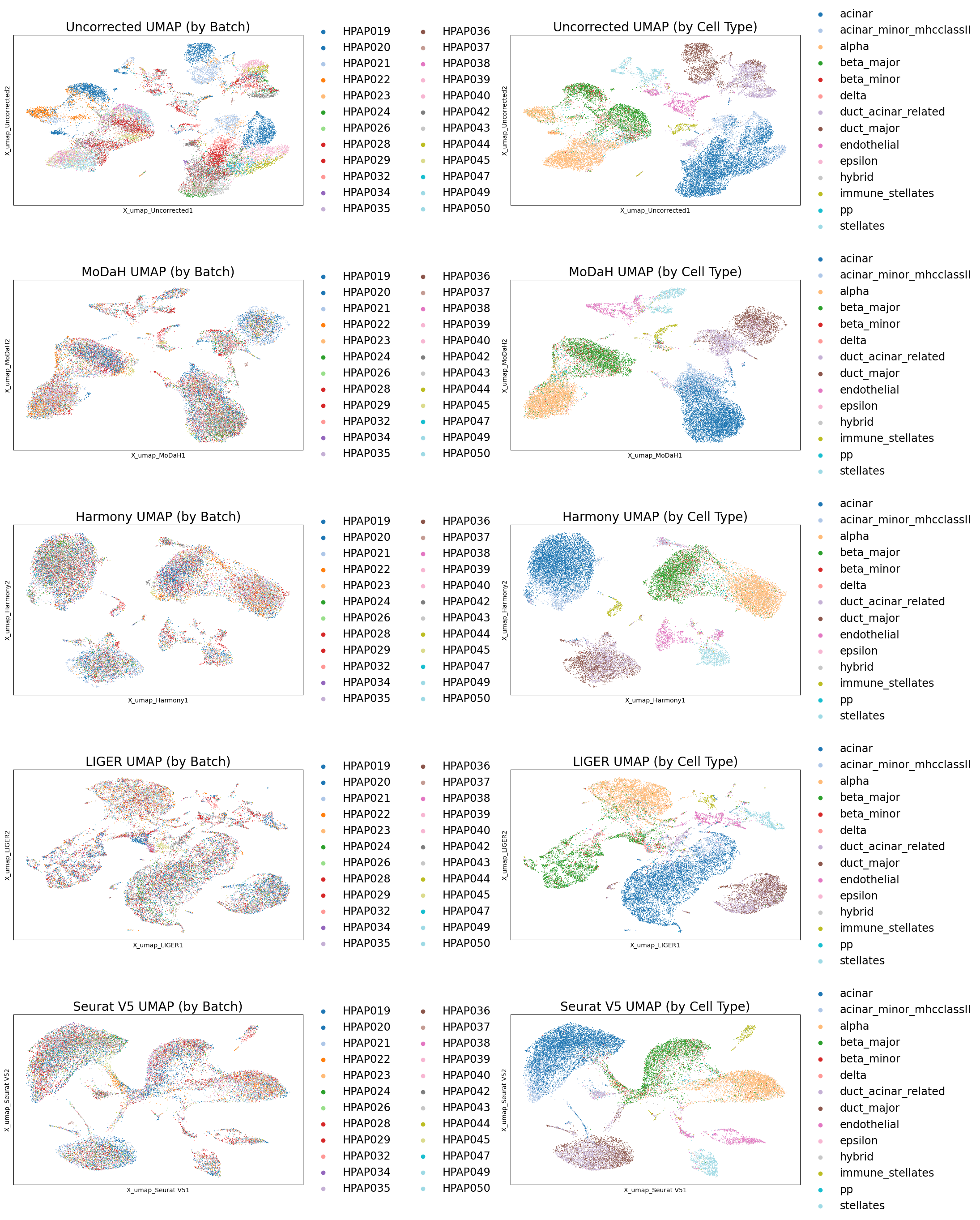}
\caption{UMAP visualizations of the human T1D dataset before and after batch correction by different methods. 
Each row displays the result for a different method, ordered from top to bottom: Uncorrected (i.e.~before any batch correction), \method, Harmony, LIGER, and Seurat V5. Plots in the left column show cells colored by batch identity to visualize the extent of batch correction. 
Plots in the right column show cells colored by annotated cell types to visualize the conservation of biological structure.}
\label{fig: t1d umap}
\end{figure}

\begin{figure}[ht]
\centering
\includegraphics[width=\linewidth]{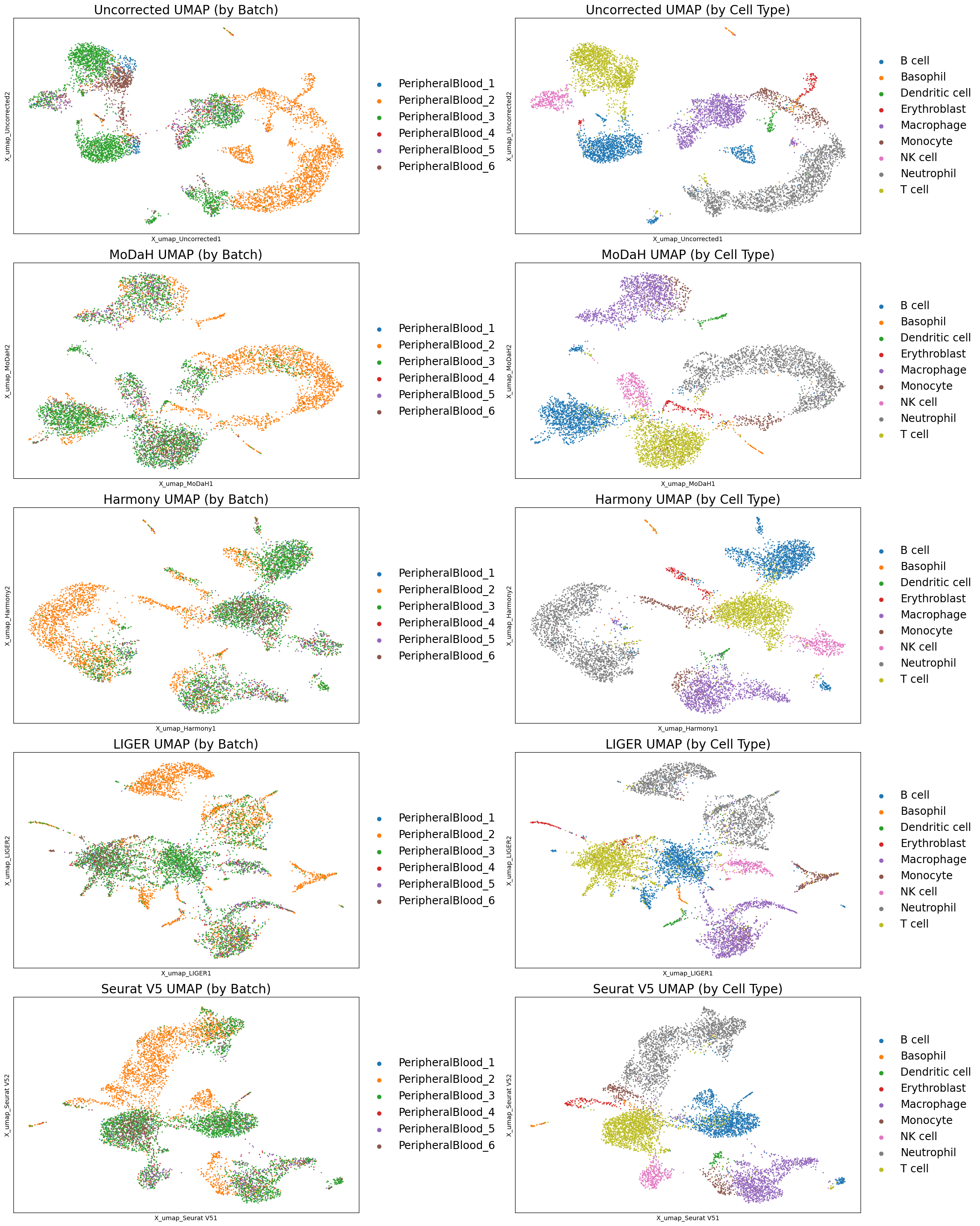}
\caption{UMAP visualizations of the mouse PBMC dataset before and after batch correction by different methods. 
Each row displays the result for a different method, ordered from top to bottom: Uncorrected (i.e.~before any batch correction), \method, Harmony, LIGER, and Seurat V5. Plots in the left column show cells colored by batch identity to visualize the extent of batch correction. 
Plots in the right column show cells colored by annotated cell types to visualize the conservation of biological structure.
}
\label{fig: pbmc umap}
\end{figure}

\begin{figure}[ht]
\centering
\includegraphics[width=\linewidth]{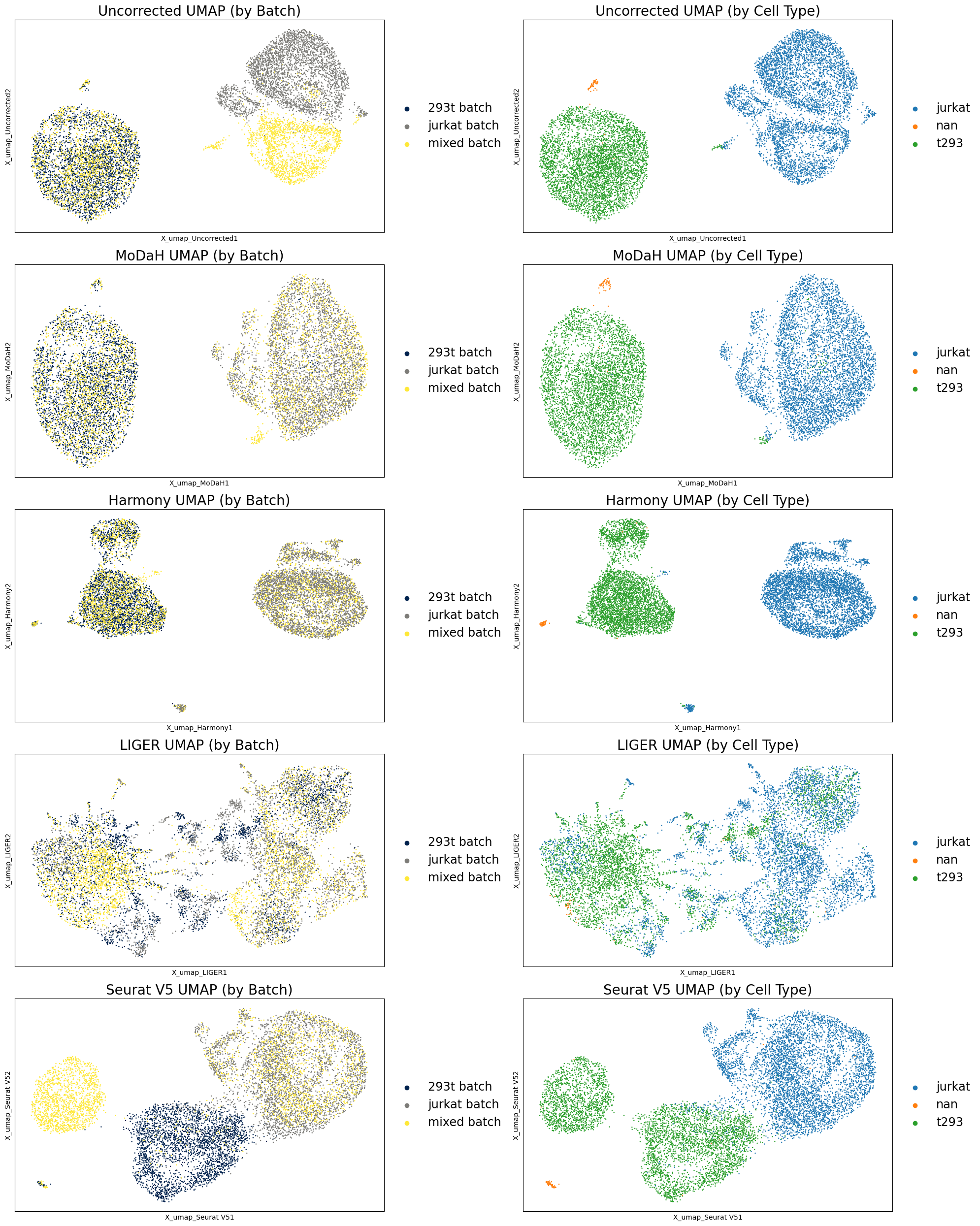}
\caption{UMAP visualizations of the cell line dataset before and after batch correction by different methods. 
Each row displays the result for a different method, ordered from top to bottom: Uncorrected (i.e.~before any batch correction), \method, Harmony, LIGER, and Seurat V5. Plots in the left column show cells colored by batch identity to visualize the extent of batch correction. 
Plots in the right column show cells colored by annotated cell types to visualize the conservation of biological structure.
}
\label{fig: cell line umap}
\end{figure}

\begin{figure}[ht]
\centering
\includegraphics[width=\linewidth]{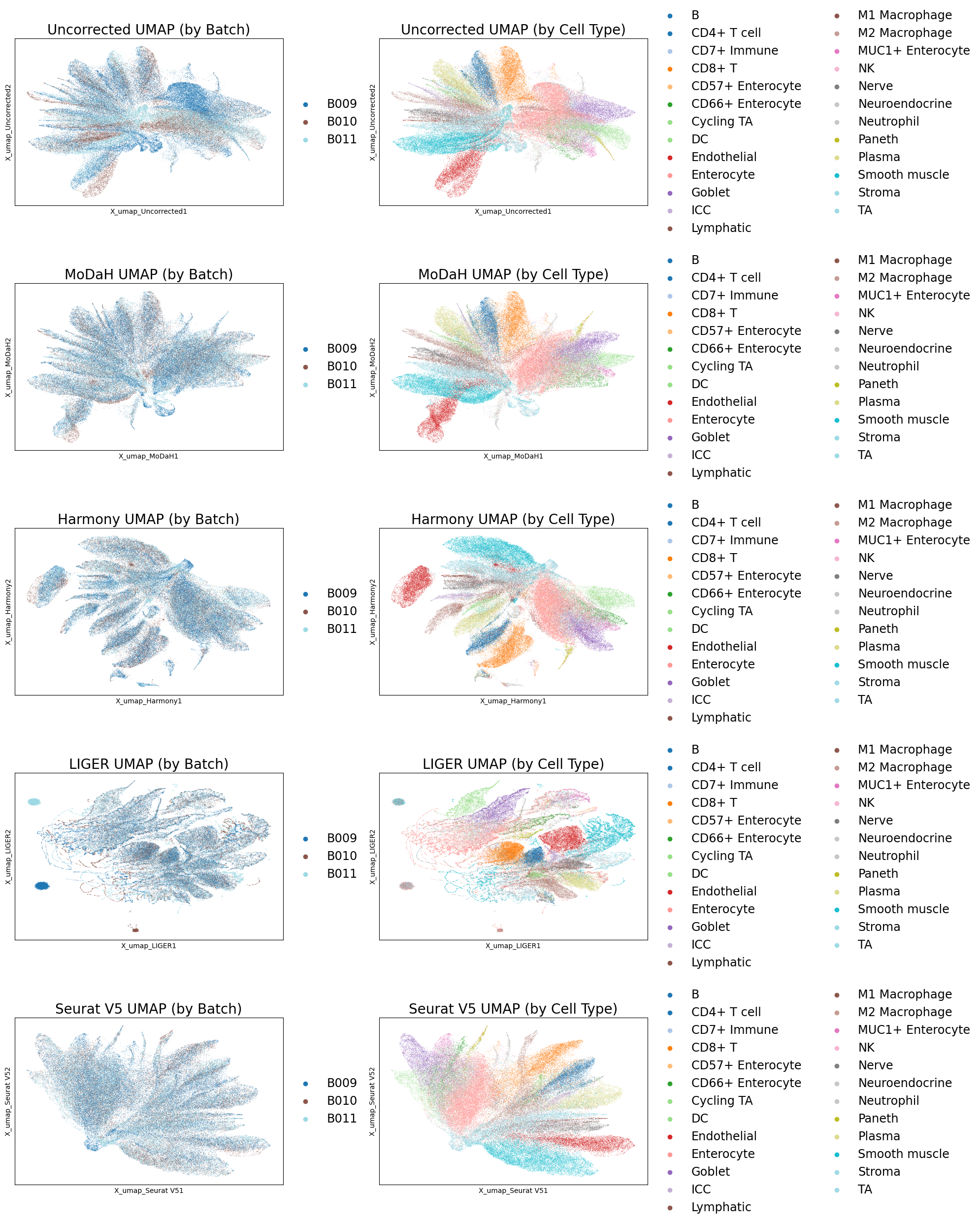}
\caption{UMAP visualizations of the healthy human intestine CODEX dataset before and after batch correction by different methods. 
Each row displays the result for a different method, ordered from top to bottom: Uncorrected (i.e.~before any batch correction), \method, Harmony, LIGER, and Seurat V5. Plots in the left column show cells colored by batch identity to visualize the extent of batch correction. 
Plots in the right column show cells colored by annotated cell types to visualize the conservation of biological structure.
}
\label{fig: codex umap}
\end{figure}

\begin{figure}[ht]
\centering
\includegraphics[width=\linewidth]{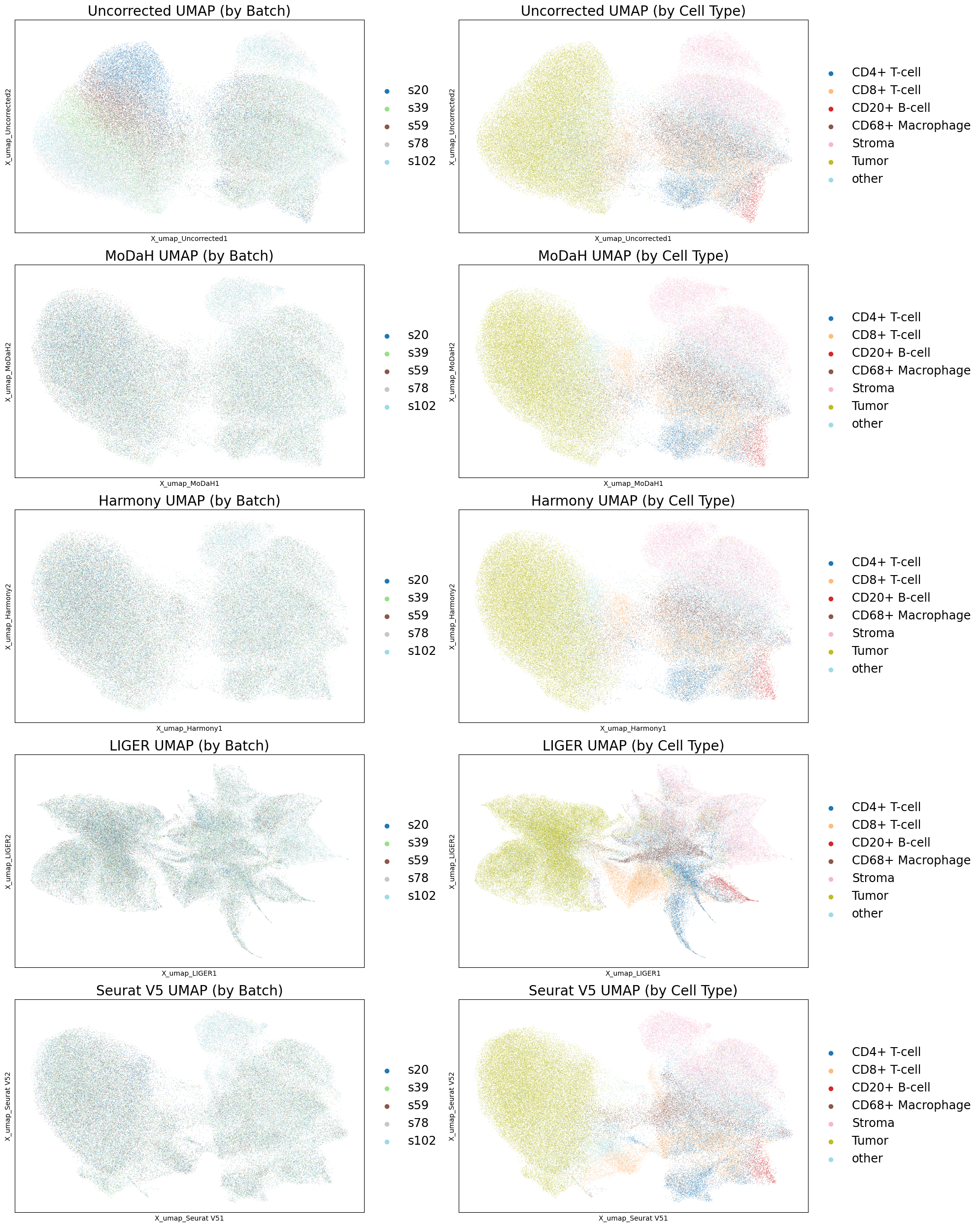}
\caption{UMAP visualizations of the human colorectal cancer CyCIF dataset before and after batch correction by different methods. 
Each row displays the result for a different method, ordered from top to bottom: Uncorrected (i.e.~before any batch correction), \method, Harmony, LIGER, and Seurat V5. Plots in the left column show cells colored by batch identity to visualize the extent of batch correction. 
Plots in the right column show cells colored by annotated cell types to visualize the conservation of biological structure.
}
\label{fig: cycif umap}
\end{figure}

Figures \ref{fig: t1d umap}--\ref{fig: cycif umap} plot UMAP visualizations of the five single-cell datasets before and after batch correction by different methods, providing a qualitative assessment of each method's performance and complementing the quantitative assessment in Section \ref{sec: real world}.
These plots show that {\method} consistently  achieves an excellent balance between batch correction and bio-conservation, comparable to state-of-the-art empirical methods. 

\end{document}